\documentclass[twocolumn]{aastex631}

\usepackage{epsfig}
\usepackage{booktabs}
\usepackage{microtype}

\submitjournal{ApJ}

\shorttitle{Star Formation History of NGC\,6822}
\shortauthors{Khatamsaz et al. 2025}

\begin{document}

\correspondingauthor{Atefeh Javadi}
\email{atefeh@ipm.ir}

\title{A Study on the Evolution of NGC\,6822. I. Investigating the Star Formation History through Evolved Stellar Populations}

\author[0009-0007-6135-1145]{Fatemeh Khatamsaz}
\affil{School of Astronomy, Institute for Research in Fundamental Sciences (IPM), Tehran, 19568-36613, Iran}

\author[0009-0006-3280-5622]{Mahdi Abdollahi}
\affil{School of Astronomy, Institute for Research in Fundamental Sciences (IPM), Tehran, 19568-36613, Iran}

\author[0000-0001-8392-6754]{Atefeh Javadi}
\affil{School of Astronomy, Institute for Research in Fundamental Sciences (IPM), Tehran, 19568-36613, Iran}

\author[0009-0002-9957-5818]{Hamidreza Mahani}
\affil{School of Astronomy, Institute for Research in Fundamental Sciences (IPM), Tehran, 19568-36613, Iran}

\author[0000-0002-1272-3017]{Jacco Th. van Loon}
\affiliation{Astrophysics Group, Lennard-Jones Laboratories, Keele University, Staffordshire ST5 5BG, United Kingdom}

\author[0000-0002-7823-7169]{Hedieh Abdollahi}
\affil{School of Astronomy, Institute for Research in Fundamental Sciences (IPM), Tehran, 19568-36613, Iran}
\affil{Konkoly Observatory, HUN-REN Research Centre for Astronomy and Earth Sciences, MTA Centre of Excellence, Konkoly-Thege Mikl\'os\'ut 15-17, H-1121, Budapest, Hungary}

\author[0000-0001-5855-846X]{Sepideh Ghaziasgar}
\affil{School of Astronomy, Institute for Research in Fundamental Sciences (IPM), Tehran, 19568-36613, Iran}

\begin{abstract}

NGC\,6822 is an isolated dwarf irregular galaxy in the Local Group at a distance of  $\sim$ 490 kpc . In this paper, we derived the star formation history (SFH) of NGC\,6822 employing a method based on evolved asymptotic giant branch (AGB) stars, known as long-period variable (LPV) stars. We utilized a dataset of 329 LPVs in JHK$_s$ bands to estimate the star formation rate (SFR) over time and reconstructed the SFH of the galaxy. In addition to obtaining the SFH assuming constant metallicity values within the range of 0.0001 $<$ Z $<$ 0.012, we also adopted three distinct age-metallicity relations (AMRs) to account for variations in the chemical content of the galaxy throughout its lifetime. The SFH has been investigated in two regions. The bar region encloses a central area of 189 arcmin$^{2}$, and the outer region encircles a field beyond the bar region, extending to the radial distance of 3 kpc. In the bar region, we identified three episodes of star formation peaking at $t \sim$ 4.7 Gyr (log $t = 9.67$), $t \sim$ 1.6 Gyr (log $t = 9.21$), and $t \sim$ 48 Myr (log $t = 7.63$), where $t$ is the look-back time. In the outer region, we found one episode of star formation occurring at $t \sim$ 4.6 Gyr. The significant peak observed $\sim$ 4--5 Gyr ago further supports the hypothesis suggested by the previous research, indicating that NGC\,6822 passed through the virial radius of the Milky Way in the past.
\end{abstract}


\section{Introduction} \label{sec:sec1}

The Local Group is home to over 80 galaxies, mainly consisting of small dwarfs. These dwarfs can be divided into two main groups: gas-rich and active dwarfs, including dwarf irregulars (dIrrs) and blue compact dwarfs (BCDs); and gas-poor and quenched dwarfs, containing dwarf spheroidals (dSphs), dwarf ellipticals (dEs) and ultra-faint dwarfs (UFDs) (\citealp{arp1975properties}; \citealp{hunter1982global}; \citealp{geha2012stellar}). DIrrs are ideal subjects for detailed studies on galaxy evolution, star formation, and galactic interactions due to their vicinity and the absence of the complexity of spiral arms and density waves, which make them relatively simple systems compared to large spiral galaxies (\citealp{mateo1998dwarf}; \citealp{cannon2011m81}).

Among the Local Group dIrrs, NGC 6822, also known as Barnard’s galaxy, has been the focus of interest because of the recent, pronounced episode of star formation detected in its central regions \citep{wyder2001star, gallart1996local3,efremova2011recent}. Given the galaxy’s well-known isolation, the origin of this activity has been the subject of considerable debate. The morphology of NGC 6822, characterized by a central bar, an irregular HI envelope, and an extended halo, has been interpreted as possible evidence of a past interaction \citep{zhang2021panoramic}. As a nearby dwarf irregular system, NGC 6822 provides a valuable opportunity for a detailed examination of these processes, owing to its proximity and well-resolved stellar populations.

\begin{figure*}[ht!]
\hspace{25mm}
	{\hbox{
        \epsfig{figure=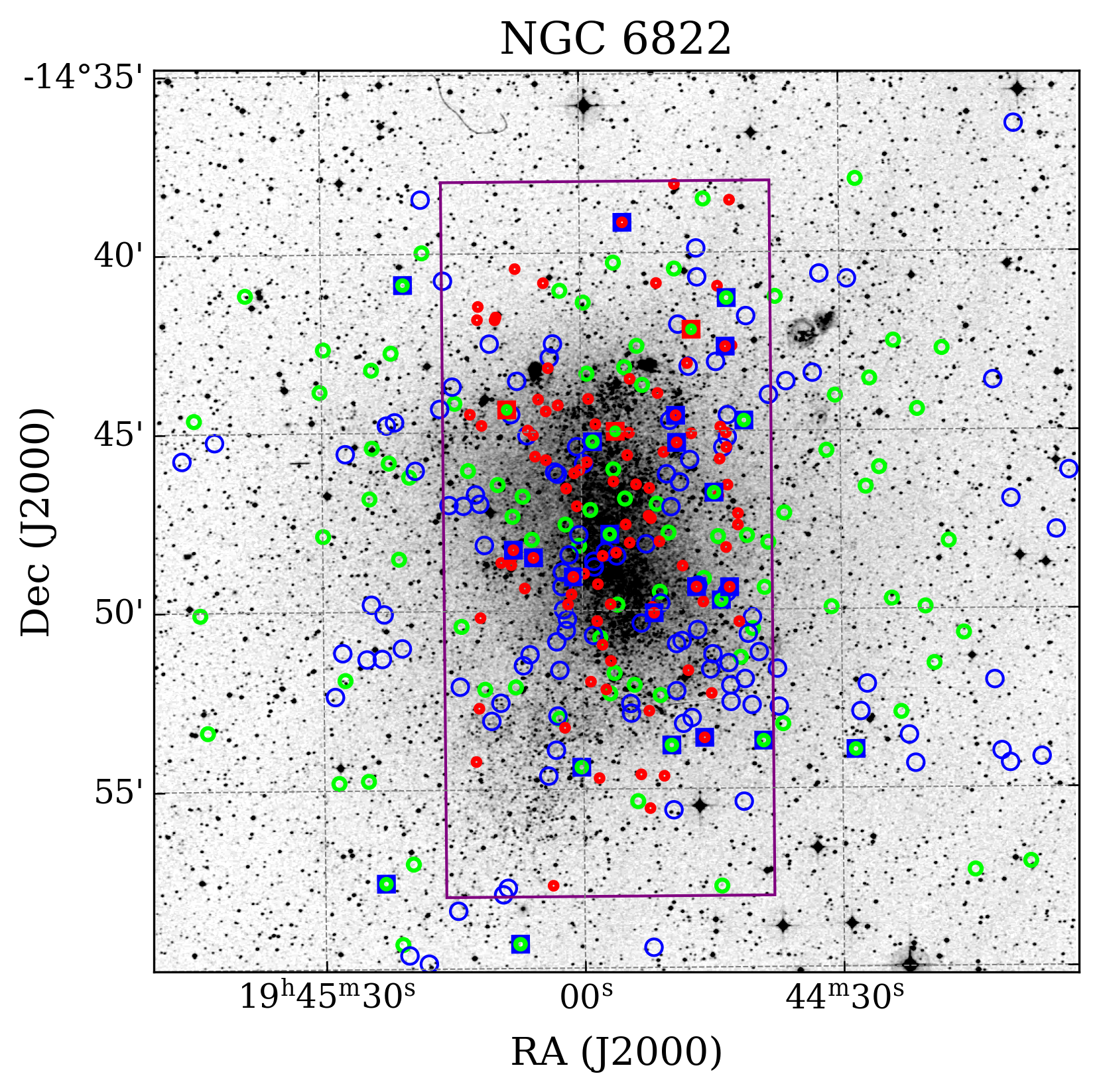,width=120mm,height=120mm}
   \centering
	}}
	\caption{The Optical image of NGC\,6822 taken by \textit{UK Schmidt Telescope} (Digitized Sky Survey). The red dots signify the 97 LPV stars collected from \cite{whitelock2013local}. The purple rectangle encloses the bar region, where the 97 LPVs from \cite{whitelock2013local} and 131 spectroscopy-confirmed carbon stars from \cite{sibbons2015spectral} (green circles) and \cite{kacharov2012spectra} (blue circles) are located. The remaining 101 spectroscopy-confirmed C-rich stars are scattered beyond the defined bar region, extended to a radius of 3 kpc.  Stars that appear in more than one catalog are highlighted with square markers in the corresponding colors of each dataset.} 
	\label{fig:fig1} 
\end{figure*}

NGC\,6822, situated at $\alpha = 19^{h}$ $44^{m}$ and $\delta = -14^{\circ}$ $48^{\prime}$, is $\sim$ 490 kpc away with a distance modulus of $\mu = 23.45 \pm 0.15 $ mag (\citealp{lee1993tip}; \citealp{mateo1998dwarf}). It is located in the constellation Sagittarius with the low galactic latitude of $b = -18.39^{\circ}$, which affects our line of sight by the foreground reddening of E(B$-$V) $=$ 0.24 mag (\citealp{gallart1996local}; \citealp{schlegel1998maps}). The stellar mass of the galaxy is estimated to be $\sim 1-2 \times 10^8$ M$_\sun$ (\citealp{carigi2006chemical}; \citealp{jones2019young}; \citealp{belland2020ngc}), while its total mass falls within the range of $\sim 0.3-4 \times 10^{10}$ M$_\sun$. This substantial dissimilarity between the stellar and total mass suggests that NGC\,6822 is a dark matter-dominated galaxy (\citealp{carigi2006chemical}; \citealp{valenzuela2007there}; \citealp{veljanoski2015globular}). However, if the galaxy is not in a dynamical equilibrium state, such high measures of total mass would be an overestimation.

\begin{table*}
\centering
\begin{tabular}{llll}
\toprule
Tracer                  & Age (Gyr)      & {[}Fe/H{]} (dex)   & Reference                \\ \midrule
A supergiant spectra    & $\leq$ 0.05 & $-$0.49 $\pm$ 0.22 & \cite{venn2001first}     \\ [0.1cm]
B supergiant spectra    & $\leq$ 0.05 & $-$0.50         & \cite{muschielok1999vlt} \\ [0.1cm]
RSG (J band spectra)      & $\leq$ 0.05 & $-$0.52 $\pm$ 0.21 & \cite{patrick2015red}    \\ [0.1cm]
Slope of RGB (J, K bands) & $\sim$ 3.00  & $-$1.00 $\pm$ 0.30   & \cite{davidge2003metallicity}          \\ [0.1cm]
RGB Ca II triplet spectra       & $\sim$ 2.00 -- 5.00 & $-$1.00 $\pm$ 0.50   & \cite{tolstoy2001using}     \\ [0.1cm]
RGB Fe I spectra        & $\sim$ 10.00 & $-$1.05 $\pm$ 0.01 & \cite{kirby2013universal}     \\ [0.1cm]
CMD fitting of Globular Clusters        & $\sim$ 10.00 & $-$2.30         & \cite{hwang2011extended}       \\ [0.1cm] 
Spectroscopy of Extended Star Cluster        & $\sim$ 10.00 & $-$2.53         & \cite{hwang2014spectroscopic}       \\ [0.1cm] \bottomrule
\end{tabular}
\caption{\small Summary of metallicity measurements reported by previous studies, utilizing various tracers that cover different age ranges and different components of NGC~6822. The young stellar populations (e.g., supergiants) trace the metallicity at recent epochs and are mainly concentrated on the central bar and HI envelope. Whereas the intermediate and old populations (e.g.,  RGB stars and globular clusters) trace the metallicity at early epochs and are distributed within the halo (\citealp{zhang2021panoramic}). Since metallicity changes over time with the evolution of the galaxy, and the parameters we used to calculate the SFR are sensitive to these changes, we adopted a range of metallicities based on the highest and lowest reported values.}
\label{tab:tab1}
\end{table*}

NGC\,6822, with an optical radius of $\sim$ 1.2 kpc, similar to the Small Magellanic Cloud (SMC), has an optically bright bar-shaped center, approximately oriented north-south that harbors most of the active star-forming complexes (\citealp{hodge1977structure}, \citeyear{hodge1991cosmos}). The optical bar is embedded in an HI envelope, which appears like an asymmetric disk-like structure extending from northwest (NW) to southeast (SE), where a tidal arm is present. The HI region extends up to a radius of  $\sim$  6 kpc, well beyond the optical structure of the galaxy. The young stellar population is mainly distributed along the bar and the HI region, while the old- and intermediate-age populations are located in an elliptical spheroid oriented perpendicular to the major axis of the HI region (\citealp{de2000evidence}; \citealp{zhang2021panoramic}). 

Previously, \cite{battinelli2006photometric} estimated that this spheroid extends to a scale length of $\sim$ 5.2 kpc, based on the surface density profile of red-giant branch (RGB) stars. However, \cite{hwang2005discovery} discovered distant globular clusters at $\sim$ 12 kpc from the center, indicating that the spheroid is much more extended than previously measured. NGC\,6822 is classified as a barred irregular galaxy, IB(s)m (\citealp{de1991third}); nonetheless, the significant similarities in its structural characteristics with other types of galaxies cannot be overlooked. The presence of a central bar, an elongated bulge with a scale length of $\sim$ 560 pc (\citealp{battinelli2006photometric}), and an extensive halo, which harbors the old- to intermediate-age stars, suggests that NGC\,6822 might be a barred spiral galaxy; however, the absence of prominent spiral arms and the low metallicity of the galaxy cast doubt on such classification (\citealp{hubble1982realm}; \citealp{de1991third}). The exotic and twisted structure of NGC\,6822 raises questions regarding its dynamical evolution and whether the irregularity is intrinsic to the galaxy or may have been caused by past interactions.

NGC\,6822 is a galaxy known for its relatively low metallicity. The mean value of its metallicity, [Fe/H]$=-$1.0 dex ($Z=0.003$) (\citealp{navabi2025outside}), is comparable to the mean metallicity for the SMC, [Fe/H]$=-$0.94 dex ($Z=0.004$), derived from the RGB population (\citealp{choudhury2018photometric}), with no significant radial metallicity gradient observed across the galaxy (\citealp{venn2001first}; \citealp{sibbons2012agb}; \citealp{hwang2014spectroscopic}).
The principal adopted quantity in this work is the iron abundance, [Fe/H]; however, the corresponding Z value is also reported (see Appendix \ref{appnx:metallicity}) for readers who prefer this notation.
\cite{davidge2003metallicity} derived a mean value of [Fe/H] = $-$1.0 $\pm$ 0.3 dex (Z $\approx$ 0.003), confirmed by \cite{tolstoy2001using}’s and \cite{kirby2013universal}’s estimations. \cite{hwang2011extended} determined a value of [Fe/H] = $-$0.7 dex (Z $\approx$ 0.006) for the present-day metallicity, which is roughly consistent with the Z = 0.004 reported by \cite{gallart1996local}, estimated based on the HII region. The highest reported value of metallicity for NGC\,6822 is $\sim$ [Fe/H] = $-$0.49 dex (Z $\approx$ 0.012) (\citealp{muschielok1999vlt}; \citealp{tolstoy2001using}; \citealp{venn2001first}; \citealp{patrick2015red}). 

Table\,1 presents a summary of the aforementioned metallicity values reported by previous studies. As shown, values are derived using several stellar tracers associated with stellar populations of different ages, demonstrating the variation of metallicity throughout the evolutionary history of the galaxy. A number of these measurements were obtained using spectroscopic techniques applied to different stellar groups and evolutionary branches, such as A and B supergiants (\citealp{venn2001first}; \citealp{muschielok1999vlt}), red supergiants (RSGs; \citealp{patrick2015red}), and RGB stars through Ca II triplet and Fe I spectroscopy (\citealp{tolstoy2001using}; \citealp{kirby2013universal}).  Metallicities are also available for the
four extended star clusters SC1--SC4 of NGC\,6822, so called because their
half-light radii of 7.5--14 pc are larger than those of typical Galactic
globular clusters; these clusters lie in the halo of the galaxy, at
projected galactocentric distances of 1.5--11 kpc. Their metallicities were
first obtained by fitting isochrones to their colour--magnitude diagrams
\citep{hwang2011extended}, and the clusters were later observed
spectroscopically by \citet{hwang2014spectroscopic}, who derived older ages
and metallicities of [Fe/H] $\lesssim -$1.5 dex. Additionally, \citet{davidge2003metallicity} estimated the metallicity based on the slope of the RGB in the J- and K-band.

Early attempts in this millennium to recover SFHs from integrated light relied on parametric or compressed-data methods. These approaches invert spectra or broad-band spectral energy distributions to trace how stellar mass assembled over time. One of the first widely used techniques was the Multiple Optimized Parameter Estimation and Data compression (MOPED) framework, introduced by \cite{Reichardt01}. MOPED compresses galaxy spectra into a small set of numbers that still preserve the information needed to recover non-parametric SFHs and related parameters such as age, metallicity, and dust content. This method was successfully applied to real galaxy spectra soon after \citep{Panter03}.

In the following decade, several full-spectral-fitting tools were developed. STARLIGHT (\citealp{Fernandes05}) fits observed spectra with combinations of single-stellar-population templates, producing light- and mass-fraction SFHs for large surveys such as SDSS. While powerful, these methods also revealed limitations due to degeneracies between age, metallicity, and dust (\citealp{Richards09}). Another example is VErsatile SPectral Analysis (VESPA), which introduced improved age-binning strategies and error propagation, allowing more controlled SFH reconstructions for large samples (\citealp{Tojeiro07}).

In this work, we take a different approach, proposed by \cite{Javadi11, Javadi17} to obtain the SFH. This method considers the relation between various parameters such as luminosity, birth mass, age, and pulsation duration of evolved asymptotic giant branch (\citealp{marigo2008evolution}) stars to estimate the star formation rate (SFR) and reconstruct the SFH of NGC\,6822.

AGB stars and RSGs, with masses up to $\sim$ 30 M$_\sun$, represent the final stages of stellar evolution. Spanning a wide range of ages, from $\sim$ 10 Myr to 10 Gyr, these stars are valuable tools for tracing the recent to old SFH in galaxies (\citealp{ekstrom2013red}). AGBs experience significant radial pulsation in their cool outer layers, which makes them distinguishable as long-period variable (LPV) stars, exhibiting periods ranging from a hundred days to a few years (\citealp{iben1983asymptotic}; \citealp{whitelock2003obscured}; \citealp{2018AJ156112Y}; \citealp{2019ApJ...877...49G}; \citealp{Abdollahi26} ).

Due to the low extinction level in the infrared (IR) wavelengths, AGBs, characterized by their high luminosity ($\sim$ 10$^{3}-$10$^{4}$ L$_\sun$; \citealp{hofner2018mass}) and low temperature (3000$-$4500 K), are more conveniently observed in the IR spectrum compared to the optical wavelengths. 

AGB stars are considered to be one of the most important sources of dust within the interstellar medium (ISM), enriching it with elements such as Li, C, N, F, and s-process elements (\citealp{boyer2009spitzer}; \citealp{Javadi13}; \citealp{karakas2014dawes}; \citealp{hofner2020explaining}). The rate at which these stars lose mass typically falls within the range of 10$^{-8}$ to 10$^{-5}$ M$_\sun$\,yr$^{-1}$, although in some cases, it reaches the higher value of 10$^{-4}$ M$_\sun$\,yr$^{-1}$ (\citealp{hofner2018mass}; \citealp{2025ApJ...992...94M}). Moreover, AGBs could also be potential sources of dust production in the early universe. Studies like DUSTiNGS (DUST in Nearby Galaxies with {\it Spitzer}) have shown that these stars can efficiently enrich the environment with dust even at extremely low metallicities (\citealp{boyer2015infrared}; \citealp{2017ApJ...851..152B}; \citealp{2025ApJ...991...24B}).

In summary, the method adopted in this manuscript, based on \citet{Javadi11}, is an intrinsically resolved-stellar method: it uses individual stars rather than integrated spectra, thereby avoiding many degeneracies that affect integrated-light inversions. The method was first applied to M33 ({\citealp{Javadi11, 2011ASPC..445..497J, Javadi17}}) and later extended to other galaxies (\citealp{2019MNRAS.483.4751H}; \citealp{2023ApJ...948...63A, Mahani23, 2024ApJ...972...47A}). 

Moreover, integrated-light methods such as MOPED, STARLIGHT, and VESPA are best suited for large samples of unresolved or distant galaxies, but degeneracies impose significant limits on their age resolution. The LPV method, by contrast, can only be applied to nearby, spatially resolved galaxies, but it yields SFHs with higher accuracy for intermediate ages and, importantly, also provides direct measurements of dust production and mass return from evolved stars (\citealp{Javadi13, 2025ApJ...992...94M}).

\begin{figure}[t!]
	{\hbox{
		\epsfig{figure=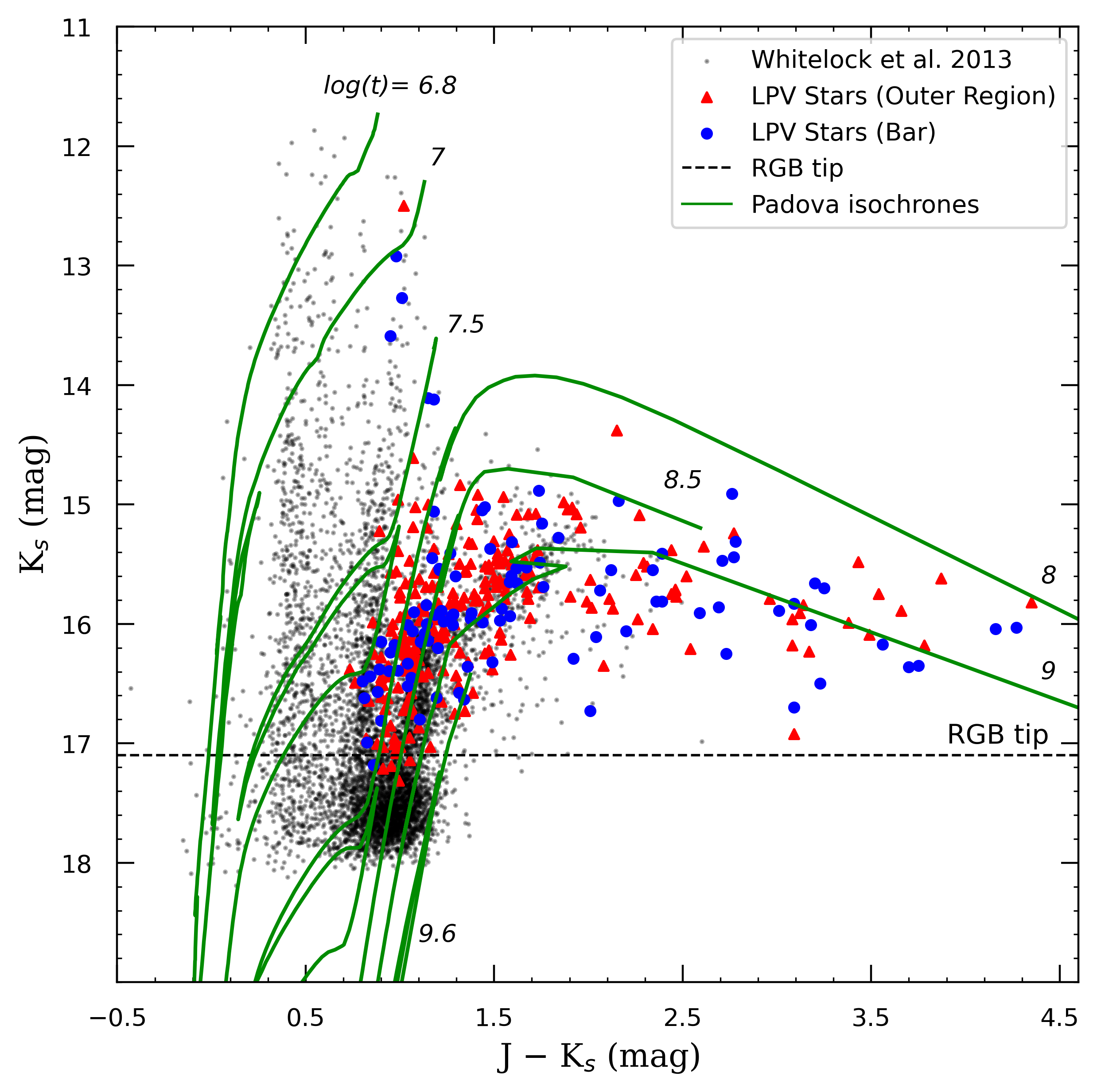,width=85mm,height=85mm}     
   \centering
	}}
	\caption{\small The CMD of NGC\,6822 in the K$_{s}$ band vs. J $-$ K$_{s}$ color. Black dots signify 6257 stars from \cite{whitelock2013local}'s survey, blue circles represent the 228 stars of the bar region, and red triangles display the 101 stars of the outer region. The red giant branch tip (RGB tip) of 17.1 mag (\citealp{cioni2005near}) marks the luminosity limit above which the stars evolved into the AGB phase reside. Additionally, isochrones from \cite{marigo2008evolution} for the metallicity of Z $=$ 0.003 and a distance modulus of 23.45 mag are overplotted.}
	\label{fig:fig2}
\end{figure}

This is Paper I, which presents the investigation on SFH of NGC\,6822 through evolved AGB stars within a radial distance of $<$ 3 kpc. The subsequent paper will study the mass loss rate of these stars and the dust production in NGC\,6822 (Paper II). This study is organized as follows: Section\,\ref{sec:sec2} provides an overview of the catalogs we used. The method is explained in Section\,\ref{sec:sec3}. We present our results in Section\,\ref{sec:sec4}. Finally, we discuss our findings in Section\,\ref{sec:sec5} and summarize our study in Section\,\ref{sec:sec6}.

\section{Data} \label{sec:sec2}

\subsection{Variable stars catalog } \label{subsec:sec2.1}

As previously mentioned, our method to derive the SFH relies on evolved AGB stars, which either have known long periods or large amplitudes or are C-rich AGBs. To ensure that our sample excludes any potential less-evolved O-rich AGBs with short periods, we only select C-rich AGBs whose classification is confirmed through spectroscopy. Our dataset consists of 329 stars collected from several published catalogs at near-IR wavelengths, J, H, and K$_s$ bands. We collected 97 stars, including 27 large-amplitude variables (LAVs) and 70 LPVs from \cite{whitelock2013local}. The survey conducted by \cite{whitelock2013local} was carried out in JHK$_s$ bands, covering three overlapping fields confined to the optical bar of NGC\,6822, using the Japanese–South African Infrared Survey Facility (IRSF) equipped with the SIRIUS camera over a period of 3.5 years. Based on the spatial distribution of the 97 LPVs, which are located along the optical bar of the galaxy, we defined a rectangular field that encloses these stars, referred to as the bar region (see Fig.\ref{fig:fig1}). Additionally, we have defined the area encompassing the stars dispersed outside the bar region as the outer region, which extends to a radial distance of 3 kpc.

\begin{figure}[t!]
	{\hbox{ \epsfig{figure=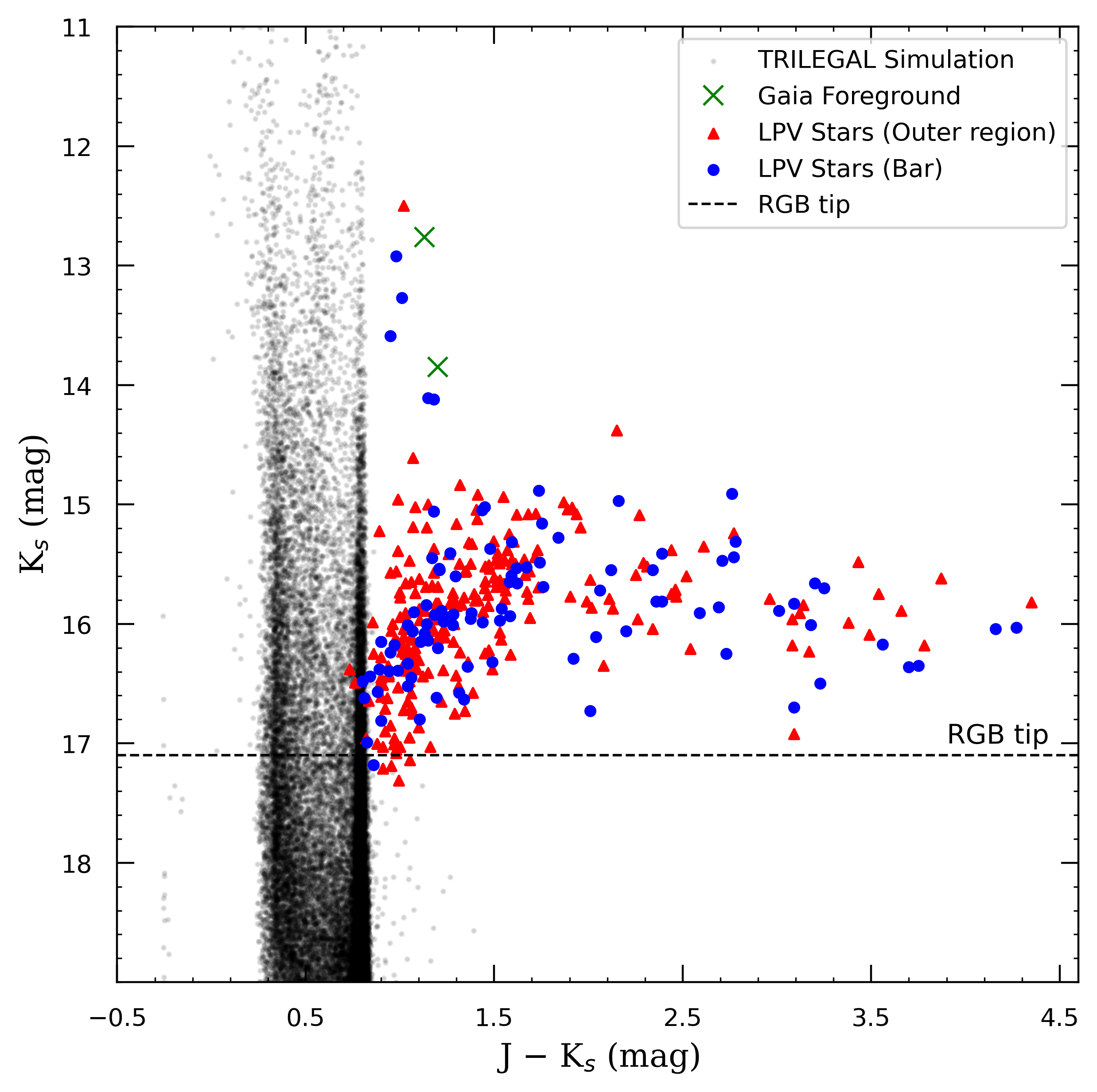,width=85mm,height=85mm}
   \centering
	}}
	\caption{\small The foreground contamination stars in the line of sight of NGC\,6822 are simulated using {\sc trilegal} (black dots). Blue circles and red triangles represent the LPVs in the bar and the outer regions, respectively. The stars in common with \textit{Gaia} DR\,3 are marked by green crosses.}
	\label{fig:fig3}
\end{figure}

In addition to the 97 LPVs, we collected 101 spectroscopy-confirmed carbon stars from \cite{sibbons2015spectral}, of which three were in common with \cite{whitelock2013local} and were subsequently excluded. Out of these 98 carbon stars, 52 are located in the bar region, and the remaining 46 stars are scattered in the outer region. \cite{sibbons2012agb} used the Wide-Field Camera (WFCam) on the 3.8\,m UK InfraRed Telescope (UKIRT) to obtain JHK photometry of a 3\,deg$^2$ area at the center of the galaxy. Later, \cite{sibbons2015spectral} used the AAOmega multi-fiber spectrograph at the 3.9\,m Anglo-Australian Telescope (AAT) to determine the spectral type of some potential AGB candidates at a radial distance of $\le$ 3 kpc and $\geq$ 5 kpc, primarily selected from \cite{sibbons2012agb}. Additionally, we selected 168 spectroscopy-confirmed C-rich AGBs from \cite{kacharov2012spectra}. They used the VIMOS instrument at the ESO VLT to obtain the low-resolution multi-object spectroscopy of approximately $800$ stars in seven fields centered on NGC\,6822. Among the 168 stars, 10 and 14 were in common with \cite{whitelock2013local} and \cite{sibbons2015spectral}, respectively. After removing the identical stars, we were left with a total of 134 carbon stars, 79 of which are in the bar region, and the remaining 55 stars are dispersed in the outer region.

 In summary, the central bar of NGC\,6822 contains most of its stellar population; the stars in our catalog are similarly distributed: 228 out of the total 329 stars are located in the bar region, including 97 LPVs and 131 spectroscopy-confirmed carbon stars. Meanwhile, the outer region harbors a total of 101 spectroscopy-confirmed carbon stars. The color-magnitude diagram (CMD) of our catalog, along with the 6257 stars from the survey by \citet{whitelock2013local}, is illustrated in Fig.\,\ref{fig:fig2}. The tip of the red giant branch (TRGB) marks the maximum luminosity
reached by low-mass stars ($M \lesssim 2\,M_{\odot}$) along the RGB,
immediately before the onset of core helium ignition. In these stars,
helium ignites under degenerate conditions in the helium core, producing
the helium flash \citep{schwarzschild1962red}. Since helium
ignition occurs at a nearly constant core mass, the TRGB luminosity is set
by that core mass and depends only weakly on age \citep{salaris2002}. After
central helium exhaustion, stars over a wide range of masses evolve onto
the AGB, where hydrogen and helium burning in two shells around a
degenerate carbon--oxygen core drives them to luminosities above the RGB
tip \citep{iben1983asymptotic}. In this work, following
\citet{cioni2005near}, an RGB tip of 17.1 mag is adopted. As expected, our
sample of evolved AGB stars populates the region above the RGB tip.

\subsection{Foreground Contamination} \label{subsec:sec2.2}

To identify and eliminate the foreground contamination of the Milky Way stars, we cross-matched our catalog with the \textit{Gaia} Data Release\,3 dataset (\textit{Gaia} DR3; \citealp{brown2021gaia}). For a star to be considered foreground contamination, it must either satisfy a criterion of $\sqrt{(\mu_\alpha)^{2} + (\mu_\delta)^{2}} > 0.28$ mas yr$^{-1}$ $+$ 2.0 $error$ for the proper motion measurement of \textit{Gaia} DR\,3, similar to the one outlined by \cite{van2019first}, or a parallax-error criterion, $\frac{Pa}{error\;of\;Pa} \geq $ 2$\sigma$ level. As \cite{2020ApJ...894..135S} illustrated, adjusting the parallax-error level, such as 3$\sigma$, will have a minor impact on the number of potential foreground stars identified. Only two stars from the \cite{whitelock2013local}  survey met the criteria to be foreground contamination stars, which were located in the bar region (see Fig.\,\ref{fig:fig3}) and consequently excluded from the sample. 

The completeness limit of \textit{Gaia} DR\,3 has slightly improved compared to the earlier releases, reaching G $\approx$ 19--21 mag on the faint end. To identify foreground stars fainter than this threshold, we simulated the foreground stars via {\sc trilegal}  web interface (\citealp{girardi2005star}). We performed the simulation for the bar and outer regions separately. We considered a field with the size of 0.4 deg$^{2}$ to encompass all the stars in our sample, centered at the galactic direction of $l =$ 25.339$^{\circ}$ and $b = -$18.399$^{\circ}$. The simulated population, the 228 LPVs in the bar, the 101 LPVs in the outer region, and the foreground stars in common with \textit{Gaia} DR\,3 are shown in Fig.\,\ref{fig:fig3}. As the dusty nature of LPVs moves them toward the redder part of the diagram, the narrow common area indicates that the foreground stars are insignificant numbers.

\section{Method} \label{sec:sec3}

The method we used to calculate the SFH is based on LPVs, originally developed by \cite{Javadi11}, and has since been applied in several studies (\citealp{2014MNRAS.445.2214R}; \citealp{Javadi17}; \citealp{2021ApJ...910..127N}; \citealp{2021ApJ...923..164S}; \citealp{2023ApJ...948...63A}; \citealp{2024ApJ...972...47A} ). We calculated the SFR, defined as the rate at which the gas mass is converted to stars as a function of time, $\xi(t)$ (M$_\sun$\,yr$^{-1}$). If the created stellar mass in the time interval between $t$ and $t+\mathrm{d}t$ is given by:

\begin{equation}
\label{eq:1}
 \mathrm{d}M(t) = \xi(t) \: \mathrm{d}t,
\end{equation}
the number of the formed stars, $N$, over the given time interval is obtained from the following equation:
\begin{equation}
\label{eq:2}
  \mathrm{d}N(t) = \frac{\int_{\mathrm{min}}^{\mathrm{max}} f_{\mathrm{IMF}}(m)\, \mathrm{d}m}%
           {\int_{\mathrm{min}}^{\mathrm{max}} f_{\mathrm{IMF}}(m)\, m\, \mathrm{d}m} \: \mathrm{d}M(t),
\end{equation}
where the $f_{\mathrm{IMF}}$ is the initial mass function (IMF), which \cite{kroupa2001variation} defined as $f_\mathrm{{IMF}} = A m^{-\alpha}$. $A$ is a normalization constant, and $\alpha$ is a factor that depends on the mass range:

\begin{equation}
\label{eq:3}
\alpha = \left\{ \begin{array}{rcl}
+0.3 \pm 0.7 & \mbox{for} & \mathrm{min} < \frac{\mathrm{m}}{\mathrm{M_\sun}} < 0.08 \\ +1.3 \pm 0.5 & \mbox{for} & 0.08 < \frac{\mathrm{m}}{\mathrm{M_\sun}} < 0.5 \\ +2.3 \pm 0.3 & \mbox{for} & 0.5 < \frac{\mathrm{m}}{\mathrm{M_\sun}} < \mathrm{max}
\end{array}\right. .
\end{equation}

The minimum mass of 0.02 M$_\sun$ and the maximum mass of 200 M$_\sun$ were considered in this paper. Then, we need to find out how many of these stars, $n$, are variables around the time $t$ in which they are observed. Assuming that stars with a mass falling within the range of $m(t)$ to $m(t + \mathrm{d}t)$ fulfill this condition, then the number of the variables born in the time interval between $t$ and $t+\mathrm{d}t$ is given by:

\begin{equation}
\label{eq:4}
  \mathrm{d}n(t) = \frac{\int_{m(t)}^{m(t+\mathrm{d}t)} f_{\mathrm{IMF}}(m)\, \mathrm{d}m}%
           {\int_{\mathrm{min}}^{\mathrm{max}} f_{\mathrm{IMF}}(m)\, \mathrm{d}m} \: \mathrm{d}N(t),
\end{equation}
combining equations \ref{eq:1} and \ref{eq:2} in equation \ref{eq:4} gives:

\begin{equation}
\label{eq:5}
  dn(t) = \frac{\int_{m(t)}^{m(t+\mathrm{d}t)} f_{\mathrm{IMF}}(m)\, \mathrm{d}m}%
           {\int_{\mathrm{min}}^{\mathrm{max}} f_{\mathrm{IMF}}(m)\, m \, \mathrm{d}m} \: \xi(t) \: \mathrm{d}t.
\end{equation}

Given an age bin of $\mathrm{d}t$, the number of the observed LPV stars in the bin, $\mathrm{d}n^{\prime}$, is defined as a function of pulsation duration (the period that the stars are in the LPV stage):

\begin{equation}
\label{eq:6}
  \mathrm{d}n^{\prime}(t) = \frac{\delta(t)}%
           {\mathrm{d}t} \: \mathrm{d}n(t).
\end{equation}

At last, by substituting equation \ref{eq:6} in relation \ref{eq:5}, we obtain the SFR based on the number of the observed LPV stars in a chosen age bin:

\begin{equation}
\label{eq:7}
  \xi(t) = \frac{\int_{\mathrm{min}}^{\mathrm{max}} f_{\mathrm{\mathrm{IMF}}}(m)\, m\, \mathrm{d}m}%
           {\int_{m(t)}^{m(t+\mathrm{d}t)} f_{\mathrm{IMF}}(m)\, \mathrm{d}m}  \: \frac{\mathrm{d}n^\prime(t)}{\delta (t)}.
\end{equation}

Moreover, we considered a statistical error for each age bin based on the Poisson statistics, which is defined as follows:

\begin{equation}
\label{eq:8}
  \sigma_{\xi(t)} = \frac{\sqrt{\mathrm{N}}}%
           {\mathrm{N}} \: \xi(t),
\end{equation}
where N is the number of stars in each age bin.

As the stars evolve and accumulate circumstellar dust, the isochrones bend and shift towards fainter magnitudes. Hence, we used the isochrones slopes for C-rich and O-rich AGB stars to de-redden the K$_s$-band magnitude of the LPVs. The corrected {K$_s$}-band magnitude is given by:

  \begin{equation}
  \label{eq:9}
     \mathrm{{K_{s_{cor}}} = K_{s} + a_{{iso}}\,[(J-K_{s}) - (J-K_{s})_{0}]},
  \end{equation}
where a$_{iso}$ is the slope of the isochrones, which varies depending on the chosen metallicity value, and (J $-$ K$_{s}$)$_0$ is the color where the isochrones peak and bend towards the fainter magnitude and redder color. 
  
Furthermore, LPV stars are assumed to be at the most luminous phase of their evolution, where their brightness reaches a maximum at near-IR wavelengths, allowing us to relate their birth mass to luminosity (\citealp{Javadi11}). The general fitting equation that connects birth mass to the K$_s$-band magnitude of the stars is formulated based on Padova’s theoretical isochrones (\citealp{marigo2008evolution, marigo2017new}):

  \begin{equation}
  \label{eq:10}
   \log\left (\frac{\mathrm{M}}{\mathrm{M_\sun}} \right) = \mathrm{a\,K_{s} + b}.
  \end{equation}

Having the birth mass of the stars, the age of the stars is obtained using the following equation:

  \begin{equation}
  \label{eq:11}
    \mathrm{\log}(t) = \mathrm{a\log\left(\frac{M}{M_\sun} \right) + b},
  \end{equation}
consequently, the pulsation duration is derived from:

  \begin{equation}
  \label{eq:12}
      \mathrm{\log}\left (\frac{\delta(t)}{t} \right) = \mathrm{D} + \mathrm{\Sigma_{i=1}^{4} a_{i}\,exp\left[\frac{-(log M[M_\sun] - b_{i})^2}{c_{i} ^2}\right]}.
  \end{equation}

The coefficients of these relations are obtained utilizing Padova isochrones (see the Appendix \ref{sec:apndix}). We customized our universal models for NGC\,6822 by adopting the distance modulus of $\mu = 23.45 \pm 0.15$ mag and several metallicity values within the range of $-$2.53 $<$ [Fe/H] $<$ $-$0.49 dex (see Section\,\ref{sec:sec4}).

\begin{figure}[t]
	{\hbox
    { \epsfig{figure=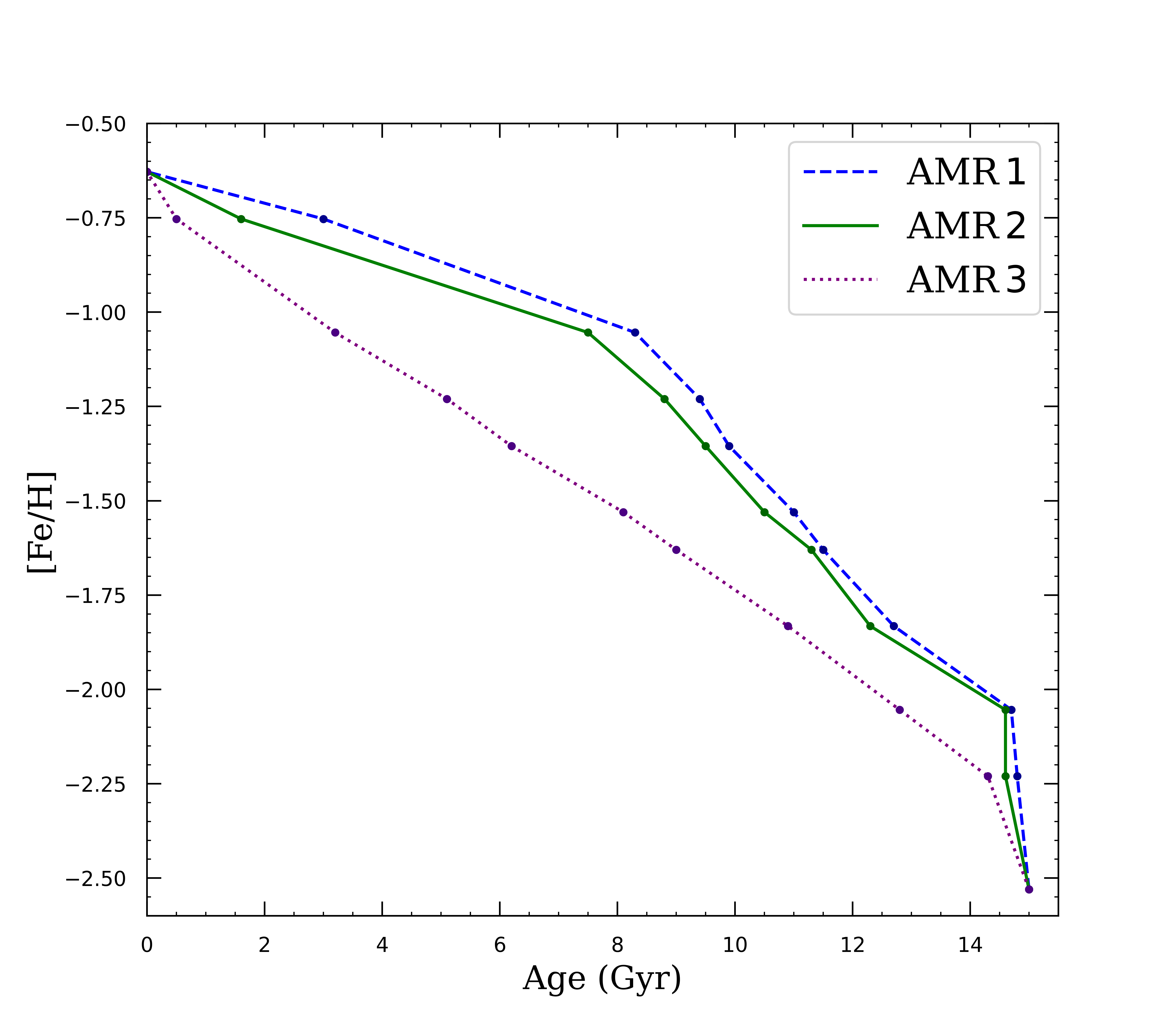,width=85mm,height=80mm}
   \centering
	}}
	\caption{\small The plot demonstrates the three age-metallicity relations used to derive SFH, adopted from \cite{wyder2001star}. Theses age-metallicity relations have the initial metallicity of [Fe/H] $=$ $- 2.53$ dex and the final metallicity of [Fe/H] $=$ $- 0.49$ dex. The blue (dashed), green (solid), and purple (dotted) curves are throughout this paper referred to as age-metallicity relations (AMRs)1, 2, and 3, respectively.}
	\label{fig: Fig. 4}
\end{figure}

Hence, to derive the SFR in different age bins, the following steps are taken:

\begin{itemize}
\item The attenuation by circumstellar dust is corrected using
Eq.\,\ref{eq:9}.

\item Using the corrected K$_{s}$-band magnitude and the
mass--magnitude relation (Eq.\,\ref{eq:10}), with the coefficients
presented for each metallicity in Table \ref{tab:tab9}, the birth mass of
each LPV is estimated.

\item Using the mass--age relation (Eq.\,\ref{eq:11}) and the
coefficients presented in Table \ref{tab:tab12}, the age of each LPV is
estimated.

\item Using the mass--pulsation duration relation
(Eq.\,\ref{eq:12}) and the coefficients presented in Table
\ref{tab:tab18}, the pulsation duration of each star is estimated. This is
used in Eq.\,\ref{eq:6} to estimate the total number of variable stars
formed, $\mathrm{d}n$, in a given time bin from the number of observed
ones, $\mathrm{d}n^{\prime}$.

\item The stars are then binned, and Eq.\,\ref{eq:3} (for the IMF),
Eq.\,\ref{eq:6} (for applying a weight to each star to estimate the total
number of variable stars formed; more detail can be found in
\citealp{2024ApJ...972...47A}), and Eq.\,\ref{eq:7} are used to derive the
SFR in each age bin. A not unimportant detail related to the presentation
of a SFH is the way it is binned in age. The younger, massive variable
stars are many times fewer than the old, low-mass variable stars, and
inadequate binning can either lead to spurious peaks in the SFR or mask any
such real bursts. From a statistical point of view there is an advantage in
ensuring that each bin contains the same number of stars, so that the SFR
values have uniform uncertainties. To accomplish this, we first ordered the
stars by mass, then started counting until we reached a given number, at
which moment we started counting stars for the subsequent bin. Eventually,
the IMF correction is applied based on the minimum and maximum mass of each
bin. The IMF therefore enters only as a normalisation, converting the
number of variables formed within a narrow birth-mass interval into the
total mass formed in that bin, while the age of each bin is set by the
individual stars themselves.

\item Finally, we present the SFH in two ways. In the first, shown
in Section\,\ref{subsec:sec4.1} and in Figs.\,\ref{fig:Apendix_Fig_1} to
\ref{fig:Apendix_Fig_3}, we derive a separate SFH for each constant
metallicity value. Since the galaxy has been chemically enriched over time,
the highest metallicity is appropriate only for the most recent epochs and
not for the older ones, as the age ranges listed in Table\,\ref{tab:tab1}
show. The advantage of this representation is that, guided by an
age--metallicity relation, the SFR value appropriate to each age bin can be
selected from the corresponding panel. In the second approach, discussed in
Section\,\ref{subsec:sec4.2}, the age--metallicity relation is applied
directly when deriving the SFH. In this case we obtain a single SFH for
each adopted AMR, as shown in Fig.\,\ref{fig: Fig. 7}, in which the
chemical evolution of the galaxy is already accounted for when deriving the
birth mass, age, and pulsation duration of each star.

\end{itemize}

As explained above, deriving the star formation rate requires the pulsation duration of the LPVs, $\delta t$, as a function of stellar birth mass and age. $\delta t$ is obtained from stellar evolutionary models and is sensitive to
the adopted prescriptions for mass loss, convection, and the treatment of thermal pulses along the AGB phase \citep{Marigo07}. Comparisons between different sets of isochrones \citep[e.g.,][]{Marigo07,Girardi10} indicate that $  \delta t  $ for intermediate-mass stars, and particularly for low-mass stars evolving onto the AGB, can vary by up to a factor of two. Since the derived SFR scales inversely with $  \delta t  $, these variations translate into systematic uncertainties of up to $\sim$0.3 dex in the normalization of the star formation history.

Additional uncertainties arise from the choice of initial mass function. Adopting the canonical IMF of \citet{Kroupa01} as our standard reference, and comparing it with a single power-law Salpeter IMF \citep{Salpeter55} (both normalized over the mass range $0.1$--$100 M_{\odot}$), yields a difference in mean stellar mass of approximately 45\%. Because the SFR scales inversely with the mean stellar mass, this introduces a systematic offset of $\approx$0.26 dex in the inferred SFR.

Further systematic uncertainties arise from photometric calibration, distance modulus, and extinction by circumstellar dust. LPVs are often heavily enshrouded in dust, which can lead to an underestimation of their intrinsic luminosities (and hence their initial masses) if not properly corrected. To quantify the combined effect of photometric and distance uncertainties, we repeated the analysis after applying systematic magnitude offsets of $  \pm0.2  $ mag. The resulting variations in the inferred SFR are typically 20--40\% ($\approx$0.1--0.2 dex), with the largest relative differences occurring in age bins with low SFR. These systematics primarily affect the overall normalization rather than the shape and timing of the main star formation episodes.

A more detailed discussion of these effects, including the full set of equations and supporting figures, is provided in Appendix~\ref{appnx:systematic}.


\section{Results} \label{sec:sec4}

\begin{figure*}[t!]
	{\hbox
    { \epsfig{figure=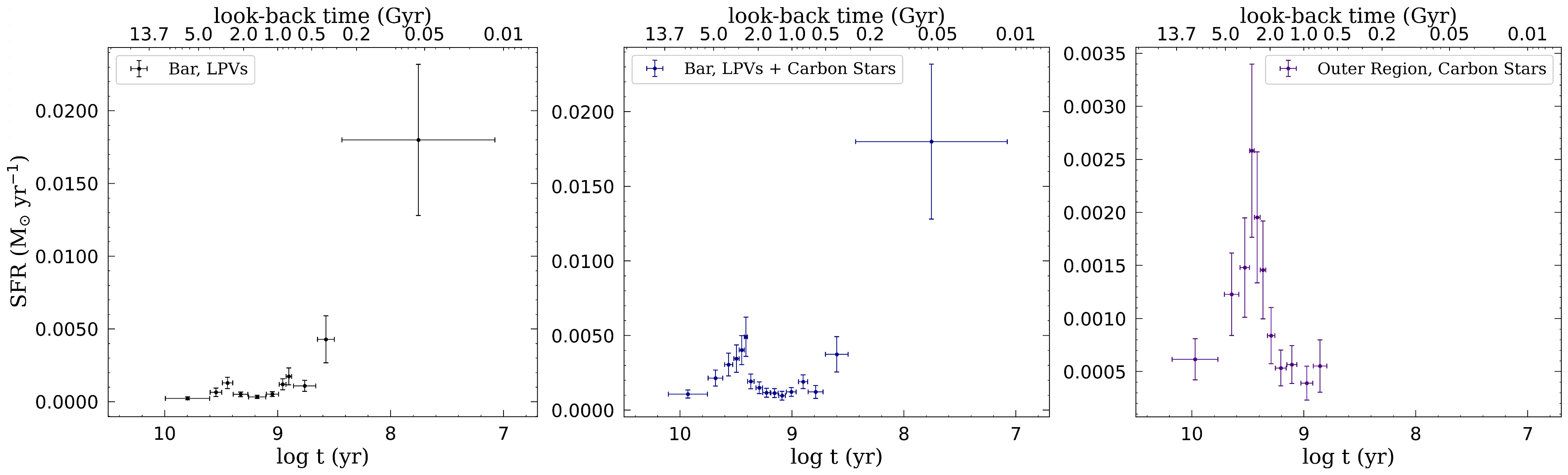,width=180mm,height=60mm}
   \centering
	}}
	\caption{\small The SFH results are presented for the mean metallicity of Z $\approx$ 0.003, assuming that the metallicity remains constant over time. Left panel: The SFH for the bar region is derived using the 97-LPVs catalog. Middle panel: The SFH for the bar region is determined using the bar region catalog, which includes 97 LPVs and 131 spectroscopy-confirmed carbon stars. Right panel: The SFH for the outer region is obtained utilizing the outer region catalog, which consists solely of 101 spectroscopy-confirmed carbon stars.}
	\label{fig: Fig. 5}
\end{figure*}

This section presents the numerical result derived by implementing our method to the selected dataset, where we estimated the SFR using the LPVs and reconstructed the SFH of NGC\,6822 through the cosmic time ranging from $\sim$ 13 Gyr to $\sim$ 10 Myr ago. We investigate the SFH within two distinct regions of the galaxy, as shown in Fig.\,\ref{fig:fig1}. The first is the bar region, which encloses a central rectangular area of 189 arcmin$^2$. The region contains 97 LPVs collected from \cite{whitelock2013local} and 131 spectroscopy-confirmed carbon stars selected from \cite{sibbons2015spectral} and \cite{kacharov2012spectra} (see Section\,\ref{subsec:sec2.1} for details). The second region is the outer region that encloses an area outside the bar region, extending to a radius of 3 kpc. It includes 101 spectroscopy-confirmed carbon stars collected from the same two catalogs from which the carbon stars in the bar region were selected. We first derived the SFH only using the 97 LPV stars in the bar region, which we refer to as the 97-LPVs catalog. However, detecting carbon stars in surveys is challenging due to their dusty nature. Hence, to enhance the completeness of our sample, we added the 131 spectroscopy-confirmed C-rich stars located in the bar region to the 97-LPVs catalog and created the bar region catalog, which contains a total of 228 stars. Additionally, we calculated the SFR for the 101 spectroscopy-confirmed carbon stars distributed in the outer region, which we refer to as the outer region catalog.

The models we used to derive the parameters to estimate the SFR are sensitive to the chosen metallicity; therefore, in addition to considering three different catalogs, we also adopted a range of metallicity values. Initially, we derived the results assuming that the metallicity remains constant throughout the entire lifetime of the galaxy. To trace the variation of metallicity, we assumed 12 constant metallicities, falling within the range of 0.0001 $<$ Z $<$ 0.012 ( $-$2.53 $<$ [Fe/H] $<$ $-$0.49 dex). This range is obtained based on the lowest and highest metallicity values reported for NGC\,6822 by various studies (see Table \ref{tab:tab1}). Furthermore, as the metallicity of the ISM changes over time, the metallicity of the formed stars during different epochs also varies. In other words, the old stellar populations were formed in a low metallicity environment, while the younger populations were formed in a relatively metal-rich environment. Therefore, in addition to deriving the SFH under the assumption of constant metallicity, we reconstructed the SFH considering age-metallicity relations (AMRs) to take into account the variation of metallicity over time. We adopted the AMRs suggested by \cite{wyder2001star}, including two non-linear and one linear AMRs, which we refer to as AMR\,1, AMR\,2, and AMR\,3 throughout this paper, as illustrated in Fig.\,\ref{fig: Fig. 4}.

\subsection{Constant metallicity} \label{subsec:sec4.1}

We initially obtained the SFH for the bar and outer regions, assuming that the metallicity remains constant over the lifetime of the galaxy. The results for the constant mean metallicity of [Fe/H] $=$ $-$1.05 dex (Z $\approx$ 0.003) are presented in Fig.\,\ref{fig: Fig. 5}. The SFH of the bar region is first obtained using the 97-LPVs catalog, as shown in the left panel of Fig.\,\ref{fig: Fig. 5}, where three episodes of star formation can be observed. The first episode occurred $\sim$ 3 Gyr ago (log $t = 9.5$; SFR $=$ 1.3 $\pm$ 0.4 $\times$ $10^{-3}$ M$_\sun$\,yr$^{-1}$), while the second is identified at $\sim$ 800 Myr ago (log $t = 8.9$; SFR $=$ 1.7 $\pm$ 0.5 $\times$ $10^{-3}$ M$_\sun$\,yr$^{-1}$). The third episode has burst over the past $\sim$ 250 Myr (log $t = 8.4$;  SFR $=$ 18.0 $\pm$ 5.2 $\times$ $10^{-3}$ M$_\sun$\,yr$^{-1}$). 

Moreover, as previously mentioned, 131 spectroscopy-confirmed carbon stars were added to the 97-LPVs catalog to enhance the completeness of the dataset in the bar region, defined as the bar region catalog. However, since only AGBs with a mass range of 1 $<$ M/ M$_\sun$ $<$ 4 become carbon stars (\citealp{groenewegen1995evolution}; \citealp{karakas2014dawes}), the corresponding age range that these stars cover is restrained. Consequently, we cannot retrieve results for both the early and recent epochs. 

With that in mind, the SFH of NGC\,6822 derived using the bar region catalog and adopting the mean metallicity is presented in the middle panel of Fig.\,\ref{fig: Fig. 5}. The oldest stellar population that can be retrieved is $\sim$ 12.6 Gyr (log $t = 10.1$), whereas the recovered population using the 97-LPVs catalog is no older than 10 Gyr old (log $t = 10.0$).

The overall SFH pattern exhibits no significant variations. However, the SFR of the first episode increases by a factor of $\sim$ 4, while the SFR of the second episode is more or less the same. Moreover, as expected, the SFR of the recent episode remains unchanged after the inclusion of carbon stars, as the youngest carbon star in this catalog is $\sim$ 500 Myr old (log $t = 8.7$), and no carbon star is present in the bins younger than this age to affect the SFR.

The SFH of the outer region, assuming the mean metallicity (Z $\approx$ 0.003), is illustrated in the right panel of Fig.\,\ref{fig: Fig. 5}. Given that the outer region catalog solely consists of carbon stars, the discussed age restriction has affected the results. Consequently, we cannot retrieve data for ages younger than $\sim$ 600 Myr old (log $t = 8.8$). Furthermore, similar to the bar region, the outer region has experienced a burst of star formation $\sim$ 3 Gyr ago (log $t = 9.5$), reaching a maximum SFR of $\sim$ 2.6 $\pm$ 0.8 $\times$ $10^{-3}$ M$_\sun$\,yr$^{-1}$. The absence of the two recent star formation episodes in the outer region might be due to incompleteness and insufficient data for this region. Consequently, we cannot make a certain statement regarding whether the recent star formation activities have propagated beyond the bar region.

In addition to the mean metallicity, we have obtained the SFH for the other 11 metallicity values (see Table\,\ref{tab:tab1}) to see the effect of the selected initial metallicity on the SFH, as it is a crucial parameter in determining the birth mass, age, and pulsation duration. The results for the 97-LPVs, the bar region, and the outer region catalogs are presented in Fig.~\ref{fig:Apendix_Fig_1}, Fig.~\ref{fig:Apendix_Fig_2} and Fig.~\ref{fig:Apendix_Fig_3} in Appendix \ref{sec:apndix}, respectively. 

Based on our findings, as the selected initial metallicity value increases, the SFH shifts towards older epochs, and the magnitude of both the mass and SFR decreases. Nevertheless, the overall pattern of the SFH remains almost the same regardless of the chosen metallicity. Furthermore, the calculated stellar mass for the bar region falls within the range of 0.68 $\pm$ 0.22 $\times$ $10^{7}$  $<$ M/M$_\sun$  $<$ 3.10 $\pm$ 1.00 $\times$ $10^{7}$ based on the 97-LPVs catalog (see Table\,\ref{tab:tab2}). However, the addition of carbon stars and enhancement of completeness of the dataset for the bar region lead to derivation of a higher range of 1.95 $\pm$ 0.50 $\times$ $10^{7}$  $<$ M/M$_\sun$  $<$ 7.30 $\pm$ 1.30 $\times$ $10^{7}$ (see Table\,\ref{tab:tab3}). Moreover, the stellar mass derived for the outer region is naturally smaller as the stellar populations are mainly located along the bar, 0.83 $\pm$ 0.27 $\times$ $10^{7}$  $<$ M/M$_\sun$  $<$ 4.80 $\pm$ 1.51 $\times$ $10^{7}$ (see Table\,\ref{tab:tab4}). We must note that, the stellar mass range obtained in the outer region is affected by the incompleteness of the dataset for this region, likely lower than the actual mass present in this area.

\begin{figure*}[t!]
	{\hbox
    { \epsfig{figure=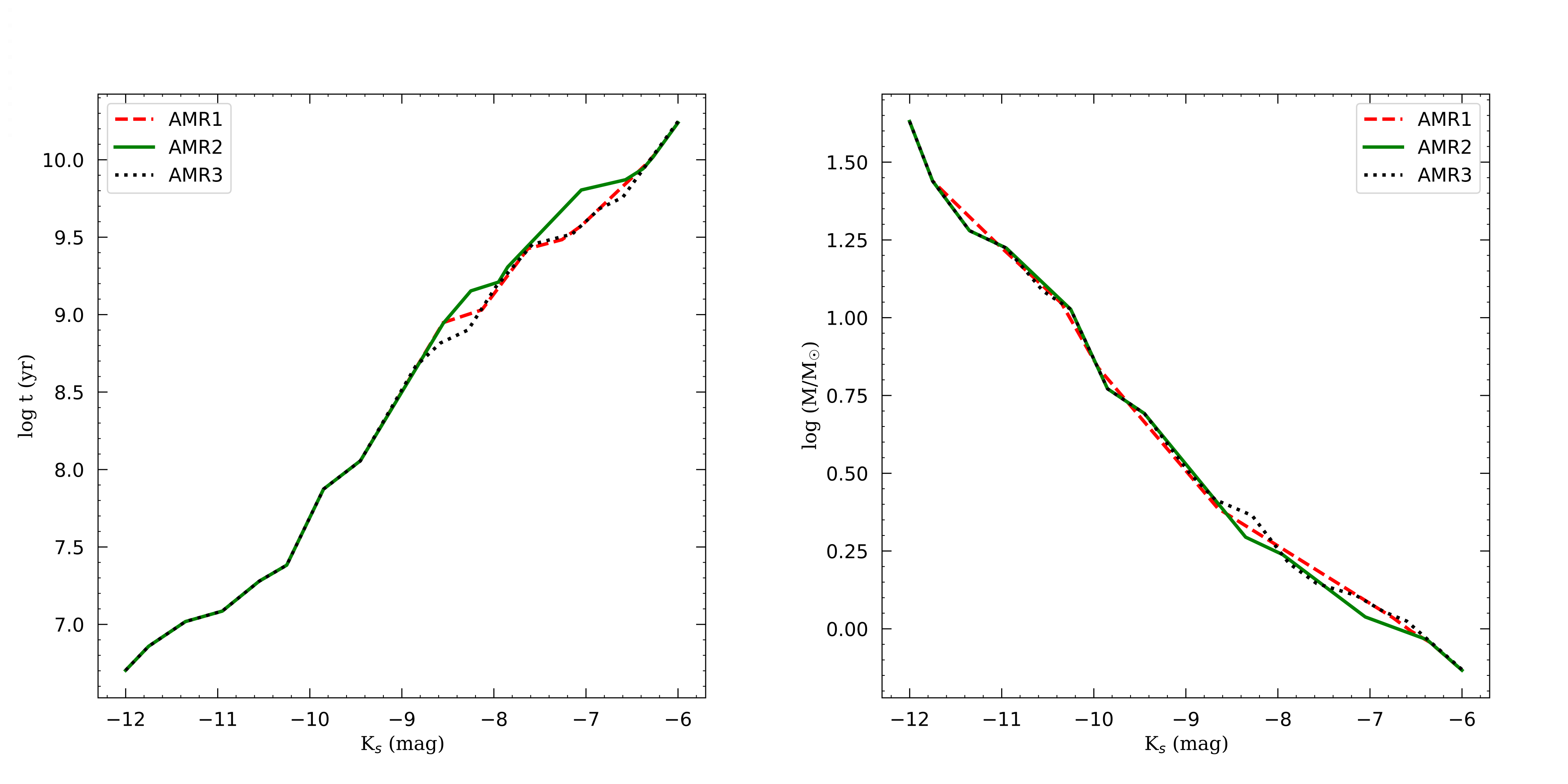,width=180mm,height=90mm}
   \centering
	}}
	\caption{\small Age-luminosity (left panel) and mass-luminosity (right panel) relations based on the three AMRs for the K$_s$-band magnitude (red squares). The solid lines signify the best linear fits, from which the coefficients of the equations are obtained.}
	\label{fig: Fig. 6}
\end{figure*}

\subsection{Age-Metallicty Relation} \label{subsec:sec4.2}

\begin{figure*}[t]
	{\hbox
    { \epsfig{figure= 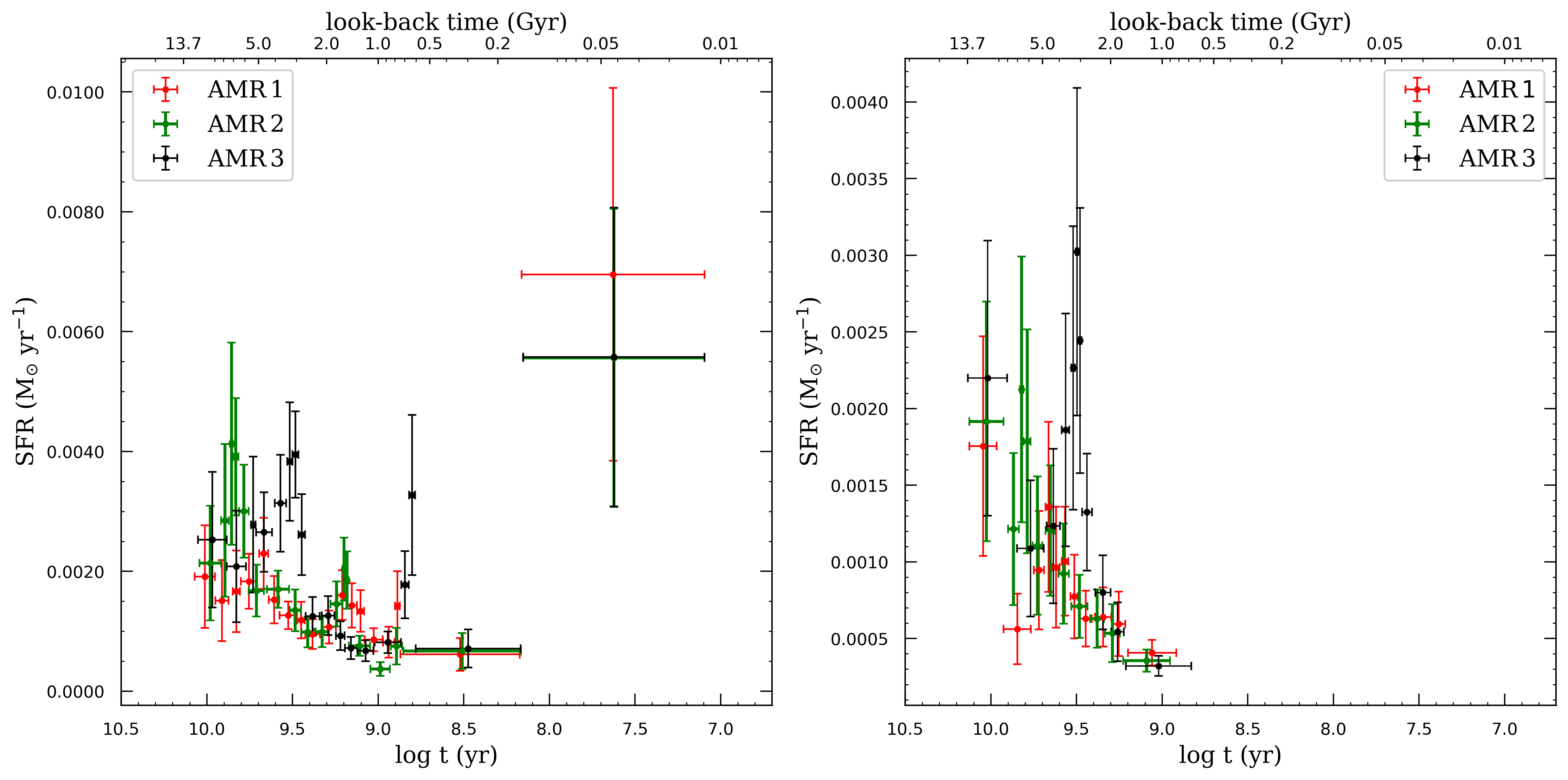,width=180mm,height=90mm}
   \centering
	}}
	\caption{\small The SFH for the bar region (left panel) and the outer region (right panel), assuming variable metallicity by time based on the three AMRs from Fig.\,\ref{fig: Fig. 4}. Individual plots can be found in the Appendix\,\ref{sec:apndix}.}
	\label{fig: Fig. 7}
 \end{figure*}
 
Stellar evolution in a galaxy is closely tied to its chemical evolution. The mass loss of evolved stars and the death of massive stars enrich the ISM. This process leads to an overall increase in metallicity over time (\citealp{audouze1976chemical}). Therefore, to determine the accurate SFH and path of evolution in NGC\,6822, we will derive the SFH by employing the AMRs suggested by \cite{wyder2001star}, as illustrated in Fig.\,\ref{fig: Fig. 4}. Based on these AMRs, the life span of the galaxy is divided into time intervals, and a specific metallicity is assigned to each one. Then, the associated Padova model with the chosen metallicity is selected for each interval accordingly. Subsequently, we utilized the models to derive the age-luminosity and mass-luminosity relations. These relations are presented in Fig.\,\ref{fig: Fig. 6}. Next, we obtained the fitting equations necessary to calculate the SFRs. In the age--luminosity relation, the three AMRs almost overlap for stars
brighter than K$_{s} \simeq -$8.8 mag ($\log t \simeq 8.7$). AMR\,3 is the
first to diverge from the other two, followed by AMR\,1 and AMR\,2, which
separate from each other at K$_{s} \simeq -$8.5 mag ($\log t \simeq 8.9$).
Towards fainter magnitudes the relations partially overlap again in some
age bins, either for two or for all three of them. The mass--luminosity
relation shows the same behaviour, with magnitude bins in which the three
relations overlap and others in which they diverge, although the
differences between them are smaller and do not occur at the same
magnitudes as in the age--luminosity relation.
Using these equations, we reconstructed the SFH of the galaxy with respect
to the variation in metallicity over time. As can be seen in
Fig.\,\ref{fig: Fig. 7}, the differences in the estimated ages and masses
obtained with the different AMRs result in different star formation rates.
As noted above, the overlaps between the relations do not occur at the same
magnitude ranges in the two panels. Consequently, an overlap in the
age--luminosity relation alone does not lead to similar star formation
rates, since the masses assigned to the stars still differ. Only where both
the age--luminosity and the mass--luminosity relations overlap over the
same magnitude range do the three AMRs give the same rate. An example is
the bin at $\log t \simeq 8.5$, where the three relations coincide at the
corresponding magnitude, K$_{s} \simeq -$9.0 mag, as shown in
Fig.\,\ref{fig: Fig. 6}.

The SFH is derived using the bar region and the outer region catalogs, illustrated in the  left and the right panel of Fig.\,\ref{fig: Fig. 7}, respectively. The results based on AMR\,1 show that the first episode of star formation has occurred around 4.7 Gyr ago (log $t = 9.67$) in the bar region with a peak SFR of 2.3 $\pm$ 0.6 $\times$ $10^{-3}$ M$_\sun$\,yr$^{-1}$ ( 0.6 $\pm$ 0.1 $\times$ $10^{-9}$ M$_\sun$\,yr$^{-1}$\,pc$^{-2}$). A similar episode is observed in the outer region $\sim$ 4.6 Gyr ago (log $t = 9.66$) with a peak SFR of 1.4 $\pm$ 0.6 $\times$ $10^{-3}$ M$_\sun$\,yr$^{-1}$ ( 0.6 $\pm$ 0.2 $\times$ $10^{-10}$ M$_\sun$\,yr$^{-1}$\,pc$^{-2}$). The simultaneous burst of star formation in both regions suggests that the star formation has been triggered all over the galaxy. We will discuss the possible instigators of the star formation bursts throughout the galaxy in Section\,\ref{sec:sec5}.

Moreover, our derived star formation rates can be compared with those reported by \cite{wyder2001star} for the five regions H\,IV, H\,VI, H\,VII, H\,VIII, and C\,25, which were originally introduced and labeled in the works of \citet{hubble1925ngc} and \citet{wyder2001star}, and we adopt the same terminology in this study. They investigated the SFH of these regions through CMD modeling and using the $\it{Hubble}$ $\it{Space}$ $\it{Telescope}$ ($\it{HST}$) data. They estimated an SFR of 0.1 $^{+0.5}_{-0.1}$  $\times$ $10^{-9}$ M$_\sun$\,yr$^{-1}$\,pc$^{-2}$ for the H VI region at the center of the bar, and a rate of 0.1 $^{+0.2}_{-0.1}$  $\times$ $10^{-9}$ M$_\sun$\,yr$^{-1}$\,pc$^{-2}$ for the H VIII region, located to the north of the bar, over the past 3 -- 5 Gyr. Additionally, they have determined a rate of 0.2 $\pm$ 0.2 $\times$ $10^{-9}$ M$_\sun$\,yr$^{-1}$\,pc$^{-2}$ for the C\,25 region, located on the east side of the bar within the outer region defined in this work, over the same period.

\begin{figure*}[t]
	{\hbox
    { \epsfig{figure=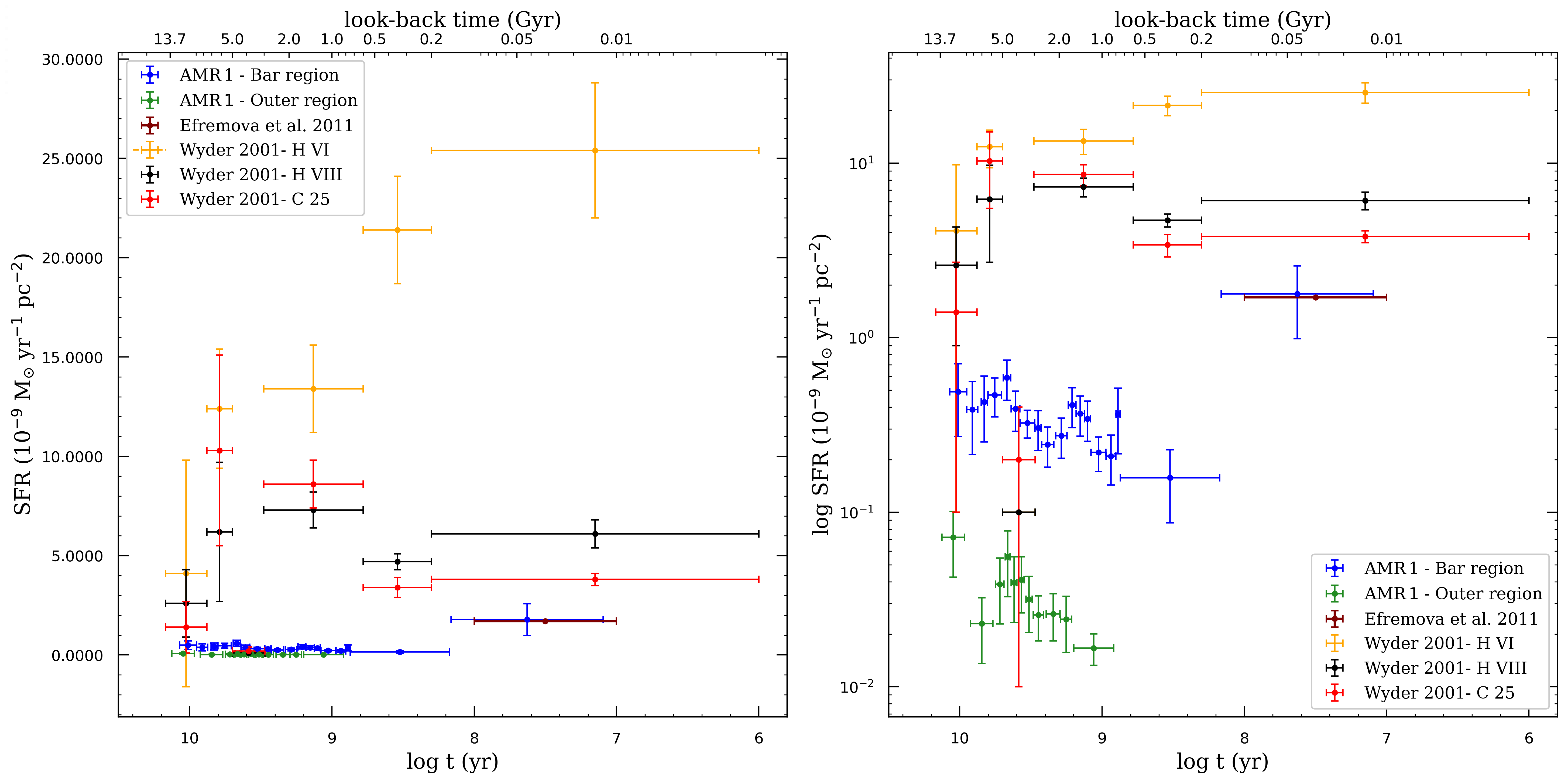, width=2\columnwidth}
   \centering
	}}
	\caption{{\small The comparison of the derived SFRs from this work (AMR1) with those from \citet{wyder2001star} and \citet{efremova2011recent} is presented with a linear SFR scale (left panel), and a logarithmic scale (right panel). This adjustment is purely for better visibility of the SFH trends, and does not represent any physical rescaling. Additionally, the log $t = 9.47$--log $t = 9.70$ bins of regions H\,VI and H\,VIII overlap entirely due to their nearly identical SFR values on the logarithmic scale.}}
	\label{fig: Fig. 8}
 \end{figure*}

Following a decline in star formation in NGC\,6822 lasting approximately 2.7 Gyr, it is once more triggered noticeably in the bar region. The second epoch of star formation burst begins around 1.7 Gyr ago (log $t = 9.24$) and peaks at $\sim$ 1.6 Gyr (log $t = 9.21$) in this area with a rate of 1.6 $\pm$ 0.4 $\times$ $10^{-3}$ M$_\sun$\,yr$^{-1}$ ( 0.4 $\pm$ 0.1 $\times$ $10^{-9}$ M$_\sun$\,yr$^{-1}$\,pc$^{-2}$).


\citet{mcquinn2010nature} derived the SFH of NGC 6822 using three deep HST fields across the bar and identified an enhanced star formation episode within the past roughly 1.5 to 1 Gyr, with an integrated rate of about $\sim 6.8 \pm 0.9 \times 10^{-3}$~M$_\odot$ yr$^{-1}$. This value is a factor of about 3 higher than our measurement for the bar region. This offset may come from three different assumptions. The first is
the adopted IMF: \citet{mcquinn2010nature} used a single power-law Salpeter
IMF over 0.1--100 M$_\sun$, while we use the Kroupa IMF. As shown in
Appendix\,\ref{appnx:systematic}, this contributes a factor of about 1.8
(0.26 dex) to the derived SFR, and therefore accounts for more than half of
the discrepancy. The second is the metallicity. We adopt the AMRs of
\citet{wyder2001star}, which fix the metallicity at the start and at the
end of the galaxy's history, so the metallicity of each age bin is set in
advance. \citet{mcquinn2010nature} instead left the metallicity free for
NGC\,6822 and let the CMD fit determine it, which their photometry allows
because it reaches two magnitudes below the red clump. The metallicity
assigned to a given epoch can therefore differ between the two analyses,
and since the birth mass and age we derive for each star depend on the
adopted metallicity, this propagates directly into the SFR. The size of the
effect can be seen in Fig.\,\ref{fig: Fig. 7}: between
$\log t \simeq 9.0$ and 9.6 the rates derived from the three AMRs differ by
up to a factor of $\sim$3 in individual age bins, comparable to the offset
between our rates and those of \citet{mcquinn2010nature}, while the overall
shape of the SFH and the total stellar mass formed are preserved. The third is the spatial coverage. \citet{mcquinn2010nature}
studied the SFH over three small regions across the bar, while the bar
region we define is a relatively large region, which covers both dense
star-forming and less dense diffuse regions. Consequently, the estimated
SFR is an average over the whole area, which results in a lower SFR per
unit area. \citet{wyder2001star} reports significant variation among his
individual fields, which emphasises the strong spatial dependence of the
SFR within NGC\,6822. Despite the differences in normalisation, all these
studies consistently identify an episode of enhanced star formation at
similar look-back times.

Furthermore, after $\sim$ 1.5 Gyr of low star formation activity, NGC\,6822 undergoes a drastic burst of star formation in the bar region approximately over the past 150 Myr. Such significant enhancement in SFR in NGC\,6822 over the past few hundred million years is reported by various other studies (e.g., \citealp{hodge1980recent}; \citealp{gallart1996local3};   \citealp{mcquinn2010nature}). Additionally, Based on \citet{wyder2001star}'s findings the SFR in the HIV, H VI, H VIII, and C\,25 fields has increased over the past 200 Myr. 

Moreover, we derived a recent SFR of $\sim$ 7.0 $\pm$ 3.1 $\times$ $10^{-3}$ M$_\sun$\,yr$^{-1}$ (1.8 $\pm$ 0.8 $\times$ $10^{-9}$ M$_\sun$\,yr$^{-1}$\,pc$^{-2}$). This rate is in good agreement with the rate of 1.7 $\times$ $10^{-9}$ M$_\sun$\,yr$^{-1}$\,pc$^{-2}$ over the past 100 Myr, derived and reported by \cite{efremova2011recent} through far-UV flux measurement. Fig.\,\ref{fig: Fig. 8} illustrates the comparison of the SFRs derived in this work with those reported by previous studies. The derived SFRs for the bar region are in agreement with the value reported by \citealp{efremova2011recent}. Additionally, our calculated SFRs at early epochs and at the most recent time are marginally consistent, in some cases within the error bars, with the values obtained by \citealp{wyder2001star} for the region C\,25.

We do not include the SFR from \citet{mcquinn2010nature} in Fig.\,\ref{fig: Fig. 8} because their reported value is integrated over the observed fields, whereas the values from \citet{wyder2001star} are expressed as SFR surface densities. Although the approximate footprints of the $\it{HST}$ fields are known, converting the integrated SFR of \citet{mcquinn2010nature} into a surface density would require adopting an effective area that is not directly comparable to the spatial regions considered in the other studies. We therefore retain their result as a qualitative comparison rather than converting it into a surface density for Fig.\,\ref{fig: Fig. 8}.

\begin{table*}[t]
\centering
\begin{tabular}{cccclclcl}
\hline
Z       & {[}Fe/H{]} & Stellar Mass  & Peak 1 & SFR  & Peak 2    & SFR & Peak 3    & SFR  \\
        & \scriptsize(dex) & \scriptsize($10^{7}$ M$_\Sun$)  & (Gyr) & \scriptsize($10^{-3}$ M$_\Sun$yr$^{-1}$)   & \scriptsize(Gyr)    & \scriptsize($10^{-3}$ M$_\Sun$yr$^{-1}$) & \scriptsize(Gyr)    & \scriptsize($10^{-3}$ M$_\Sun$yr$^{-1}$)  \\

\hline
0.0001  & $-$2.53 & 3.10 $\pm$ 1.00          & -             & -             & 0.490             & 20.0 $\pm$ 6.3              & 0.038             & 90.5$\pm$ 30.0            \\ [0.08cm]
0.0003  & $-$2.05 & 1.50 $\pm$ 0.50           & 1.660             & 4.2 $\pm$ 1.3             & 0.690             & 5.3 $\pm$ 1.7              & 0.049            & 32.8 $\pm$ 10.9           \\ [0.08cm]
0.0005  & $-$1.83 & 1.90 $\pm$ 0.62          & 1.660            & 2.1 $\pm$ 0.7             & 0.310             & 16.4 $\pm$ 5.5              & 0.040             & 65.8$\pm$ 21.9           \\ [0.08cm]
0.0008  & $-$1.63 & 1.40 $\pm$ 0.46           & 1.660            & 6.2 $\pm$ 1.9             & 0.480             & 9.0 $\pm$ 3.0              & 0.045              & 47.4$\pm$ 15.8           \\ [0.08cm]
0.0010   & $-$1.53 & 1.70 $\pm$ 0.55           & 1.800            & 10.2 $\pm$ 3.2             & 1.550              & 3.7 $\pm$ 1.2              & 0.050              & 46.2 $\pm$ 15.4           \\ [0.08cm]
0.0015  & $-$1.36 & 1.93 $\pm$ 0.63          & 1.930            & 1.6 $\pm$ 0.5             & -              & -              & 0.053             & 61.6 $\pm$ 20.5           \\ [0.08cm]
0.0020   & $-$1.23 & 0.76 $\pm$ 0.25           & 2.400             & 2.4 $\pm$ 0.8             & -              & -              & 0.050              & 14.7 $\pm$ 4.9           \\ [0.08cm]
0.0030   & $-$1.05 & 0.84 $\pm$ 0.27           & 2.754             & 1.3 $\pm$ 0.4             & 0.800             & 1.7 $\pm$ 0.8              & 0.060              & 18.0 $\pm$ 5.2           \\ [0.08cm]
0.0060   & $-$0.75 & 0.79 $\pm$ 0.25           & 4.000             & 6.5 $\pm$ 0.2             & -              & -              & 0.071              & 4.7 $\pm$ 1.6           \\ [0.08cm]
0.0080   & $-$0.63 & 1.30 $\pm$ 0.41           & 5.631             & 1.0 $\pm$ 0.3             & -              & -               & 0.076             & 3.2 $\pm$ 1.0           \\ [0.08cm]
0.0100    & $-$0.53 & 1.50 $\pm$ 0.46          & 5.725             & 1.6 $\pm$ 0.5             & -              & -              & 0.078             & 2.8 $\pm$ 1.0           \\ [0.08cm]
0.0120   & $-$0.45 & 1.50 $\pm$ 0.49           & 5.865             & 1.2 $\pm$ 0.4             & 1.071              & 0.8 $\pm$ 0.3              & 0.087              & 2.0 $\pm$ 0.6           \\ [0.08cm] \bottomrule
\hspace{20mm}
\end{tabular}
\caption{\small The calculated masses for the set of metallicity values and the star formation peaks along with their associated SFR for the 97 LPV stars from \cite{whitelock2013local} located in the bar region are listed.}
\label{tab:tab2}
\end{table*}

As previously explained, the limitation of the outer region catalog to spectroscopy-confirmed carbon stars puts constraints on the age of the obtained results for this region. Thus, the stellar populations we have retrieved for this region are no younger than $\sim$ 830 Myr (log $t = 8.92$), $\sim$ 900 Myr (log $t = 8.96$), and $\sim$ 680 Myr (log $t = 8.83$) based on AMR\,1, AMR\,2, and AMR\,3, respectively. Since the second episode of star formation in the bar region is not detected in the outer region, the extent of this burst was likely limited to the central bar of the galaxy. However, we cannot draw the same conclusion for the most recent episode of star formation burst in this area since no data is available for the recent epochs.

According to the findings from AMR\,1, the stellar mass formed in the bar region is estimated to be 1.80 $\pm$ 0.63 $\times$ 10$^{7}$ M$_\sun$. However, when considering the results from AMR\,2 and AMR\,3, slightly higher values of 2.10 $\pm$ 0.70 $\times$ 10$^{7}$ M$_\sun$ and 2.40 $\pm$ 0.87 $\times$ 10$^{7}$ M$_\sun$ are derived, respectively. Additionally, the stellar mass formed in the outer region amounts to 1.20 $\pm$ 0.46 $\times$ 10$^{7}$ M$_\sun$ (AMR\,1), 1.56 $\pm$ 0.61 $\times$ 10$^{7}$ M$_\sun$ (AMR\,2), and 1.85 $\pm$ 0.73 $\times$ 10$^{7}$ M$_\sun$ (AMR\,3), with respect to the age limit mentioned earlier. Furthermore, it is observed that approximately 3.5$\%$ of the mass present in the bar region formed within the last 200 Myr. This finding aligns with the value of 3$\%$ for the H VII region reported by \cite{wyder2001star}. Moreover, our results reveal that around 52$\%$ of the mass in the bar formed between $\sim$ 13 to 7 Gyr ago, consistent with the reported percentage of 50$\%$ for H VII region between 15--7 Gyr ago.

\begin{table*}
\centering
\begin{tabular}{cccclclcl}
\hline
Z       & {[}Fe/H{]} & Stellar Mass  & Peak 1 & SFR  & Peak 2    & SFR & Peak 3    & SFR  \\
        & \scriptsize(dex) & \scriptsize($10^{7}$ M$_\Sun$)  & (Gyr) & \scriptsize($10^{-3}$ M$_\Sun$yr$^{-1}$)   & \scriptsize(Gyr)    & \scriptsize($10^{-3}$ M$_\Sun$yr$^{-1}$) & \scriptsize(Gyr)    & \scriptsize($10^{-3}$ M$_\Sun$yr$^{-1}$)  \\

\hline
0.0001  & $-$2.53 & 4.97 $\pm$ 1.27          & 1.150            & 21.6 $\pm$ 5.6            & 0.568             & 36.5 $\pm$ 6.3              & 0.040             & 86.5$\pm$ 22.3            \\ [0.08cm]
0.0003  & $-$2.05 & 3.10 $\pm$ 0.79           & 1.932             & 19.4 $\pm$ 4.8             & 0.680             & 7.9 $\pm$ 2.0              & 0.047            & 29.0 $\pm$ 7.4           \\ [0.08cm]
0.0005  & $-$1.83 & 3.50 $\pm$ 0.89          & 2.356            & 6.8 $\pm$ 1.7             & 1.581             & 13.0 $\pm$ 3.3              & 0.042             & 63.3$\pm$ 16.3           \\ [0.08cm]
0.0008  & $-$1.63 & 2.70 $\pm$ 0.68           & 2.261            & 5.9 $\pm$ 1.5             & 1.700             & 19.5 $\pm$ 5.0              & 0.054              & 38.6$\pm$ 10.0           \\ [0.08cm]
0.0010  & $-$1.36 & 3.25 $\pm$ 0.83          & 2.370            & 5.3 $\pm$ 1.3             & 1.810              & 7.8 $\pm$ 2.0              & 0.071             & 32.2 $\pm$ 8.3           \\ [0.08cm]
0.0015  & $-$1.36 & 4.36 $\pm$ 1.11          & 2.785            & 6.0 $\pm$ 1.5             & 2.110              & 5.9 $\pm$ 1.5             & 0.069             & 46.5 $\pm$ 12.0           \\ [0.08cm]
0.0020   & $-$1.23 & 1.95 $\pm$ 0.50           & 2.370             & 6.1 $\pm$ 1.6             & -              & -              & 0.067              & 11.6 $\pm$ 3.0           \\ [0.08cm]
0.0030   & $-$1.05 & 2.39 $\pm$ 0.61           & 2.600             & 4.9 $\pm$ 1.3             & 0.800             & 1.9 $\pm$ 0.5              & 0.056              & 18.0 $\pm$ 5.2           \\ [0.08cm]
0.0060   & $-$0.75 & 3.23 $\pm$ 0.82           & 6.224             & 2.7 $\pm$ 0.7             & -              & -              & 0.090              & 3.8 $\pm$ 1.0           \\ [0.08cm]
0.0080   & $-$0.63 & 7.30 $\pm$ 1.30           & 10.983             & 3.6 $\pm$ 0.9             & 1.974              & 2.2 $\pm$ 0.5              & 0.103             & 2.3 $\pm$ 0.6           \\ [0.08cm]
0.0100    & $-$0.53 & 6.67 $\pm$ 1.68          & 9.636             & 4.4 $\pm$ 1.1             & -              & -              & 0.107             & 2.0 $\pm$ 0.5           \\ [0.08cm]
0.0120   & $-$0.45 & 6.63 $\pm$ 1.67           & 12.600             & 4.8 $\pm$ 1.2             & 6.841              & 4.4 $\pm$ 1.1              & 0.108              & 1.7 $\pm$ 0.4           \\ [0.08cm] \bottomrule
\hspace{20mm}
\end{tabular}
\caption{\small The same as Table \ref{tab:tab2} for the bar region catalog.}
\label{tab:tab3}
\end{table*}

\begin{table*}
\addtolength{\tabcolsep}{+7pt}
\centering
\begin{tabular}{ccccccl}
\hline
Z       & {[}Fe/H{]} & Stellar Mass  & Peak 1 & SFR  \\
        & \scriptsize(dex) & \scriptsize($10^{7}$ M$_\Sun$)  & (Gyr) & \scriptsize($10^{-3}$ M$_\Sun$yr$^{-1}$)   &   \\
\hline
0.0001  & $-$2.53 & 1.32 $\pm$ 0.42           & 1.073             & 13.7 $\pm$ 4.3             \\ [0.08cm]
0.0003  & $-$2.05 & 1.22 $\pm$ 0.39           & 2.058             & 8.7 $\pm$ 2.7             \\ [0.08cm]
0.0005  & $-$1.83 & 1.23 $\pm$ 0.40           & 1.655             & 7.8 $\pm$ 2.4               \\ [0.08cm]
0.0008  & $-$1.63 & 0.83 $\pm$ 0.27           & 2.124             & 2.7 $\pm$ 0.8               \\ [0.1cm]
0.0010   & $-$1.53 & 0.84 $\pm$ 0.27           & 2.197             & 3.2 $\pm$ 1.0               \\ [0.08cm]
0.0015  & $-$1.36 & 1.90 $\pm$ 0.60           & 2.603             & 3.7 $\pm$ 1.2               \\ [0.08cm]
0.0020   & $-$1.23 & 0.87 $\pm$ 0.28           & 2.422             & 4.0 $\pm$ 1.3             \\ [0.08cm]
0.0030   & $-$1.05 & 1.07 $\pm$ 0.34           & 3.000             & 2.6 $\pm$ 0.8             \\ [0.08cm]
0.0060   & $-$0.75 & 1.75 $\pm$ 0.56           & 5.542             & 1.3 $\pm$ 0.4              \\ [0.08cm]
0.0080   & $-$0.63 & 4.80 $\pm$ 1.51           & 5.804             & 1.8 $\pm$ 0.5               \\ [0.08cm]
0.0100    & $-$0.53 & 3.92 $\pm$ 1.24          & 8.865             & 2.2 $\pm$ 0.7               \\ [0.08cm]
0.0120   & $-$0.45 & 3.62 $\pm$ 1.15           & 11.546             & 1.8 $\pm$ 0.6              \\ [0.08cm] \bottomrule
\hspace{10mm}
\end{tabular}
\caption{\small The same as Table \ref{tab:tab2} for the outer region catalog.}
\label{tab:tab4}
\end{table*}


\section{Discussion} \label{sec:sec5}

\subsection{Variation of assumptions} \label{subsec:sec5.1}

In this subsection, we derive the variations in the SFH results considering different assumptions.  We first examine the effect of changing the adopted distance module, $\mu$, and the extinction values, A$_{k}$ and A$_{J}$. Subsequently, we investigate the SFH using BaSTI stellar tracks and isochrones as an alternative to the Padova models.

\subsubsection{Distance and Extinction} \label{subsec:sec5.1.1}

We utilize the distance modulus, $\mu$, to obtain the absolute magnitude of the stars in our sample, which we use to derive their mass and age, hence the SFR and the SFH. In this study, we have adopted a distance of $\mu =$ 23.450 mag, obtained by observing the $I$ magnitude of the RGB tip (\citealp{lee1993tip}). 

To see the effect of a different distance modulus, we adopt the value of $\mu =$ 23.312 $\pm$ 0.021 mag derived from Cepheid variables (\citealp{gieren2006araucaria}), which has the most substantial deviation from our initial chosen value. Furthermore, interstellar extinction is another factor affecting the stars' magnitude and color. Consequently, it might affect the SFH as well. The infrared light is less influenced by the dust compared to shorter wavelengths. However, even a minor effect must be considered. 

\begin{figure*}[ht!]
\hspace{20mm}
	{\hbox
    { \epsfig{figure=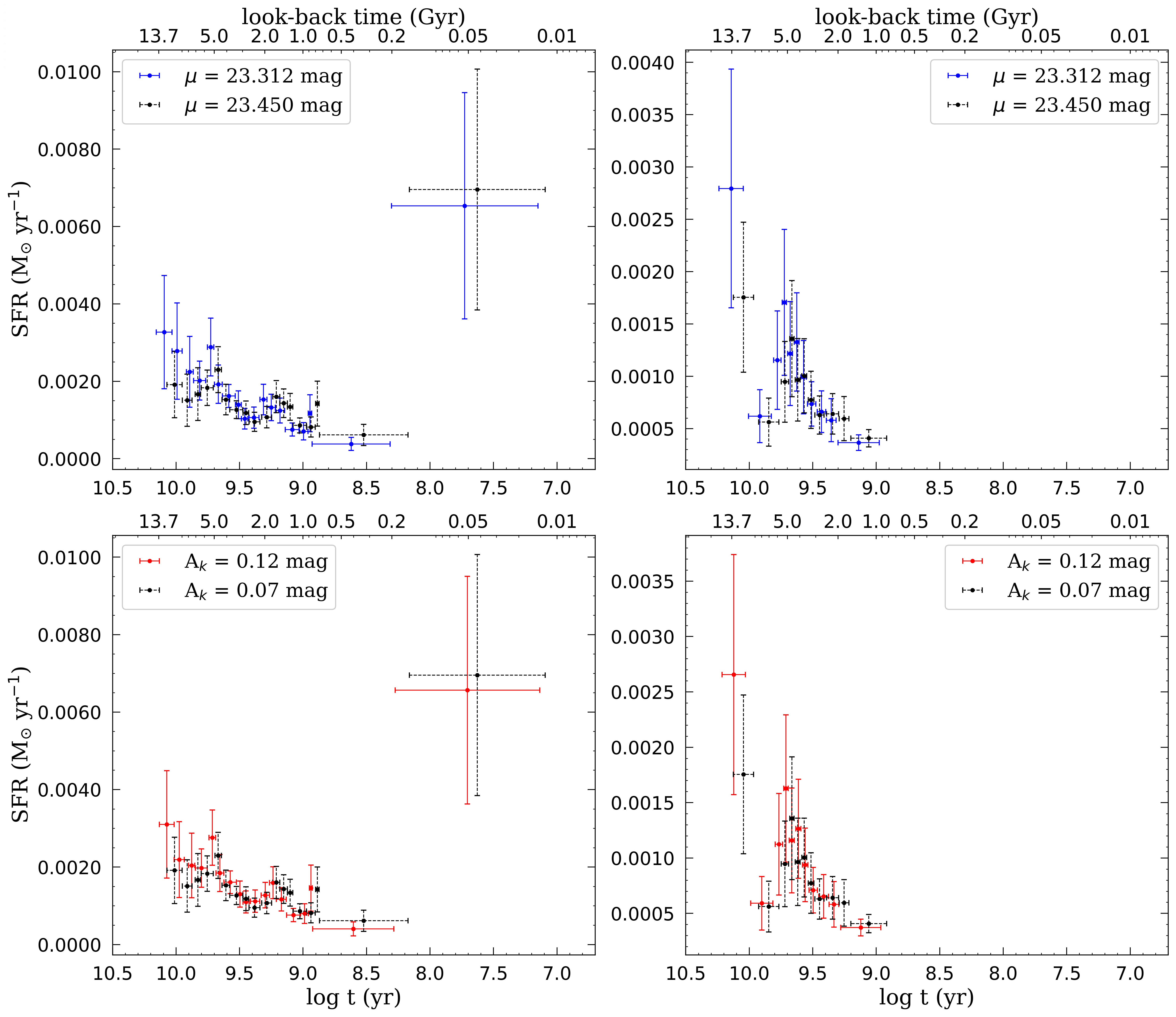,width=140mm,height=140mm}
   \centering
	}}
	\caption{\small SFH derived for the bar (left panels) and the outer regions (right panels) after adopting alternative values for distance modulus ($\mu$), shown in the upper panels (in blue/solid), and galactic extinction (A$_{k}$) illustrated in the lower panels (in red/solid), compared to the original values (in black dashed).}
	\label{fig: Fig. 9}
\end{figure*}

In this work, similar to \cite{whitelock2013local}, we have adopted the extinction values of A$_{k} =$ 0.07 mag and A$_{J} =$ 0.20 mag (\citealp{clementini2003rr}; \citealp{schlegel1998maps}), and by using alternative values of A$_{k} =$ 0.12 mag and A$_{J} =$ 0.30 mag, considering the \textit{E(B$-$V)} $=$ 0.35 $\pm$ 0.04 mag (\citealp{rich2014new}) and using the information of \cite{schlegel1998maps}, we aim to investigate how the SFH varies. As shown in Fig.\,\ref{fig: Fig. 9}, the results are derived considering the AMR\,1 for the bar region (left column) and the outer region (right column). The overall pattern of SFH remains the same in both cases; however, the SFH shifts slightly backward in time, resulting in the onset of star formation $\sim$ 2 Gyr and $\sim$ 3 Gyr earlier in the bar and the outer region, respectively.

\subsubsection{Stellar evolutionary models} \label{subsec:sec5.1.2}

As explained in the previous sections, we discussed the method employed in this study, which is based on the relations between the stars' birth mass, age, and pulsation duration. We derived these parameters using the Padova stellar evolutionary tracks and isochrones. The Padova models cover a wide range of birth mass (0.8 $<$ M/ M$_\sun$ $<$ 30 ) and are available for numerous optical and infrared photometric systems (for more details see \citealp{Javadi11}). 

While various models are available, the Padova models are the most comprehensive for our method. However, we also considered the Bag of Stellar Tracks and Isochrones (BaSTI; \citealp{pietrinferni2013basti}; \citealp{cassisi2006basti}) models to compare them with the Padova models and examine the sensitivity of our results to the choice of model. BaSTI contains a set of theoretical stellar evolutionary models and isochrones based on the scaled-solar and $\alpha$-enhanced heavy-element distributions. These models include the core convective overshooting during the H-burning phase and cover the stellar mass range of 0.5  $<$ M/M$_\sun$ $<$ 15  (\citealp{hidalgo2018updated}), a lower range compared to Padova. The BaSTI models are computed for several metallicity values, but only a few are consistent with our assumed values. In this subsection, we compare the models adopting the mean metallicity value of [Fe/H] $= -$ 1.05 dex (Z $\approx$ 0.003).

\begin{figure*}[ht!]
\hspace{20mm}
	{\hbox
    { \epsfig{figure=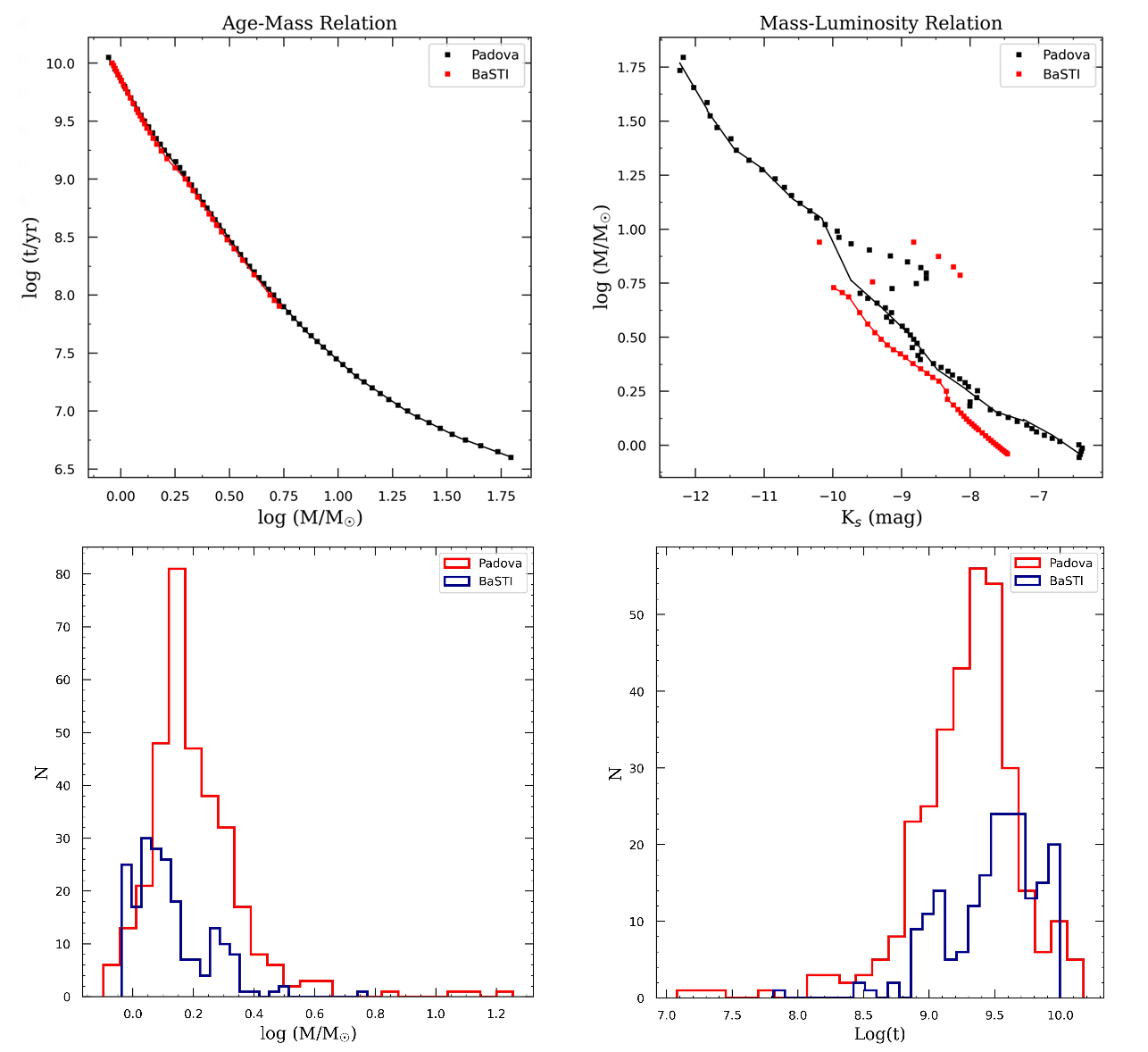,width=140mm,height=140mm}
   \centering
	}}
	\caption{Upper panels: Age--mass  (left) and Mass--luminosity (right) relations for the $K_s$-band magnitude, assuming a mean metallicity of $Z \approx 0.003$ based on the BaSTI models. The solid lines indicate the best-fitting linear relations, $\log{\mathrm{t}} = a\,\log{\mathrm{(M/M_\sun)}} + b$ and $\log{\mathrm{(M/M_\sun})} = a\,\mathrm{K_s} + b$, respectively. Lower panels: Comparison of the stellar properties derived from the BaSTI and Padova models. Left panel: stellar mass distributions; right panel: stellar age distributions.}
	\label{fig: Fig. 10}
\end{figure*}

The mass-luminosity and age-mass relations derived from the Padova and BaSTI models are illustrated in Fig.\,\ref{fig: Fig. 10}. The key difference is that for the stars fainter than $\sim$ $-$10 mag, the BaSTI models yield a lower mass and older age. However, as shown in the upper-left panel of Fig.\,\ref{fig: Fig. 10}, the age obtained using BaSTI models does not differ significantly from that derived using the Padova. Furthermore, as evident in the upper-right panel of Fig.\,\ref{fig: Fig. 10}, the mass range covered by BaSTI models is limited and does not include the mass of RSGs. 

The lower panels of Fig. \ref{fig: Fig. 10} compare the stellar properties inferred from the BaSTI and Padova models. The left panel presents the stellar mass distributions, while the right panel shows the corresponding age distributions. The BaSTI-based results exhibit a sharp decline in the number of stars, with several sources entirely absent at the youngest ages due to the limited age coverage of the models. 

Additionally, as explained in Section\,\ref{sec:sec3}, pulsation duration is one of the parameters we require to calculate the SFR; however, the BaSTI models lack the necessary information to calculate this parameter. Hence, we are unable to derive SFH utilizing the BaSTI models.



\subsection{Star Formation History of NGC 6822} \label{subsec:sec5.2}

Having derived the SFH of NGC\,6822, we identified three major episodes of star formation burst throughout the lifetime of the galaxy (see Section\,\ref{sec:sec4}). \cite{weisz2014star} have shown that dIrrs experience an initial, modest burst of star formation early in their lives ($>$ 12$-$10 Gyr), which is followed by a constant or slightly increasing SFR. However, despite expectations for a typical dIrr, the presence of such noticeable bursts during the lifespan of NGC\,6822 suggests that the galaxy may have undergone interactions despite its significant isolation at the time. \cite{gallart1996local2} studied the early SFH of the galaxy by comparing the observed and model CMDs for the old- and intermediate-age star populations. Their findings suggest that the galaxy's star formation began approximately 15--12 Gyr ago from low metallicity gas (Z $=$ 0.0001), consistent with our findings for the onset of the star formation about  13 Gyr, adopting constant metallicities and AMRs. Additionally, detecting 20 RR Lyrae stars, which are typically older than 10 Gyr, by \cite{baldacci2003variable} using the Very Large Telescope (VLT) confirms the early star formation activation in the galaxy.

According to our results, the first major episode of star formation peaks $\sim$ 4--5 Gyr ago in the bar and outer regions. For a galaxy as isolated as NGC\,6822, experiencing a disturbance that significantly stimulates star formation is not quite common. Consequently, each burst of star formation must be associated with a remarkable event, such as a collision or a tidal interaction with neighboring galaxies.
 
\cite{zhang2021panoramic} used the Gaia Early Data Release\,3 (Gaia EDR\,3) to obtain measurements for the galaxy's proper motion, position, distance, and line of sight velocity. The orbital history of NGC~6822 provides an independent indication that a past encounter with the Milky Way is dynamically possible. Using Gaia EDR3 proper motions together with the present-day distance, line-of-sight velocity, and position of NGC~6822, \citet{zhang2021panoramic} integrated its orbit backwards in a Milky Way potential with a total mass of $1.3\times10^{12}\,M_{\odot}$. Their preferred orbital solution indicates that NGC~6822 approached the Milky Way from distances of approximately $500$--$600$~kpc and may have passed within the Milky Way virial radius, assumed to be $\sim300$~kpc, approximately $3$--$4$~Gyr ago. In their Monte Carlo analysis, $10^5$ orbital realizations were generated by sampling the observational uncertainties in distance, proper motion, and line-of-sight velocity. Approximately $57\%$ of these realizations passed within $300$~kpc of the Milky Way, with a median pericentric time of $\sim3.1$~Gyr ago \citep{zhang2021panoramic}. These results constrain the timing of a possible close passage, but do not by themselves establish the duration of the interaction. Therefore, the results are more appropriately interpreted as indicating a possible pericentric passage approximately $3$--$4$~Gyr ago, followed by the subsequent dynamical evolution of NGC~6822.

The similarity between the inferred epoch of the possible pericentric passage and the star formation enhancement identified in our SFH further motivates the possibility of a connection between the two events. In addition, \citet{battaglia2022gaia} examined the proper motions of Local Group dwarf irregular galaxies using Gaia EDR3 and found results consistent with a past passage of NGC~6822 within the Milky Way virial radius. However, establishing a causal connection between this possible encounter and the observed HI structures requires dedicated dynamical modelling.

Hydrodynamical $N$-body simulations using codes such as AREPO \citep{weinberger2020arepo} or GIZMO \citep{hopkins2015new}, initialized with the allowed orbital parameters, could test whether a pericentric passage $\sim3$--$4$~Gyr ago can reproduce the observed HI asymmetry and extended arm. The simulated HI morphology and velocity field could then be compared with observations to quantify the impact of the Milky Way encounter. Deeper HI observations with facilities such as FAST or MeerKAT would further constrain the low-column-density gas and its kinematics, providing additional observational tests of the interaction scenario.

\cite{williamson2021evolution} have performed a series of chemodynamical simulations of interacting Magellanic-like dwarfs. Their results show that the tidal forces between the interacting dwarfs can indirectly affect and trigger star formation, increasing it by growing instabilities and internal gravitational collapse. This is in agreement with \cite{ellison2013galaxy}'s study, which shows that star formation is triggered in the early stage of encounter in merging systems; such tidal interactions can turn non-star-forming galaxies into star-forming ones. Additionally, the coeval star formation burst in both the Large and Small Magellanic Clouds has been attributed to tidal interactions between the clouds or with the Milky Way (e.g. \citealp{bekki2005formation}; \citealp{indu2011recent}; \citealp{2014MNRAS.445.2214R}).

In addition to initiation of star formation bursts, the tidal forces caused by interactions can affect and distort the morphology of interacting galaxies (\citealp{williamson2021evolution}). Tidal features such as streams, bridges, and tidal arms can be seen in interacting galaxies, for instance, the streams and bridge between the Magellanic Clouds (e.g. \citealp{nidever2008origin}; \citealp{belokurov2017clouds}).

Moreover, the recent JWST/NIRCam observation of the interacting system of NGC\,4485 and NGC\,4490 further studies the tidal bridge connecting the two galaxies (\citealp{bortolini2025feast}). Additionally, \cite{williamson2021evolution} have shown that even in the simulations that do not result in a collision between the interacting dwarfs, the tidal features are still apparent; after the close encounter, the systems regain their fairly regular disk structure, however, with a relative inclination. 
The complex HI morphology of NGC~6822 provides possible evidence for past gravitational perturbations. The extended HI distribution is characterized by an asymmetric structure, a prominent HI arm, and a large HI hole in the southeastern part of the galaxy \citep{de2000evidence}. These features, including the extended HI distribution and the main disturbed structures, are presented in Figs.~1 and 2 of \citet{de2000evidence} and Fig.~1 of \citet{cannon2012origin}. However, the physical origin of these structures is not uniquely established. In particular, \citet{cannon2012origin} showed that stellar feedback can plausibly account for the formation and persistence of the large HI hole, while the nature of the putative companion or interacting HI complex remains uncertain. Thus, the present HI morphology may be indicative of past external perturbations.

As explained in the results section, the second episode of star formation occurred around 1.6 Gyr ago, mainly concentrated on the central part of the galaxy. The identification of similar star formation episodes in the galaxy by two other studies, \cite{wyder2001star} and \cite{mcquinn2010nature}, dismisses the possibility of a false peak.

Additionally, our analysis shows that NGC\,6822 has experienced a significant increase in its SFR over the past 250 Myr, indicating that the galaxy has probably undergone tidal interactions or a merger event. Similarly, \cite{ren2024star} estimated that an interaction event has occurred $\sim$ 75$-$200 Myr ago. \cite{de2000evidence} used the Australia Telescope Compact Array to map the HI envelope of NGC\,6822. They observed a velocity difference at the interface between the NW cloud and the main body of the galaxy, and with the asymmetric distribution of the HI envelope, they suggested that the HI complex might be a separate system interacting with NGC\,6822 and triggering the star formation in recent epochs. Additionally, they have derived a time scale of $\sim$ 300 Myr for the interaction of the NW cloud and the galaxy, similar to the time scale we derived for the recent star formation burst. 

Several other studies have also observed and reported the recent significant enhancement in the SFR. \citet{gallart1996local} derived that the SFR has increased in the past 400 Myr, with a higher rate in the galaxy's central region than its outer regions. Moreover, \cite{park2022gas} have found that the SFR increases as the radius decreases, likely because the SFR is higher toward the center of the galaxy. Furthermore, \cite{wyder2001star, wyder2003star} studied five fields alongside the bar and four regions perpendicular to it, and they also observed a similar gradient in star formation activity within NGC\,6822.  We also find that the SFR is notably high in the bar region at recent times, which covers the galaxy’s central area. However, because the outer-region catalog is limited to carbon stars, we cannot recover the recent star formation history of the outer regions; consequently, the radial gradient of recent star formation activity cannot be investigated in this study.

\section{Summary} \label{sec:sec6}

In this work, we studied the evolution and SFH of the dwarf irregular galaxy, NGC\,6822, using the data of 329 LPV stars in JHK$_s$ bands. We have derived the SFH for two regions: the bar region covers a central rectangular area of 189 arcmin$^{2}$, and the outer region encircles a field beyond the bar region up to  a radial distance of $\sim$ 3 kpc. The SFH is derived considering constant metallicities and adopting AMRs to account for the variations in the metallicity over time. In summary, our results can be outlined as follows:

\begin{itemize}
  \item In the bar region, two star formation episodes were identified $\sim$ 4.6 Gyr and 1.6 Gyr ago. The most recent episode occurred over the past $\sim$ 150 Myr, with a SFR of 7.0 $\pm$ 3.1 $\times$ $10^{-3}$ M$_\sun$\,yr$^{-1}$.
  \item In the outer region, the limitation of the catalog to spectroscopy-confirmed carbon stars restricts the age range of our results; hence, we could not retrieve any stellar populations for recent epochs and only identified 1 episode of star formation at $\sim$ 4.6 Gyr ago. 
  \item The first episode of star formation has occurred in both the bar
and outer regions $\sim$ 4$-$5 Gyr ago. Despite the noticeable isolation of
NGC\,6822, the galaxy has undergone episodes of star formation. These
bursts could be due to possible interactions with other distant galaxies,
like the Milky Way. Other studies find that such an interaction between
NGC\,6822 and the Milky Way may have taken place within a similar time
period, which is consistent with it having triggered these bursts.
  \item The short-lasting increase in SFR $\sim$ 1.6 Gyr ago in the bar region and the recent burst of star formation in NGC\,6822, could both plausibly be associated with tidal interactions. 
\end{itemize}

The principal systematic uncertainties affecting our derived star formation histories are associated with the pulsation duration of AGB stars ($  \delta t  $), the choice of initial mass function, and photometric calibration/distance modulus. Variations in $  \delta t  $ (up to a factor of $\sim$2, depending on the adopted stellar evolution models) introduce uncertainties of up to $\sim$0.3 dex in the SFR normalization. The choice between a Kroupa (\citealp{Kroupa01}) and Salpeter (\citealp{Salpeter55}) IMF produces a systematic offset of $\approx$0.26 dex. Finally, magnitude shifts of $\pm$0.2 mag (encompassing photometric and distance uncertainties) result in mean variations in the inferred SFR of 20--40\% ($\approx$0.1--0.2 dex), with the largest relative differences occurring in low-SFR age bins. Importantly, these systematics primarily affect the overall normalization of the star formation rate while leaving the shape and timing of the main star formation episodes largely unchanged.

Investigating the influence of interactions on the SFH of NGC\,6822 is challenging. Nonetheless, deep high-resolution observations of HI regions with radio telescopes, e.g. the Five-hundred-meter Aperture Spherical radio Telescope (FAST), allow detailed maps of the faint diffuse gas and the structure of the HI envelope to be obtained. Highly resolved maps can reveal the spatial distribution of gas across the galaxy and tidal features, formed due to possible interactions with neighboring objects such as the Milky Way. Additionally, by implementing techniques such as High-resolution Adaptive Mesh Refinement (AMR) in hydrodynamical simulations, the gas kinematics and turbulence, like the shock experienced by the ISM due to tidal interactions or stellar feedback, can be studied more accurately. By distinguishing the internal and external physical processes that initiate the star formation bursts, it may be possible to shed more light on the evolution of NGC\,6822.


\section*{Acknowledgments} \label{sec:sec8}

The authors thank the School of Astronomy at the Institute for Research in Fundamental Sciences (IPM) and
the Iranian National Observatory (INO) for supporting
this project. HA was supported by the `SeismoLab' KKP-137523 \'Elvonal grant of the Hungarian Research, Development and Innovation Office (NKFIH).


\bibliography{Bibliography.bib}

@ARTICLE{Marigo07,
       author = {{Marigo}, P. and {Girardi}, L.},
        title = "{Evolution of asymptotic giant branch stars. I. Updated synthetic TP-AGB models and their basic calibration}",
      journal = {\aap},
         year = 2007,
        month = jul,
       volume = {469},
       number = {1},
        pages = {239-263},
          doi = {10.1051/0004-6361:20066772},
archivePrefix = {arXiv},
       eprint = {astro-ph/0703139},
 primaryClass = {astro-ph},
       adsurl = {https://ui.adsabs.harvard.edu/abs/2007A&A...469..239M}
}

@ARTICLE{Javadi11,
       author = {{Javadi}, Atefeh and {van Loon}, Jacco Th. and {Mirtorabi}, Mohammad Taghi},
        title = "{The UK Infrared Telescope M33 monitoring project - II. The star formation history in the central square kiloparsec}",
      journal = {\mnras},
         year = 2011,
        month = jul,
       volume = {414},
       number = {4},
        pages = {3394-3409},
          doi = {10.1111/j.1365-2966.2011.18638.x},
archivePrefix = {arXiv},
       eprint = {1103.0755},
 primaryClass = {astro-ph.SR},
       adsurl = {https://ui.adsabs.harvard.edu/abs/2011MNRAS.414.3394J}
}

@ARTICLE{Javadi17,
       author = {{Javadi}, Atefeh and {van Loon}, Jacco Th. and {Khosroshahi}, Habib G. and {Tabatabaei}, Fatemeh and {Hamedani Golshan}, Roya and {Rashidi}, Maryam},
        title = "{The UK Infrared Telescope M 33 monitoring project - V. The star formation history across the galactic disc}",
      journal = {\mnras},
         year = 2017,
        month = jan,
       volume = {464},
       number = {2},
        pages = {2103-2119},
          doi = {10.1093/mnras/stw2463},
archivePrefix = {arXiv},
       eprint = {1610.00334},
 primaryClass = {astro-ph.GA},
       adsurl = {https://ui.adsabs.harvard.edu/abs/2017MNRAS.464.2103J}
}

@ARTICLE{Kroupa01,
       author = {{Kroupa}, Pavel},
        title = "{On the variation of the initial mass function}",
      journal = {\mnras},
         year = 2001,
        month = apr,
       volume = {322},
       number = {2},
        pages = {231-246},
          doi = {10.1046/j.1365-8711.2001.04022.x},
archivePrefix = {arXiv},
       eprint = {astro-ph/0009005},
 primaryClass = {astro-ph},
       adsurl = {https://ui.adsabs.harvard.edu/abs/2001MNRAS.322..231K}
}

@ARTICLE{Salpeter55,
       author = {{Salpeter}, Edwin E.},
        title = "{The Luminosity Function and Stellar Evolution.}",
      journal = {\apj},
         year = 1955,
        month = jan,
       volume = {121},
        pages = {161},
          doi = {10.1086/145971},
       adsurl = {https://ui.adsabs.harvard.edu/abs/1955ApJ...121..161S}
}

@ARTICLE{Reichardt01,
       author = {{Reichardt}, Christian and {Jimenez}, Raul and {Heavens}, Alan F.},
        title = "{Recovering physical parameters from galaxy spectra using MOPED}",
      journal = {\mnras},
         year = 2001,
        month = nov,
       volume = {327},
       number = {3},
        pages = {849-867},
          doi = {10.1046/j.1365-8711.2001.04768.x},
archivePrefix = {arXiv},
       eprint = {astro-ph/0101074},
 primaryClass = {astro-ph},
       adsurl = {https://ui.adsabs.harvard.edu/abs/2001MNRAS.327..849R}
}

@article{hofner2020explaining,
  title={Explaining the winds of AGB stars: Recent progress},
  author={H{\"o}fner, Susanne and Freytag, Bernd},
  journal={Proceedings of the International Astronomical Union},
  volume={16},
  number={S366},
  pages={165--172},
  year={2020},
  publisher={Cambridge University Press}
}

@article{boyer2009spitzer,
  title={A Spitzer study of asymptotic giant branch stars. III. Dust production and gas return in local group Dwarf irregular galaxies},
  author={Boyer, Martha L and Skillman, Evan D and van Loon, Jacco Th and Gehrz, Robert D and Woodward, Charles E},
  journal={The Astrophysical Journal},
  volume={697},
  number={2},
  pages={1993},
  year={2009},
  publisher={IOP Publishing}
}

@article{karakas2014dawes,
  title={The Dawes review 2: nucleosynthesis and stellar yields of low-and intermediate-mass single stars},
  author={Karakas, Amanda I and Lattanzio, John C},
  journal={Publications of the Astronomical Society of Australia},
  volume={31},
  pages={e030},
  year={2014},
  publisher={Cambridge University Press}
}

@article{boyer2015infrared,
  title={An infrared census of DUST in nearby galaxies with spitzer (DUSTiNGS). II. Discovery of metal-poor dusty AGB stars},
  author={Boyer, Martha L and McQuinn, Kristen BW and Barmby, Pauline and Bonanos, Alceste Z and Gehrz, Robert D and Gordon, Karl D and Groenewegen, MAT and Lagadec, Eric and Lennon, Daniel and Marengo, Massimo and others},
  journal={The Astrophysical Journal},
  volume={800},
  number={1},
  pages={51},
  year={2015},
  publisher={IOP Publishing}
}

@article{hodge1977structure,
  title={The structure and content of NGC 6822.},
  author={Hodge, Paul W},
  journal={Astrophysical Journal, Suppl. Ser., Vol. 33, p. 69-82, plates 13-16},
  volume={33},
  pages={69--82},
  year={1977}
}

@article{ekstrom2013red,
  title={Red supergiants and stellar evolution},
  author={Ekstr{\"o}m, Sylvia and Georgy, Cyril and Meynet, Georges and Groh, Jos{\'e} and Granada, Anah{\i}},
  journal={EAS Publications Series},
  volume={60},
  pages={31--41},
  year={2013},
  publisher={EDP Sciences}
}

@article{iben1983asymptotic,
  title={Asymptotic giant branch evolution and beyond},
  author={Iben Jr, Icko and Renzini, Alvio},
  journal={IN: Annual review of astronomy and astrophysics. Volume 21 (A84-10851 01-90). Palo Alto, CA, Annual Reviews, Inc., 1983, p. 271-342.},
  volume={21},
  pages={271--342},
  year={1983}
}

@article{whitelock2003obscured,
  title={Obscured asymptotic giant branch variables in the Large Magellanic Cloud and the period--luminosity relation},
  author={Whitelock, Patricia A and Feast, Michael W and Loon, Jacco Th van and Zijlstra, Albert A},
  journal={Monthly Notices of the Royal Astronomical Society},
  volume={342},
  number={1},
  pages={86--104},
  year={2003},
  publisher={Blackwell Science Ltd Oxford, UK}
}

@article{hodge1991cosmos,
  title={A COSMOS study of the structure and content of NGC 6822},
  author={Hodge, Paul and Smith, Toby and Eskridge, Paul and MacGillivray, Harvey and Beard, Steven},
  journal={Astrophysical Journal, Part 1 (ISSN 0004-637X), vol. 379, Oct. 1, 1991, p. 621-630.},
  volume={379},
  pages={621--630},
  year={1991}
}

@article{de2000evidence,
  title={Evidence for tidal interaction and a supergiant hi shell in the local group dwarf galaxy ngc 6822},
  author={De Blok, WJG and Walter, F},
  journal={The Astrophysical Journal},
  volume={537},
  number={2},
  pages={L95},
  year={2000},
  publisher={IOP Publishing}
}

@article{jones2019young,
  title={The young stellar population of the metal-poor galaxy NGC 6822},
  author={Jones, Olivia C and Sharp, Michael J and Reiter, Megan and Hirschauer, Alec S and Meixner, M and Srinivasan, Sundar},
  journal={Monthly Notices of the Royal Astronomical Society},
  volume={490},
  number={1},
  pages={832--847},
  year={2019},
  publisher={Oxford University Press}
}

@article{wyder2001star,
  title={The star formation history of NGC 6822},
  author={Wyder, Ted K},
  journal={The Astronomical Journal},
  volume={122},
  number={5},
  pages={2490},
  year={2001},
  publisher={IOP Publishing}
}

@article{belland2020ngc,
  title={NGC 6822 as a probe of dwarf galactic evolution},
  author={Belland, Brent and Kirby, Evan and Boylan-Kolchin, Michael and Wheeler, Coral},
  journal={The Astrophysical Journal},
  volume={903},
  number={1},
  pages={10},
  year={2020},
  publisher={IOP Publishing}
}

@article{gallart1996local,
  title={The Local Group dwarf irregular galaxy NGC 6822. I. The stellar content},
  author={Gallart, C and Aparicio, A and Vilchez, JM},
  journal={The Astronomical Journal},
  volume={112},
  pages={1928},
  year={1996}
}

@article{gallart1996local2,
  title={The Local Group dwarf irregular galaxy NGC 6822. II. The old and intermediate-age star formation history},
  author={Gallart, C and Aparicio, A and Bertelli, G and Chiosi, C},
  journal={Astronomical Journal v. 112, p. 1950},
  volume={112},
  pages={1950},
  year={1996}
}

@article{gallart1996local3,
  title={The local group dwarf irregular galaxy NGC 6822. III. The recent star formation history},
  author={Gallart, C and Aparicio, A and Bertelli, G and Chiosi, C},
  journal={Astronomical Journal v. 112, p. 2596},
  volume={112},
  pages={2596},
  year={1996}
}

@article{whitelock2013local,
  title={The Local Group galaxy NGC 6822 and its asymptotic giant branch stars},
  author={Whitelock, Patricia A and Menzies, John W and Feast, Michael W and Nsengiyumva, Francois and Matsunaga, Noriyuki},
  journal={Monthly Notices of the Royal Astronomical Society},
  volume={428},
  number={3},
  pages={2216--2231},
  year={2013},
  publisher={Oxford University Press}
}

@article{sibbons2012agb,
  title={The AGB population of NGC 6822: distribution and the C/M ratio from JHK photometry},
  author={Sibbons, LF and Ryan, Sean G and Cioni, M-RL and Irwin, M and Napiwotzki, R},
  journal={Astronomy \& Astrophysics},
  volume={540},
  pages={A135},
  year={2012},
  publisher={EDP Sciences}
}

@article{sibbons2015spectral,
  title={Spectral classification of photometrically selected AGB candidates in NGC 6822},
  author={Sibbons, LF and Ryan, Sean G and Napiwotzki, Ralf and Thompson, GP},
  journal={Astronomy \& Astrophysics},
  volume={574},
  pages={A102},
  year={2015},
  publisher={EDP Sciences}
}

@article{kacharov2012spectra,
  title={Spectra probing the number ratio of C-to M-type AGB stars in the NGC 6822 galaxy},
  author={Kacharov, Nikolay and Rejkuba, Marina and Cioni, M-RL},
  journal={Astronomy \& Astrophysics},
  volume={537},
  pages={A108},
  year={2012},
  publisher={EDP Sciences}
}

@article{cioni2005near,
  title={Near-IR observations of NGC 6822: AGB stars, distance, metallicity and structure},
  author={Cioni, M-RL and Habing, HJ},
  journal={Astronomy \& Astrophysics},
  volume={429},
  number={3},
  pages={837--850},
  year={2005},
  publisher={EDP Sciences}
}

@article{schlegel1998maps,
  title={Maps of dust infrared emission for use in estimation of reddening and cosmic microwave background radiation foregrounds},
  author={Schlegel, David J and Finkbeiner, Douglas P and Davis, Marc},
  journal={The Astrophysical Journal},
  volume={500},
  number={2},
  pages={525},
  year={1998},
  publisher={IOP Publishing}
}

@article{carigi2006chemical,
  title={Chemical and photometric evolution of the local group galaxy NGC 6822 in a cosmological context},
  author={Carigi, Leticia and Col{\'\i}n, Pedro and Peimbert, Manuel},
  journal={The Astrophysical Journal},
  volume={644},
  number={2},
  pages={924},
  year={2006},
  publisher={IOP Publishing}
}

@article{battinelli2006photometric,
  title={Photometric survey of the polar ring galaxy NGC 6822},
  author={Battinelli, P and Demers, S and Kunkel, WE},
  journal={Astronomy \& Astrophysics},
  volume={451},
  number={1},
  pages={99--108},
  year={2006},
  publisher={EDP Sciences}
}

@article{hwang2005discovery,
  title={Discovery of remote star clusters in the halo of the Irregular galaxy NGC 6822},
  author={Hwang, Narae and Lee, Myung Gyoon and Lee, Jong Chul and Park, Won-Kee and Park, Hong Soo and Park, Jang-Hyun and Sohn, Sangmo Tony and Lee, Sang-Gak and Lee, Hyung Mok and Chun, Mun-Suk and others},
  journal={Proceedings of the International Astronomical Union},
  volume={1},
  number={C198},
  pages={257--258},
  year={2005},
  publisher={Cambridge University Press}
}

@book{hubble1982realm,
  title={The realm of the nebulae},
  author={Hubble, Edwin Powell},
  volume={25},
  year={1982},
  publisher={Yale University Press}
}

@misc{de1991third,
  title={Third Reference Catalogue of Bright Galaxies, Version 3.9. Springer, New York, NY},
  author={De Vaucouleurs, G and De Vaucouleurs, A and Corwin, JHG and Buta, RJ and Paturel, G and Fouque, P},
  year={1991}
}

@article{hodge1980recent,
  title={The recent evolutionary history of the galaxies NGC 6822 and IC 1613},
  author={Hodge, PW},
  journal={The Astrophysical Journal},
  volume={241},
  pages={125--131},
  year={1980}
}

@article{audouze1976chemical,
  title={Chemical evolution of galaxies},
  author={Audouze, Jean and Tinsley, Beatrice M},
  journal={Annual Review of Astronomy and Astrophysics},
  volume={14},
  number={1},
  pages={43--79},
  year={1976},
  publisher={Annual Reviews 4139 El Camino Way, PO Box 10139, Palo Alto, CA 94303-0139, USA}
}

@article{hwang2014spectroscopic,
  title={Spectroscopic study of extended star clusters in dwarf Galaxy NGC 6822},
  author={Hwang, Narae and Park, Hong Soo and Lee, Myung Gyoon and Lim, Sungsoon and Hodge, Paul W and Kim, Sang Chul and Miller, Bryan and Weisz, Daniel},
  journal={The Astrophysical Journal},
  volume={783},
  number={1},
  pages={49},
  year={2014},
  publisher={IOP Publishing}
}

@article{venn2001first,
  title={First stellar abundances in NGC 6822 from VLT-UVES and Keck-HIRES spectroscopy},
  author={Venn, KA and Lennon, DJ and Kaufer, A and McCarthy, JK and Przybilla, N and Kudritzki, RP and Lemke, M and Skillman, ED and Smartt, SJ},
  journal={The Astrophysical Journal},
  volume={547},
  number={2},
  pages={765},
  year={2001},
  publisher={IOP Publishing}
}

@article{hwang2011extended,
  title={Extended star clusters in the remote halo of the intriguing dwarf galaxy NGC 6822},
  author={Hwang, Narae and Lee, Myung Gyoon and Lee, Jong Chul and Park, Won-Kee and Park, Hong Soo and Kim, Sang Chul and Park, Jang-Hyun},
  journal={The Astrophysical Journal},
  volume={738},
  number={1},
  pages={58},
  year={2011},
  publisher={IOP Publishing}
}

@article{davidge2003metallicity,
  title={The Metallicity of the Red Giant Branch in the Disk of NGC 6822},
  author={Davidge, TJ},
  journal={Publications of the Astronomical Society of the Pacific},
  volume={115},
  number={808},
  pages={635},
  year={2003},
  publisher={IOP Publishing}
}

@article{tolstoy2001using,
  title={Using the Ca ii triplet to trace abundance variations in individual red giant branch stars in three nearby galaxies},
  author={Tolstoy, Eline and Irwin, Michael J and Cole, Andrew A and Pasquini, L and Gilmozzi, R and Gallagher, JS},
  journal={Monthly Notices of the Royal Astronomical Society},
  volume={327},
  number={3},
  pages={918--938},
  year={2001},
  publisher={Blackwell Science Ltd Oxford, UK}
}

@article{kirby2013universal,
  title={The universal stellar mass--stellar metallicity relation for dwarf galaxies},
  author={Kirby, Evan N and Cohen, Judith G and Guhathakurta, Puragra and Cheng, Lucy and Bullock, James S and Gallazzi, Anna},
  journal={The Astrophysical Journal},
  volume={779},
  number={2},
  pages={102},
  year={2013},
  publisher={IOP Publishing}
}

@article{muschielok1999vlt,
  title={VLT FORS spectra of blue supergiants in the Local Group galaxy NGC 6822},
  author={Muschielok, B and Kudritzki, RP and Appenzeller, I and Bresolin, F and Butler, K and G{\"a}ssler, W and H{\"a}fner, R and Hess, HJ and Hummel, W and Lennon, DJ and others},
  journal={Astronomy and Astrophysics, v. 352, p. L40-L44 (1999)},
  volume={352},
  pages={L40--L44},
  year={1999}
}

@article{patrick2015red,
  title={Red Supergiant Stars as Cosmic Abundance Probes: KMOS Observations in NGC 6822},
  author={Patrick, LR and Evans, CJ and Davies, B and Kudritzki, RP and Gazak, JZ and Bergemann, M and Plez, Bertrand and Ferguson, AMN},
  journal={The Astrophysical Journal},
  volume={803},
  number={1},
  pages={14},
  year={2015},
  publisher={IOP Publishing}
}

@article{baldacci2003variable,
  title={Variable stars as tracers of stellar populations in Local Group galaxies: Leo I and NGC 6822},
  author={Baldacci, L and Matonti, F and Rizzi, L and Clementini, Gisella and Held, EV and Momany, Y and Di Fabrizio, L and Saviane, I},
  journal={arXiv preprint astro-ph/0305506},
  year={2003}
}

@article{efremova2011recent,
  title={The Recent Star Formation in NGC 6822: An Ultraviolet Study},
  author={Efremova, Boryana V and Bianchi, Luciana and Thilker, David A and Neill, James D and Burgarella, Denis and Wyder, Ted K and Madore, Barry F and Rey, Soo-Chang and Barlow, Tom A and Conrow, Tim and others},
  journal={The Astrophysical Journal},
  volume={730},
  number={2},
  pages={88},
  year={2011},
  publisher={IOP Publishing}
}

@article{marigo2008evolution,
  title={Evolution of asymptotic giant branch stars-II. Optical to far-infrared isochrones with improved tp-agb models},
  author={Marigo, Paola and Girardi, L{\'e}o and Bressan, Alessandro and Groenewegen, Martin AT and Silva, Laura and Granato, Gian Luigi},
  journal={Astronomy \& Astrophysics},
  volume={482},
  number={3},
  pages={883--905},
  year={2008},
  publisher={EDP Sciences}
}

@article{marigo2017new,
  title={A new generation of PARSEC-COLIBRI stellar isochrones including the TP-AGB phase},
  author={Marigo, Paola and Girardi, L{\'e}o and Bressan, Alessandro and Rosenfield, Philip and Aringer, Bernhard and Chen, Yang and Dussin, Marco and Nanni, Ambra and Pastorelli, Giada and Rodrigues, Tha{\'\i}se S and others},
  journal={The Astrophysical Journal},
  volume={835},
  number={1},
  pages={77},
  year={2017},
  publisher={IOP Publishing}
}

@article{kroupa2001variation,
  title={On the variation of the initial mass function},
  author={Kroupa, Pavel},
  journal={Monthly Notices of the Royal Astronomical Society},
  volume={322},
  number={2},
  pages={231--246},
  year={2001},
  publisher={Blackwell Science Ltd Oxford, UK}
}

@article{girardi2005star,
  title={Star counts in the galaxy-simulating from very deep to very shallow photometric surveys with the trilegal code},
  author={Girardi, L and Groenewegen, MAT and Hatziminaoglou, E and Da Costa, L},
  journal={Astronomy \& Astrophysics},
  volume={436},
  number={3},
  pages={895--915},
  year={2005},
  publisher={EDP Sciences}
}

@article{zhang2021panoramic,
  title={A panoramic view of the Local Group dwarf galaxy NGC 6822},
  author={Zhang, Shumeng and Mackey, Dougal and Da Costa, Gary S},
  journal={Monthly Notices of the Royal Astronomical Society},
  volume={508},
  number={2},
  pages={2098--2113},
  year={2021},
  publisher={Oxford University Press}
}

@article{battaglia2022gaia,
  title={Gaia early DR3 systemic motions of Local Group dwarf galaxies and orbital properties with a massive Large Magellanic Cloud},
  author={Battaglia, Giuseppina and Taibi, S and Thomas, GF and Fritz, TK},
  journal={Astronomy \& Astrophysics},
  volume={657},
  pages={A54},
  year={2022},
  publisher={EDP Sciences}
}

@article{wyder2003star,
  title={The star formation histories of four fields spanning the minor axis of NGC 6822},
  author={Wyder, Ted K},
  journal={The Astronomical Journal},
  volume={125},
  number={6},
  pages={3097},
  year={2003},
  publisher={IOP Publishing}
}

@article{valenzuela2007there,
  title={Is there evidence for flat cores in the halos of dwarf galaxies? The case of NGC 3109 and NGC 6822},
  author={Valenzuela, Octavio and Rhee, George and Klypin, Anatoly and Governato, Fabio and Stinson, Gregory and Quinn, Thomas and Wadsley, James},
  journal={The Astrophysical Journal},
  volume={657},
  number={2},
  pages={773},
  year={2007},
  publisher={IOP Publishing}
}

@article{mateo1998dwarf,
  title={Dwarf galaxies of the Local Group},
  author={Mateo, Mario},
  journal={Annual Review of Astronomy and Astrophysics},
  volume={36},
  number={1},
  pages={435--506},
  year={1998},
  publisher={Annual Reviews 4139 El Camino Way, PO Box 10139, Palo Alto, CA 94303-0139, USA}
}

@article{lee1993tip,
  title={The tip of the red giant branch as a distance indicator for resolved galaxies},
  author={Lee, Myung Gyoon and Freedman, Wendy L and Madore, Barry F},
  journal={Astrophysical Journal v. 417, p. 553},
  volume={417},
  pages={553},
  year={1993}
}

@article{gieren2006araucaria,
  title={The araucaria project: An accurate distance to the local group galaxy NGC 6822 from near-infrared photometry of cepheid variables},
  author={Gieren, Wolfgang and Pietrzy{\'n}ski, Grzegorz and Nalewajko, Krzysztof and Soszy{\'n}ski, Igor and Bresolin, Fabio and Kudritzki, Rolf-Peter and Minniti, Dante and Romanowsky, Aaron},
  journal={The Astrophysical Journal},
  volume={647},
  number={2},
  pages={1056},
  year={2006},
  publisher={IOP Publishing}
}

@article{clementini2003rr,
  title={RR Lyrae and short-period variable stars in the dwarf irregular galaxy NGC 6822},
  author={Clementini, Gisella and Held, Enrico V and Baldacci, Lara and Rizzi, Luca},
  journal={The Astrophysical Journal},
  volume={588},
  number={2},
  pages={L85},
  year={2003},
  publisher={IOP Publishing}
}

@article{pietrinferni2013basti,
  title={The BaSTI Stellar Evolution Database: models for extremely metal-poor and super-metal-rich stellar populations},
  author={Pietrinferni, Adriano and Cassisi, Santi and Salaris, Maurizio and Hidalgo, Sebastian},
  journal={Astronomy \& Astrophysics},
  volume={558},
  pages={A46},
  year={2013},
  publisher={EDP Sciences}
}

@article{hofner2018mass,
  title={Mass loss of stars on the asymptotic giant branch: Mechanisms, models and measurements},
  author={H{\"o}fner, Susanne and Olofsson, Hans},
  journal={The Astronomy and Astrophysics Review},
  volume={26},
  pages={1--92},
  year={2018},
  publisher={Springer}
}

@article{brown2021gaia,
  title={Gaia early data release 3-summary of the contents and survey properties},
  author={Brown, Anthony GA and Vallenari, Antonella and Prusti, T and De Bruijne, JHJ and Babusiaux, C and Biermann, M and Creevey, OL and Evans, DW and Eyer, L and Hutton, A and others},
  journal={Astronomy \& Astrophysics},
  volume={649},
  pages={A1},
  year={2021},
  publisher={EDP sciences}
}

@article{cassisi2006basti,
  title={BASTI: an interactive database of updated stellar evolution models},
  author={Cassisi, S and Pietrinferni, A and Salaris, M and Castelli, F and Cordier, D and Castellani, M},
  journal={Memorie della Societ{\`a} Astronomica Italiana, v. 77, p. 71 (2006)},
  volume={77},
  pages={71},
  year={2006}
}

@article{hidalgo2018updated,
  title={The updated BaSTI stellar evolution models and isochrones. I. Solar-scaled calculations},
  author={Hidalgo, Sebastian L and Pietrinferni, Adriano and Cassisi, Santi and Salaris, Maurizio and Mucciarelli, Alessio and Savino, Alessandro and Aparicio, Antonio and Aguirre, Victor Silva and Verma, Kuldeep},
  journal={The Astrophysical Journal},
  volume={856},
  number={2},
  pages={125},
  year={2018},
  publisher={IOP Publishing}
}

@article{mcquinn2010nature,
  title={The nature of starbursts. I. The star formation histories of eighteen nearby starburst dwarf galaxies},
  author={McQuinn, Kristen BW and Skillman, Evan D and Cannon, John M and Dalcanton, Julianne and Dolphin, Andrew and Hidalgo-Rodr{\'\i}guez, Sebastian and Holtzman, Jon and Stark, David and Weisz, Daniel and Williams, Benjamin},
  journal={The Astrophysical Journal},
  volume={721},
  number={1},
  pages={297},
  year={2010},
  publisher={IOP Publishing}
}

@article{geha2012stellar,
  title={A stellar mass threshold for quenching of field galaxies},
  author={Geha, M and Blanton, MR and Yan, R and Tinker, JL},
  journal={The Astrophysical Journal},
  volume={757},
  number={1},
  pages={85},
  year={2012},
  publisher={IOP Publishing}
}

@article{arp1975properties,
  title={Properties of two blue compact galaxies},
  author={Arp, Halton and O'Connell, Robert W},
  journal={Astrophysical Journal, vol. 197, Apr. 15, 1975, pt. 1, p. 291-296.},
  volume={197},
  pages={291--296},
  year={1975}
}

@article{hunter1982global,
  title={Global properties of irregular galaxies},
  author={Hunter, DA and Gallagher, JS and Rautenkranz, D},
  journal={Astrophysical Journal Supplement Series, vol. 49, May 1982, p. 53-88.},
  volume={49},
  pages={53--88},
  year={1982}
}

@article{cannon2011m81,
  title={The M81 Group Dwarf Irregular Galaxy DDO 165. I. High-velocity Neutral Gas in a Post-starburst System},
  author={Cannon, John M and Most, Hans P and Skillman, Evan D and Weisz, Daniel R and Cook, David and Dolphin, Andrew E and Kennicutt, Robert C and Lee, Janice and Seth, Anil and Walter, Fabian and others},
  journal={The Astrophysical Journal},
  volume={735},
  number={1},
  pages={35},
  year={2011},
  publisher={IOP Publishing}
}

@article{ellison2013galaxy,
  title={Galaxy pairs in the Sloan Digital Sky Survey--VIII. The observational properties of post-merger galaxies},
  author={Ellison, Sara L and Mendel, J Trevor and Patton, David R and Scudder, Jillian M},
  journal={Monthly Notices of the Royal Astronomical Society},
  volume={435},
  number={4},
  pages={3627--3638},
  year={2013},
  publisher={Oxford University Press}
}

@article{van2019first,
  title={First Gaia dynamics of the Andromeda system: DR2 proper motions, orbits, and rotation of M31 and M33},
  author={van der Marel, Roeland P and Fardal, Mark A and Sohn, Sangmo Tony and Patel, Ekta and Besla, Gurtina and del Pino, Andr{\'e}s and Sahlmann, Johannes and Watkins, Laura L},
  journal={The Astrophysical Journal},
  volume={872},
  number={1},
  pages={24},
  year={2019},
  publisher={IOP Publishing}
}

@article{cannon2012origin,
  title={ON THE ORIGIN OF THE SUPERGIANT H i SHELL AND PUTATIVE COMPANION IN NGC 6822},
  author={Cannon, John M and O’Leary, Erin M and Weisz, Daniel R and Skillman, Evan D and Dolphin, Andrew E and Bigiel, Frank and Cole, Andrew A and De Blok, WJG and Walter, Fabian},
  journal={The Astrophysical Journal},
  volume={747},
  number={2},
  pages={122},
  year={2012},
  publisher={IOP Publishing}
}

@article{groenewegen1995evolution,
  title={The evolution of galactic carbon stars.},
  author={Groenewegen, MAT and Van den Hoek, LB and De Jong, T},
  journal={Astronomy and Astrophysics, Vol. 293, p. 381-395 (1995)},
  volume={293},
  pages={381--395},
  year={1995}
}

@article{rich2014new,
  title={A new Cepheid distance measurement and method for NGC 6822},
  author={Rich, Jeffrey A and Persson, SE and Freedman, Wendy L and Madore, Barry F and Monson, Andrew J and Scowcroft, Victoria and Seibert, Mark},
  journal={The Astrophysical Journal},
  volume={794},
  number={2},
  pages={107},
  year={2014},
  publisher={IOP Publishing}
}

@article{veljanoski2015globular,
  title={The globular cluster system of NGC 6822},
  author={Veljanoski, J and Ferguson, AMN and Mackey, AD and Huxor, AP and Hurley, JR and Bernard, EJ and C{\^o}t{\'e}, P and Irwin, MJ and Martin, NF and Burgett, WS and others},
  journal={Monthly Notices of the Royal Astronomical Society},
  volume={452},
  number={1},
  pages={320--332},
  year={2015},
  publisher={Oxford University Press}
}

@article{weisz2014star,
  title={The star formation histories of local group dwarf galaxies. I. Hubble space telescope/wide field planetary camera 2 observations},
  author={Weisz, Daniel R and Dolphin, Andrew E and Skillman, Evan D and Holtzman, Jon and Gilbert, Karoline M and Dalcanton, Julianne J and Williams, Benjamin F},
  journal={The Astrophysical Journal},
  volume={789},
  number={2},
  pages={147},
  year={2014},
  publisher={IOP Publishing}
}

@article{ren2024star,
  title={The star formation history in local group galaxies. I. Ten dwarf galaxies},
  author={Ren, Yi and Jiang, Biwei and Wang, Yuxi and Yang, Ming and Yan, Zhiqiang},
  journal={The Astrophysical Journal},
  volume={966},
  number={1},
  pages={25},
  year={2024},
  publisher={IOP Publishing}
}

@article{park2022gas,
  title={Gas dynamics and star formation in NGC 6822},
  author={Park, Hye-Jin and Oh, Se-Heon and Wang, Jing and Zheng, Yun and Zhang, Hong-Xin and De Blok, WJG},
  journal={The Astronomical Journal},
  volume={164},
  number={3},
  pages={82},
  year={2022},
  publisher={IOP Publishing}
}

@ARTICLE{2025ApJ...992...94M,
       author = {{Mahani}, Hamidreza and {Javadi}, Atefeh and {van Loon}, Jacco Th. and {Kemper}, Francisca and {Hamedani Golshan}, Roya and {McDonald}, Iain and {Khosroshahi}, Habib G. and {Abdollahi}, Hedieh and {Mahdizadeh}, Sajjad},
        title = "{Long-period Variable Stars in NGC 147 and NGC 185. II. Their Dust Production}",
      journal = {\apj},
         year = 2025,
        month = oct,
       volume = {992},
       number = {1},
          eid = {94},
        pages = {94},
          doi = {10.3847/1538-4357/adfa2b},
       eprint = {2508.05596},
 primaryClass = {astro-ph.GA},
       adsurl = {https://ui.adsabs.harvard.edu/abs/2025ApJ...992...94M}
}

@ARTICLE{2024ApJ...972...47A,
       author = {{Aghdam}, Sima T. and {Javadi}, Atefeh and {Hashemi}, Seyedazim and {Abdollahi}, Mahdi and {van Loon}, Jacco Th. and {Khosroshahi}, Habib and {Hamedani Golshan}, Roya and {Saremi}, Elham and {Saberi}, Maryam},
        title = "{The Complex Star Formation History of the Halo of NGC 5128 (Cen A)}",
      journal = {\apj},
         year = 2024,
        month = sep,
       volume = {972},
       number = {1},
          eid = {47},
        pages = {47},
 doi = {10.3847/1538-4357/ad57c0},
 primaryClass = {astro-ph.GA},
       adsurl = {https://ui.adsabs.harvard.edu/abs/2024ApJ...972...47A}
}

@ARTICLE{2023ApJ...948...63A,
       author = {{Abdollahi}, Hedieh and {Javadi}, Atefeh and {Mirtorabi}, Mohammad Taghi and {Saremi}, Elham and {van Loon}, Jacco Th. and {Khosroshahi}, Habib G. and {McDonald}, Iain and {Khalouei}, Elahe and {Mahani}, Hamidreza and {Aghdam}, Sima Taefi and {Saberi}, Maryam and {Torki}, Maryam},
        title = "{The Isaac Newton Telescope Monitoring Survey of Local Group Dwarf Galaxies. VI. The Star Formation History and Dust Production in Andromeda IX}",
      journal = {\apj},
         year = 2023,
        month = may,
       volume = {948},
       number = {1},
          eid = {63},
        pages = {63},
doi = {10.3847/1538-4357/acbbc9},
 primaryClass = {astro-ph.GA},
       adsurl = {https://ui.adsabs.harvard.edu/abs/2023ApJ...948...63A}
}

@ARTICLE{2021ApJ...923..164S,
       author = {{Saremi}, Elham and {Javadi}, Atefeh and {Navabi}, Mahdieh and {van Loon}, Jacco Th. and {Khosroshahi}, Habib G. and {Bojnordi Arbab}, Behzad and {McDonald}, Iain},
        title = "{The Isaac Newton Telescope Monitoring Survey of Local Group Dwarf Galaxies. II. The Star-formation History of Andromeda I Derived from Long-period Variables}",
      journal = {\apj},
         year = 2021,
        month = dec,
       volume = {923},
       number = {2},
          eid = {164},
        pages = {164},
doi = {10.3847/1538-4357/ac2d96},
 primaryClass = {astro-ph.GA},
       adsurl = {https://ui.adsabs.harvard.edu/abs/2021ApJ...923..164S}
}

@ARTICLE{2021ApJ...910..127N,
       author = {{Navabi}, Mahdieh and {Saremi}, Elham and {Javadi}, Atefeh and {Noori}, Majedeh and {van Loon}, Jacco Th. and {Khosroshahi}, Habib G. and {McDonald}, Iain and {Alizadeh}, Mina and {Danesh}, Arash and {Gozaliasl}, Ghassem and {Molaeinezhad}, Alireza and {Parto}, Tahere and {Raouf}, Mojtaba},
        title = "{The Isaac Newton Telescope Monitoring Survey of Local Group Dwarf Galaxies. IV. The Star Formation History of Andromeda VII Derived from Long-period Variable Stars}",
      journal = {\apj},
         year = 2021,
        month = apr,
       volume = {910},
       number = {2},
          eid = {127},
        pages = {127},
doi = {10.3847/1538-4357/abdec1},
 primaryClass = {astro-ph.GA},
       adsurl = {https://ui.adsabs.harvard.edu/abs/2021ApJ...910..127N}
}

@ARTICLE{2020ApJ...894..135S,
       author = {{Saremi}, Elham and {Javadi}, Atefeh and {van Loon}, Jacco Th. and {Khosroshahi}, Habib and {Molaeinezhad}, Alireza and {McDonald}, Iain and {Raouf}, Mojtaba and {Danesh}, Arash and {Bamber}, James R. and {Short}, Philip and {Su{\'a}rez-Andr{\'e}s}, Lucia and {Clavero}, Rosa and {Gozaliasl}, Ghassem},
        title = "{The Isaac Newton Telescope Monitoring Survey of Local Group Dwarf Galaxies. I. Survey Overview and First Results for Andromeda I}",
      journal = {\apj},
         year = 2020,
        month = may,
       volume = {894},
       number = {2},
          eid = {135},
        pages = {135},
          doi = {10.3847/1538-4357/ab88a2},
archivePrefix = {arXiv},
       eprint = {2004.05620},
 primaryClass = {astro-ph.GA},
       adsurl = {https://ui.adsabs.harvard.edu/abs/2020ApJ...894..135S}
}

@ARTICLE{2018AJ156112Y,
       author = {{Yuan}, Wenlong and {Macri}, Lucas M. and {Javadi}, Atefeh and {Lin}, Zhenfeng and {Huang}, Jianhua Z.},
        title = "{Near-infrared Mira Period-Luminosity Relations in M33}",
      journal = {\aj},
         year = 2018,
        month = sep,
       volume = {156},
       number = {3},
          eid = {112},
        pages = {112},
doi = {10.3847/1538-3881/aad330},
 primaryClass = {astro-ph.SR},
       adsurl = {https://ui.adsabs.harvard.edu/abs/2018AJ....156..112Y}
}

@ARTICLE{2014MNRAS.445.2214R,
       author = {{Rezaeikh}, Sara and {Javadi}, Atefeh and {Khosroshahi}, Habib and {van Loon}, Jacco Th.},
        title = "{The star formation history of the Magellanic Clouds derived from long-period variable star counts}",
      journal = {\mnras},
         year = 2014,
        month = dec,
       volume = {445},
       number = {3},
        pages = {2214-2222},
doi = {10.1093/mnras/stu1807},
 primaryClass = {astro-ph.GA},
       adsurl = {https://ui.adsabs.harvard.edu/abs/2014MNRAS.445.2214R}
}

@ARTICLE{2019ApJ...877...49G,
       author = {{Goldman}, S.~R. and {Boyer}, M.~L. and {McQuinn}, K.~B.~W. and {Whitelock}, P.~A. and {McDonald}, I. and {van Loon}, J. Th. and {Skillman}, E.~D. and {Gehrz}, R.~D. and {Javadi}, A. and {Sloan}, G.~C. and {Jones}, O.~C. and {Groenewegen}, M.~A.~T. and {Menzies}, J.~W.},
        title = "{An Infrared Census of DUST in Nearby Galaxies with Spitzer (DUSTiNGS). V. The Period-Luminosity Relation for Dusty Metal-poor AGB Stars}",
      journal = {\apj},
         year = 2019,
        month = may,
       volume = {877},
       number = {1},
          eid = {49},
        pages = {49},
doi = {10.3847/1538-4357/ab0965},
 primaryClass = {astro-ph.SR},
       adsurl = {https://ui.adsabs.harvard.edu/abs/2019ApJ...877...49G}
}

@article{salaris1993alpha,
  title={The alpha-enhanced isochrones and their impact on the FITS to the Galactic globular cluster system},
  author={Salaris, Maurizio and Chieffi, Alessandro and Straniero, Oscar},
  journal={Astrophysical Journal, Part 1 (ISSN 0004-637X), vol. 414, no. 2, p. 580-600.},
  volume={414},
  pages={580--600},
  year={1993}
}

@article{vagnozzi2019new,
  title={New solar metallicity measurements},
  author={Vagnozzi, Sunny},
  journal={Atoms},
  volume={7},
  number={2},
  pages={41},
  year={2019},
  publisher={MDPI}
}

@ARTICLE{Girardi10,
       author = {{Girardi}, L{\'e}o and {Williams}, Benjamin F. and {Gilbert}, Karoline M. and {Rosenfield}, Philip and {Dalcanton}, Julianne J. and {Marigo}, Paola and {Boyer}, Martha L. and {Dolphin}, Andrew and {Weisz}, Daniel R. and {Melbourne}, Jason and {Olsen}, Knut A.~G. and {Seth}, Anil C. and {Skillman}, Evan},
        title = "{The ACS Nearby Galaxy Survey Treasury. IX. Constraining Asymptotic Giant Branch Evolution with Old Metal-poor Galaxies}",
      journal = {\apj},
         year = 2010,
        month = dec,
       volume = {724},
       number = {2},
        pages = {1030-1043},
          doi = {10.1088/0004-637X/724/2/1030},
archivePrefix = {arXiv},
       eprint = {1009.4618},
 primaryClass = {astro-ph.SR},
       adsurl = {https://ui.adsabs.harvard.edu/abs/2010ApJ...724.1030G}
}

@ARTICLE{Panter03,
       author = {{Panter}, Benjamin and {Heavens}, Alan F. and {Jimenez}, Raul},
        title = "{Star formation and metallicity history of the SDSS galaxy survey: unlocking the fossil record}",
      journal = {\mnras},
         year = 2003,
        month = aug,
       volume = {343},
       number = {4},
        pages = {1145-1154},
          doi = {10.1046/j.1365-8711.2003.06722.x},
archivePrefix = {arXiv},
       eprint = {astro-ph/0211546},
 primaryClass = {astro-ph},
       adsurl = {https://ui.adsabs.harvard.edu/abs/2003MNRAS.343.1145P}
}

@ARTICLE{Richards09,
       author = {{Richards}, Joseph W. and {Freeman}, Peter E. and {Lee}, Ann B. and {Schafer}, Chad M.},
        title = "{Accurate parameter estimation for star formation history in galaxies using SDSS spectra}",
      journal = {\mnras},
         year = 2009,
        month = oct,
       volume = {399},
       number = {2},
        pages = {1044-1057},
          doi = {10.1111/j.1365-2966.2009.15349.x},
archivePrefix = {arXiv},
       eprint = {0905.4683},
 primaryClass = {astro-ph.CO},
       adsurl = {https://ui.adsabs.harvard.edu/abs/2009MNRAS.399.1044R}
}

@ARTICLE{Fernandes05,
       author = {{Cid Fernandes}, Roberto and {Mateus}, Ab{\'\i}lio and {Sodr{\'e}}, Laerte and {Stasi{\'n}ska}, Gra{\.z}yna and {Gomes}, Jean M.},
        title = "{Semi-empirical analysis of Sloan Digital Sky Survey galaxies - I. Spectral synthesis method}",
      journal = {\mnras},
         year = 2005,
        month = apr,
       volume = {358},
       number = {2},
        pages = {363-378},
          doi = {10.1111/j.1365-2966.2005.08752.x},
archivePrefix = {arXiv},
       eprint = {astro-ph/0412481},
 primaryClass = {astro-ph},
       adsurl = {https://ui.adsabs.harvard.edu/abs/2005MNRAS.358..363C}
}

@ARTICLE{Tojeiro07,
       author = {{Tojeiro}, R. and {Heavens}, A.~F. and {Jimenez}, R. and {Panter}, B.},
        title = "{Recovering galaxy star formation and metallicity histories from spectra using VESPA}",
      journal = {\mnras},
         year = 2007,
        month = nov,
       volume = {381},
       number = {3},
        pages = {1252-1266},
          doi = {10.1111/j.1365-2966.2007.12323.x},
archivePrefix = {arXiv},
       eprint = {0704.0941},
 primaryClass = {astro-ph},
       adsurl = {https://ui.adsabs.harvard.edu/abs/2007MNRAS.381.1252T}
}

@ARTICLE{hubble1925ngc,
       author = {{Hubble}, E.~P.},
        title = "{NGC 6822, a remote stellar system.}",
      journal = {\apj},
         year = 1925,
        month = dec,
       volume = {62},
        pages = {409-433},
          doi = {10.1086/142943},
       adsurl = {https://ui.adsabs.harvard.edu/abs/1925ApJ....62..409H}
}

@article{williamson2021evolution,
  title={The Evolution of Magellanic-like Galaxy Pairs and the Production of Magellanic Stream Analogs in Simulations with Tides, Ram Pressure, and Stellar Feedback},
  author={Williamson, David and Martel, Hugo},
  journal={The Astrophysical Journal},
  volume={907},
  number={1},
  pages={9},
  year={2021},
  publisher={IOP Publishing}
}

@article{bekki2005formation,
  title={Formation and evolution of the Magellanic Clouds--I. Origin of structural, kinematic and chemical properties of the Large Magellanic Cloud},
  author={Bekki, Kenji and Chiba, Masashi},
  journal={Monthly Notices of the Royal Astronomical Society},
  volume={356},
  number={2},
  pages={680--702},
  year={2005},
  publisher={Blackwell Science Ltd Oxford, UK}
}

@article{indu2011recent,
  title={The recent star-formation history of the Large and Small Magellanic Clouds},
  author={Indu, G and Subramaniam, Annapurni},
  journal={Astronomy \& Astrophysics},
  volume={535},
  pages={A115},
  year={2011},
  publisher={EDP Sciences}
}

@article{belokurov2017clouds,
  title={Clouds, Streams and Bridges: redrawing the blueprint of the Magellanic System with Gaia DR1},
  author={Belokurov, Vasily and Erkal, Denis and Deason, Alis J and Koposov, Sergey E and De Angeli, Francesca and Evans, Dafydd Wyn and Fraternali, Filippo and Mackey, Dougal},
  journal={Monthly Notices of the Royal Astronomical Society},
  volume={466},
  number={4},
  pages={4711--4730},
  year={2017},
  publisher={Oxford University Press}
}

@article{nidever2008origin,
  title={The origin of the magellanic stream and its leading arm},
  author={Nidever, David L and Majewski, Steven R and Burton, W Butler},
  journal={The Astrophysical Journal},
  volume={679},
  number={1},
  pages={432},
  year={2008},
  publisher={IOP Publishing}
}

@article{bortolini2025feast,
  title={FEAST: JWST/NIRCam view of the resolved stellar populations of the interacting dwarf galaxies NGC 4485 and NGC 4490},
  author={Bortolini, Giacomo and Correnti, Matteo and Adamo, Angela and Cignoni, Michele and Sacchi, Elena and Tosi, Monica and {\"O}stlin, G{\"o}ran and Kapodistrias, Anastasios and Bik, Arjan and Calzetti, Daniela and others},
  journal={The Astrophysical Journal},
  volume={991},
  number={2},
  pages={212},
  year={2025},
  publisher={IOP Publishing}
}

@ARTICLE{Javadi13,
       author = {{Javadi}, Atefeh and {van Loon}, Jacco Th. and {Khosroshahi}, Habib and {Mirtorabi}, Mohammad Taghi},
        title = "{The UK Infrared Telescope M33 monitoring project - III. Feedback from dusty stellar winds in the central square kiloparsec}",
      journal = {\mnras},
         year = 2013,
        month = jul,
       volume = {432},
       number = {4},
        pages = {2824-2836},
          doi = {10.1093/mnras/stt640},
archivePrefix = {arXiv},
       eprint = {1304.3782},
 primaryClass = {astro-ph.SR},
       adsurl = {https://ui.adsabs.harvard.edu/abs/2013MNRAS.432.2824J}
}

@INPROCEEDINGS{Mahani23,
       author = {{Mahani}, Hamidreza and {Javadi}, Atefeh and {van Loon}, Jacco Th. and {Khosroshahi}, Habib and {Saremi}, Elham and {Hamedani Golshan}, Roya and {Navabi}, Mahdieh and {Hashemi}, Seyed Azim and {Gholami}, Mahtab and {Aghdam}, Sima Taefi},
        title = "{From evolved stars to the formation and evolution of galaxies}",
    booktitle = {Resolving the Rise and Fall of Star Formation in Galaxies},
         year = 2023,
       editor = {{Wong}, Tony and {Kim}, Woong-Tae},
       series = {IAU Symposium},
       volume = {373},
        month = jan,
        pages = {264-267},
          doi = {10.1017/S1743921322004951},
       adsurl = {https://ui.adsabs.harvard.edu/abs/2023IAUS..373..264M}
}

@ARTICLE{Abdollahi26,
       author = {{Abdollahi}, Hedieh and {Javadi}, Atefeh and {van Loon}, Jacco Th. and {McDonald}, Iain and {Abdollahi}, Mahdi and {Saremi}, Elham and {Khosroshahi}, Habib G. and {Moln{\'a}r}, L{\'a}szl{\'o} and {Mahani}, Hamidreza},
        title = "{The Isaac Newton Telescope Monitoring Survey of Local Group Dwarf Galaxies. VIII. A Census of Long-period Variable Stars across the Andromeda Dwarf Satellite System}",
      journal = {\apj},
         year = 2026,
        month = mar,
       volume = {1000},
       number = {1},
          eid = {69},
        pages = {69},
          doi = {10.3847/1538-4357/ae40f3},
archivePrefix = {arXiv},
       eprint = {2601.06924},
 primaryClass = {astro-ph.GA},
       adsurl = {https://ui.adsabs.harvard.edu/abs/2026ApJ..1000...69A}
}

@article{navabi2025outside,
  title={Outside-in evolution with a twist: metallicity gradients and asymmetries in the SMC},
  author={Navabi, M and Carrera, R and No{\"e}l, NED and De Leo, M},
  journal={Monthly Notices of the Royal Astronomical Society},
  volume={544},
  number={4},
  pages={3980--3993},
  year={2025},
  publisher={Oxford University Press}
}

@article{choudhury2018photometric,
  title={Photometric metallicity map of the Small Magellanic Cloud},
  author={Choudhury, Samyaday and Subramaniam, Annapurni and Cole, Andrew A and Sohn, Young-Jong},
  journal={Monthly Notices of the Royal Astronomical Society},
  volume={475},
  number={4},
  pages={4279--4297},
  year={2018},
  publisher={Oxford University Press}
}

@article{schwarzschild1962red,
  title={Red Giants of Population II. II.},
  author={Schwarzschild, M and H{\"a}rm, R},
  journal={Astrophysical Journal, vol. 136, p. 158},
  volume={136},
  pages={158},
  year={1962}
}

@ARTICLE{salaris2002,
  author  = {Salaris, Maurizio and Cassisi, Santi and Weiss, Achim},
  title   = {{Red Giant Branch Stars: The Theoretical Framework}},
  journal = {\pasp},
  year    = {2002},
  volume  = {114},
  number  = {794},
  pages   = {375},
  doi     = {10.1086/342498},
  eprint  = {astro-ph/0201387},
  archivePrefix = {arXiv},
  adsurl  = {https://ui.adsabs.harvard.edu/abs/2002PASP..114..375S}
}

@article{hopkins2015new,
  title={A new class of accurate, mesh-free hydrodynamic simulation methods},
  author={Hopkins, Philip F},
  journal={Monthly Notices of the Royal Astronomical Society},
  volume={450},
  number={1},
  pages={53--110},
  year={2015},
  publisher={The Royal Astronomical Society}
}

@article{weinberger2020arepo,
  title={The Arepo public code release},
  author={Weinberger, Rainer and Springel, Volker and Pakmor, R{\"u}diger},
  journal={The Astrophysical Journal Supplement Series},
  volume={248},
  number={2},
  pages={32},
  year={2020},
  publisher={The American Astronomical Society}
}

@ARTICLE{2025ApJ...991...24B,
       author = {{Boyer}, M.~L. and {Sloan}, G.~C. and {Nanni}, A. and {Tarantino}, E. and {McDonald}, I. and {Goldman}, S. and {Blommaert}, J.~A.~D.~L. and {Dell'Agli}, F. and {Di Criscienzo}, M. and {Garc{\'\i}a-Hern{\'a}ndez}, D.~A. and {Gehrz}, Robert D. and {Groenewegen}, M.~A.~T. and {Javadi}, A. and {Jones}, O.~C. and {Kemper}, F. and {Marengo}, M. and {McQuinn}, Kristen B.~W. and {Oliveira}, Joana M. and {Pastorelli}, Giada and {Roman-Duval}, Julia and {Sahai}, R. and {Skillman}, Evan D. and {Srinivasan}, S. and {van Loon}, J. Th. and {Weisz}, Daniel R. and {Whitelock}, Patricia A.},
        title = "{Discovery of SiC and Iron Dust around AGB Stars in the Very Metal-poor Sextans a Dwarf Galaxy with JWST: Implications for Dust Production at High Redshift}",
      journal = {\apj},
         year = 2025,
        month = sep,
       volume = {991},
       number = {1},
          eid = {24},
        pages = {24},
          doi = {10.3847/1538-4357/adf06a},
archivePrefix = {arXiv},
       eprint = {2507.16766},
 primaryClass = {astro-ph.SR},
       adsurl = {https://ui.adsabs.harvard.edu/abs/2025ApJ...991...24B}
}

@ARTICLE{2017ApJ...851..152B,
       author = {{Boyer}, M.~L. and {McQuinn}, K.~B.~W. and {Groenewegen}, M.~A.~T. and {Zijlstra}, A.~A. and {Whitelock}, P.~A. and {van Loon}, J. Th. and {Sonneborn}, G. and {Sloan}, G.~C. and {Skillman}, E.~D. and {Meixner}, M. and {McDonald}, I. and {Jones}, O.~C. and {Javadi}, A. and {Gehrz}, R.~D. and {Britavskiy}, N. and {Bonanos}, A.~Z.},
        title = "{An Infrared Census of DUST in Nearby Galaxies with Spitzer (DUSTiNGS). IV. Discovery of High-redshift AGB Analogs}",
      journal = {\apj},
         year = 2017,
        month = dec,
       volume = {851},
       number = {2},
          eid = {152},
        pages = {152},
          doi = {10.3847/1538-4357/aa9892},
archivePrefix = {arXiv},
       eprint = {1711.02129},
 primaryClass = {astro-ph.SR},
       adsurl = {https://ui.adsabs.harvard.edu/abs/2017ApJ...851..152B}
}

@INPROCEEDINGS{2011ASPC..445..497J,
       author = {{Javadi}, A. and {van Loon}, J. Th. and {Mirtorabi}, M.~T.},
        title = "{Infrared Survey of Pulsating Giant Stars in the Spiral Galaxy M 33: Dust Production, Star Formation History, and Galactic Structure}",
    booktitle = {Why Galaxies Care about AGB Stars II: Shining Examples and Common Inhabitants},
         year = 2011,
       editor = {{Kerschbaum}, F. and {Lebzelter}, T. and {Wing}, R.~F.},
       series = {Astronomical Society of the Pacific Conference Series},
       volume = {445},
        month = sep,
    publisher = {ASP},
        pages = {497},
          doi = {10.48550/arXiv.1101.5271},
archivePrefix = {arXiv},
       eprint = {1101.5271},
 primaryClass = {astro-ph.GA},
       adsurl = {https://ui.adsabs.harvard.edu/abs/2011ASPC..445..497J}
}

@ARTICLE{2019MNRAS.483.4751H,
       author = {{Hashemi}, Seyed Azim and {Javadi}, Atefeh and {van Loon}, Jacco Th},
        title = "{From evolved stars to the evolution of IC 1613}",
      journal = {\mnras},
         year = 2019,
        month = mar,
       volume = {483},
       number = {4},
        pages = {4751-4765},
          doi = {10.1093/mnras/sty3450},
archivePrefix = {arXiv},
       eprint = {1812.07230},
 primaryClass = {astro-ph.GA},
       adsurl = {https://ui.adsabs.harvard.edu/abs/2019MNRAS.483.4751H}
}
\bibliographystyle{aasjournal}


\appendix 

\renewcommand{\thefigure}{B\arabic{figure}}
\renewcommand{\thetable}{B\arabic{table}}
\setcounter{figure}{0} 
\setcounter{table}{0}  

\section{ Conversion from [Fe/H] to Metallicity (Z)}
\label{appnx:metallicity}

In this work, the principal metallicity indicator is the iron abundance, [Fe/H]. For readers who prefer the total metallicity notation, $Z$, we provide here the conversion method used.

Following \citet{salaris1993alpha}, the total metallicity $Z$ can be estimated from [Fe/H] and the $\alpha$-element enhancement [$\alpha$/Fe] using:

\begin{equation}
\log Z = \log Z_\odot + \mathrm{[Fe/H]} + \log \left(0.638 \times 10^{[\alpha/\mathrm{Fe}]} + 0.362\right),
\end{equation}

where $Z_\odot$ is the solar metallicity, which we adopt as $Z_\odot \approx 0.02$ \citep{vagnozzi2019new}, and [$\alpha$/Fe] is the $\alpha$-element enhancement, set to 0.3 dex in this work following \citet{hwang2014spectroscopic}.

Using this relation, the adopted [Fe/H] range, $-$2.53 $<$ [Fe/H] $< -$0.49 dex (\citealp{venn2001first}; \citealp{hwang2014spectroscopic}), corresponds to 0.0001 $<$ Z $<$ 0.012.

\section{Detailed Assessment of Systematic Uncertainties in the Derived Star Formation Histories}
\label{appnx:systematic}

Beyond the metallicity dependence discussed in Section 4.2, the pulsation duration is inferred from stellar evolutionary models and is highly sensitive to the adopted prescriptions for mass loss, convection, and the treatment of thermal pulses during the AGB phase \citep{Marigo07}. As shown in Fig. 3  from \citet{Girardi10}, comparisons among different sets of isochrones (e.g. \citealp{Marigo07, Girardi10}) indicate that $\delta t$ for intermediate-mass stars, and particularly for low-mass stars evolving onto the AGB, can vary by up to a factor of two. Since the derived SFR scales inversely with $\delta t$, these variations translate into systematic uncertainties in the normalization of $\xi(t)$.

Moreover, additional uncertainties in our analysis arise from the choice of IMF. As a standard reference, we adopt the canonical broken power-law IMF of \citet{Kroupa01}. For comparison, we also consider a single power-law Salpeter IMF \citep{Salpeter55}. Both IMFs are normalized over the mass range $0.1$ to $100 M_{\odot}$.

For the Salpeter IMF,

\begin{equation}
\xi(m) \propto m^{-2.35}.
\end{equation}

the mean stellar mass is

\begin{equation}
\langle m \rangle_{\mathrm{Salpeter}} =
\frac{\int_{0.1}^{100} m^{-1.35}\,dm}
{\int_{0.1}^{100} m^{-2.35}\,dm}
\simeq 0.351\,M_{\odot}.
\end{equation}

For the Kroupa IMF,

\begin{equation}
\xi(m) \propto \left\{ \begin{array}{rcl}
m^{-1.3} & \mbox{for} & 0.1 \leq \frac{\mathrm{m}}{\mathrm{M_\sun}} < 0.5 \\ m^{-2.3} & \mbox{for} & 0.5 \leq \frac{\mathrm{m}}{\mathrm{M_\sun}} < 100
\end{array}\right. ,
\end{equation}

we obtain

\begin{equation}
\langle m \rangle_{\mathrm{Kroupa}} =
\frac{
\int_{0.1}^{0.5} m^{-0.3}\,dm
+
0.5 \int_{0.5}^{100} m^{-1.3}\,dm
}{
\int_{0.1}^{0.5} m^{-1.3}\,dm
+
0.5 \int_{0.5}^{100} m^{-2.3}\,dm
}
\simeq 0.638\,M_{\odot}.
\end{equation}

The resulting difference in mean stellar mass is $\approx 45\%$. Because the SFR scales inversely with the mean stellar mass for a given number of observed stars ($  \mathrm{SFR} \propto 1/\langle m \rangle_{\rm IMF}  $), this IMF variation introduces a systematic offset of

\begin{equation}
\Delta\log_{10}(\mathrm{SFR}) \simeq \log_{10}\left(\frac{\langle m \rangle_{\rm Kroupa}}{\langle m \rangle_{\rm Salpeter}}\right) \approx 0.26~\mathrm{dex}
\end{equation}
in the inferred SFR.

\begin{figure*}
 \begin{center}
 \includegraphics[width=0.45\textwidth]{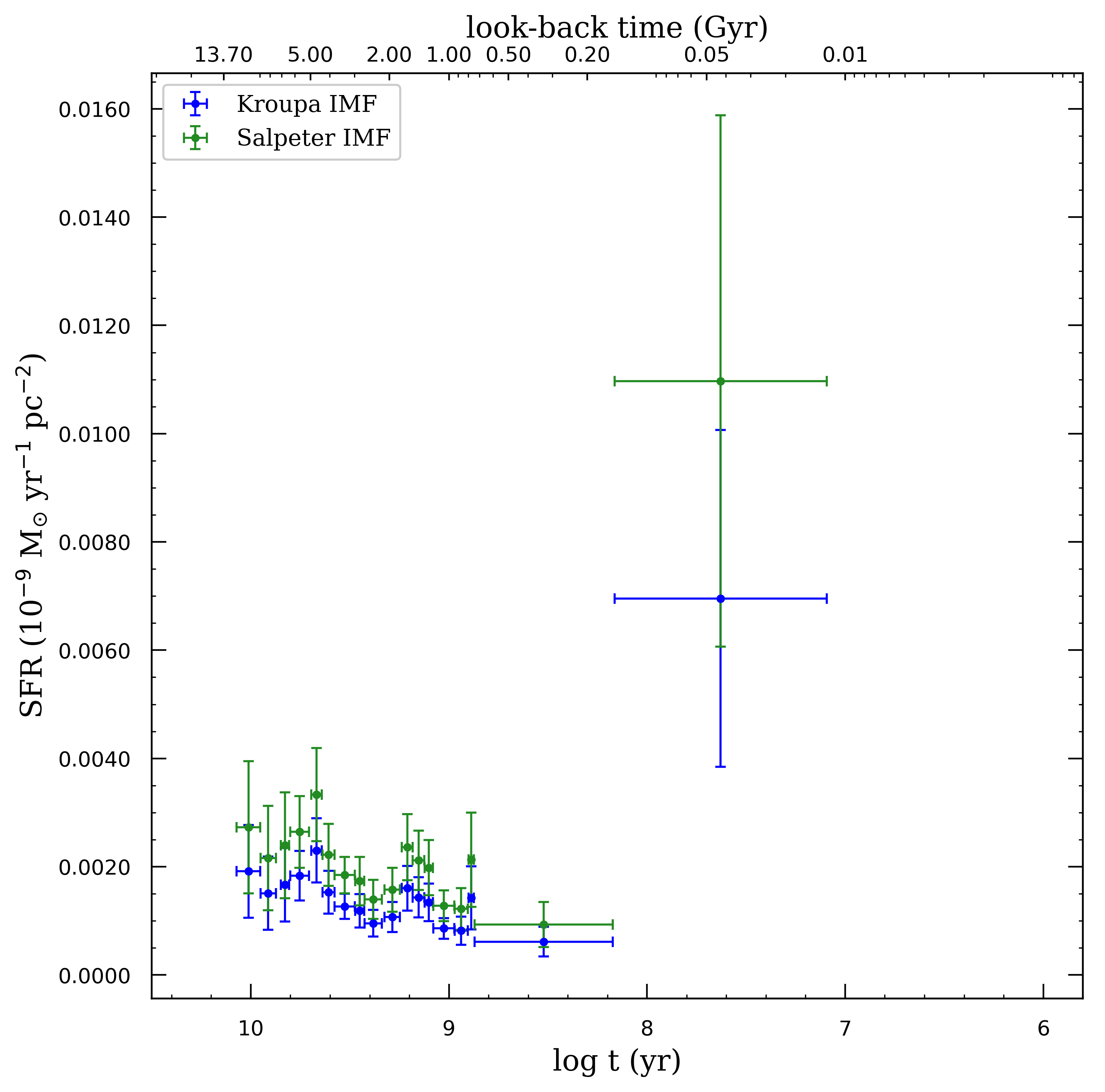}
  \includegraphics[width=0.45\textwidth]{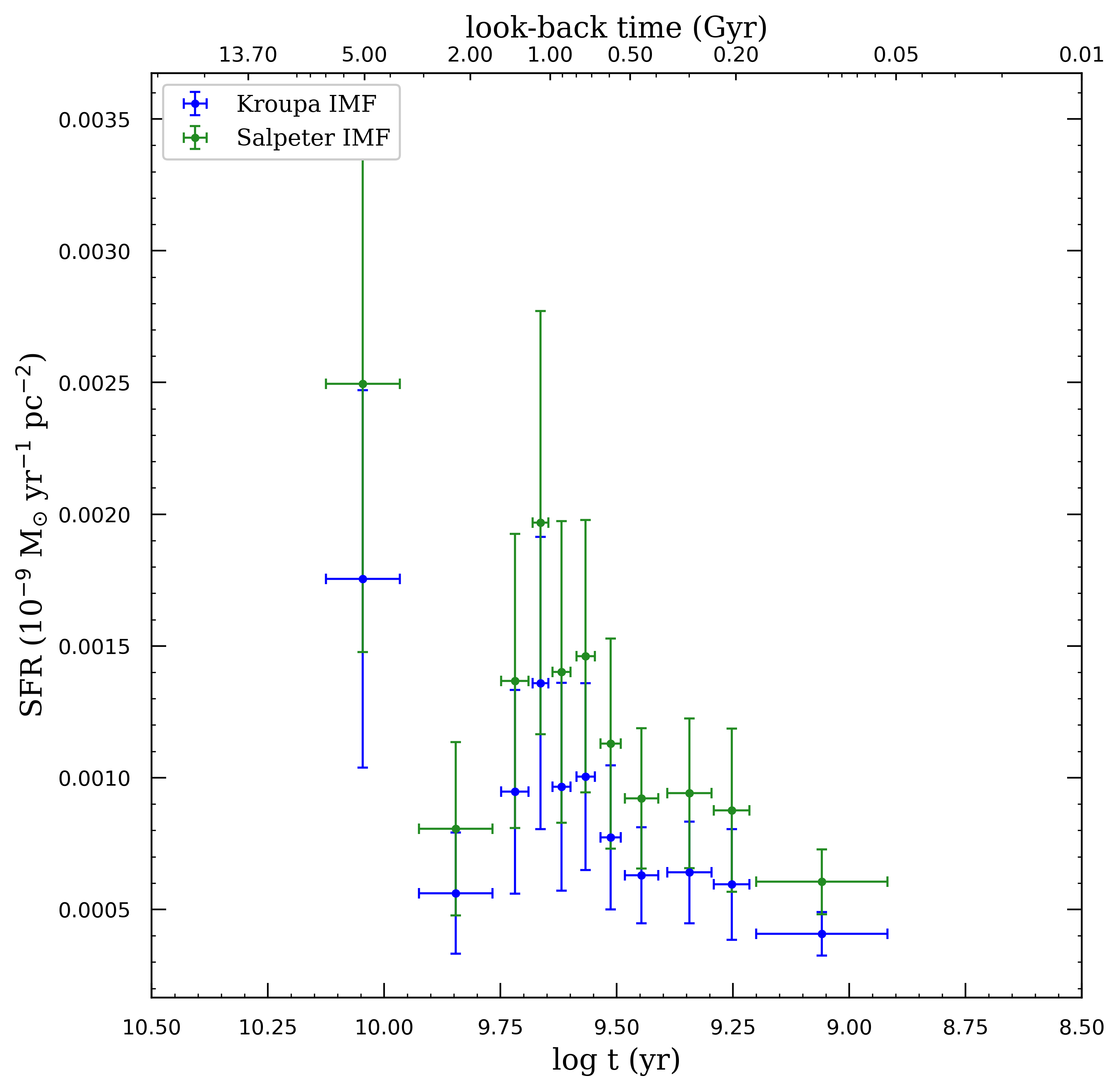}
\caption{Star formation histories derived for the bar region (left panel) and the outer region (right panel), assuming different initial mass functions.}
 \label{fig:IMF_a}
 \end{center}
 \end{figure*}

As shown in Fig.~\ref{fig:IMF_a}, the differences between the two sets of results based on the Kroupa and Salpeter IMFs are typically $0.2$--$0.4$ dex, in good agreement with the analytical expectation discussed above. This also indicates that the choice of IMF mainly affects the overall normalization of the SFR, while the shape of the star-formation history remains largely unchanged.

\mbox{}\\

\begin{figure*}
 \begin{center}
 \includegraphics[width=0.45\textwidth]{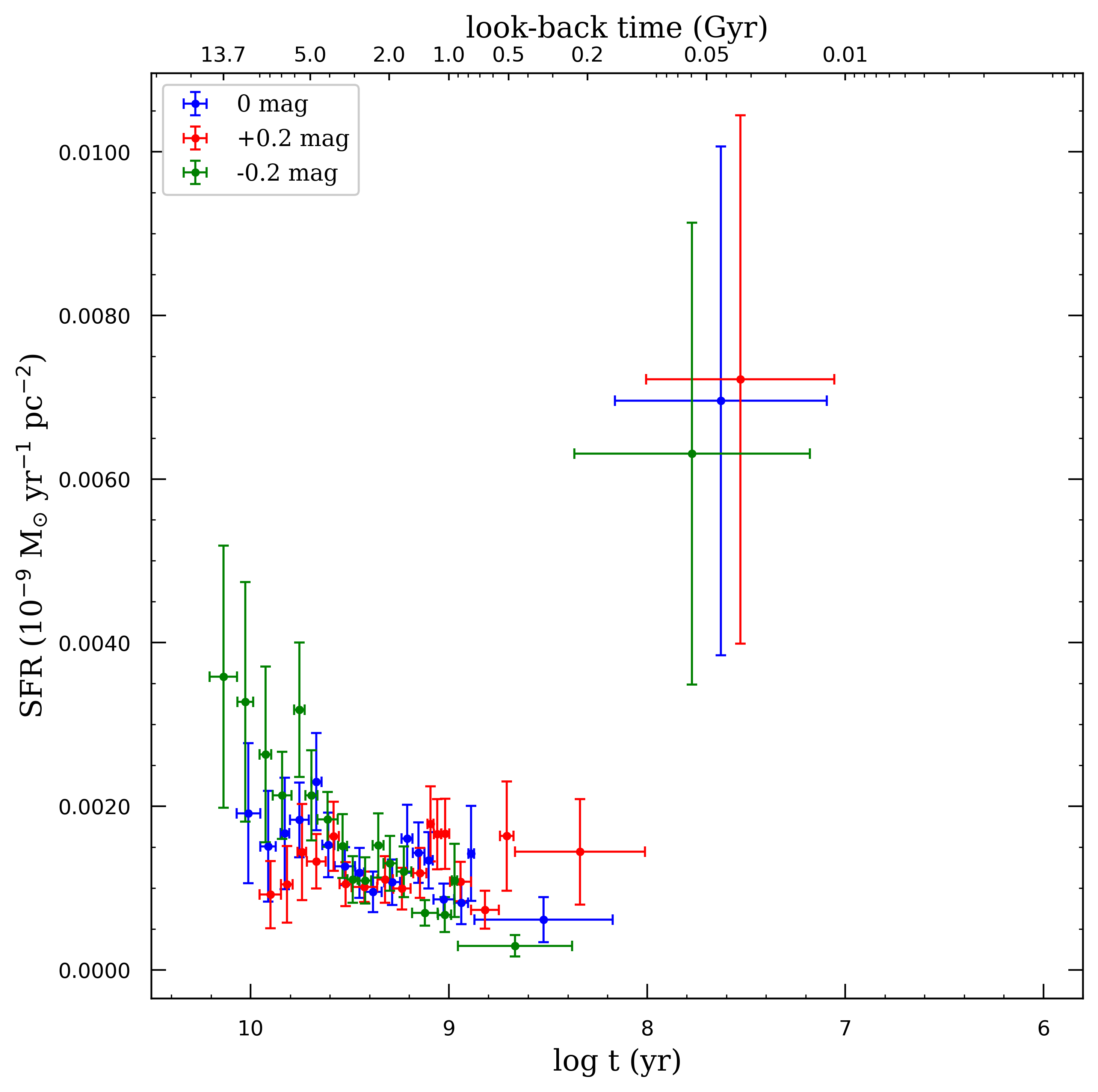}
  \includegraphics[width=0.45\textwidth]{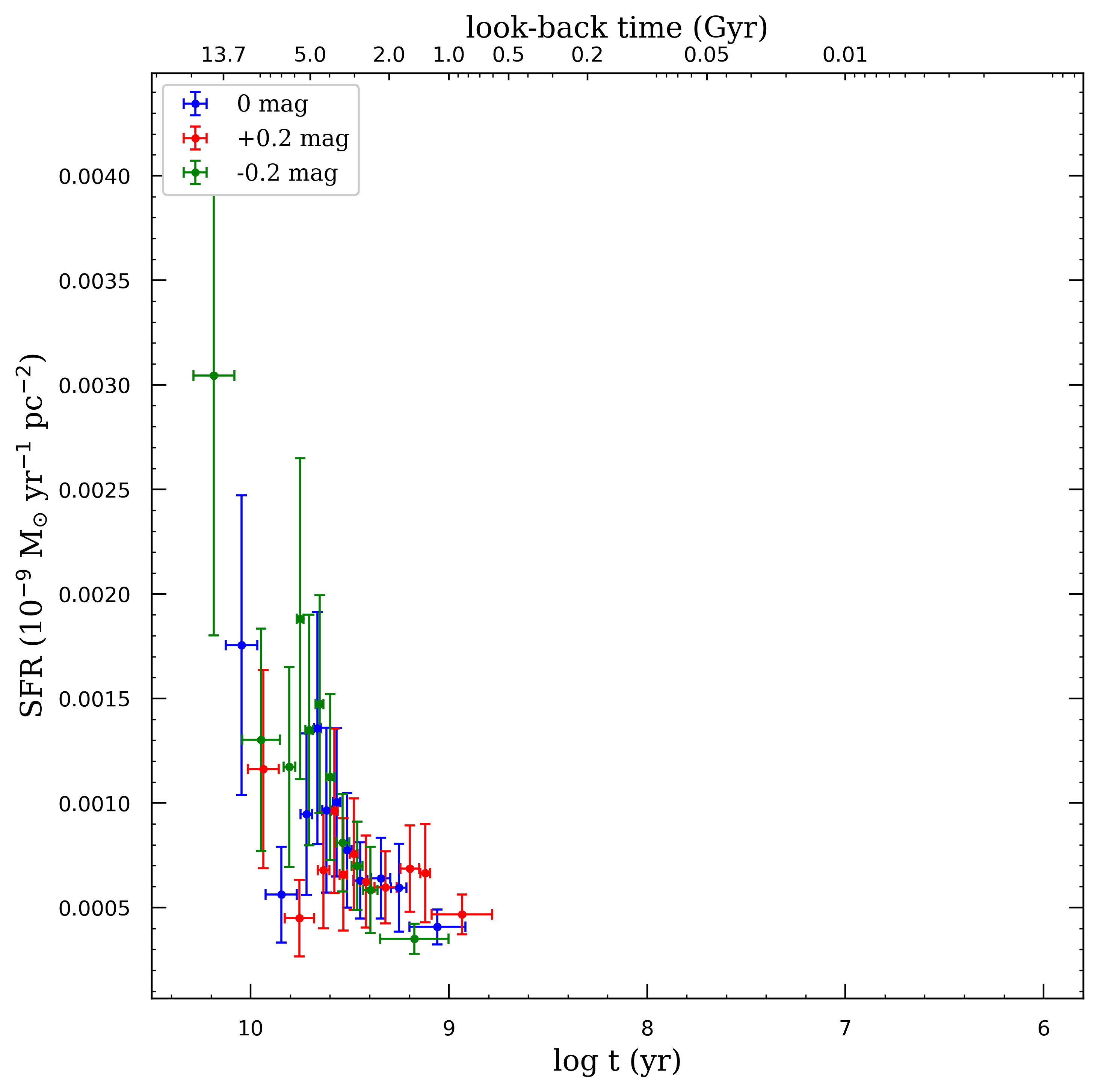}
\caption{Star formation histories of the bar region (left panel) and the outer region (right panel) derived under different assumptions for photometric and distance uncertainties. The blue markers shows the nominal solution, while the red and green symbols correspond to systematic magnitude shifts of $+0.2$ mag and $-0.2$ mag, respectively.}
 \label{fig:mags_a}
 \end{center}
 \end{figure*}

To assess the impact of photometric and distance-related uncertainties on the recovered SFHs, we repeated the analysis after applying systematic magnitude offsets of $\pm0.2$ mag to the photometric catalog. The resulting SFHs for both the Bar and Outer regions are shown in Fig.~\ref{fig:mags_a}. Comparison of the perturbed and nominal solutions indicates that the recovered SFHs are generally robust against these systematic shifts. Relative to the standard solution, the mean variation in the inferred SFR is approximately 25\% and 33\% for the $+0.2$ and $-0.2$ mag realizations in the Bar region, respectively. For the Outer region, the corresponding mean variations are approximately 21\% and 41\%. The largest fractional differences occur in age bins with intrinsically low SFRs, where small absolute changes produce large relative variations. Importantly, the principal episodes of star formation and the overall temporal evolution remain unchanged across all three realizations. These results indicate that uncertainties associated with photometric calibration and distance modulus primarily affect the absolute normalization of the SFH at the level of a few tens of percent, while leaving the main evolutionary trends of both regions intact.

\counterwithin{figure}{section}
\counterwithin{table}{section}
 \section{Supplementary Material}
\label{sec:apndix}

\begin{figure*}[h!]
\centering
	{\hbox
    { \epsfig{figure=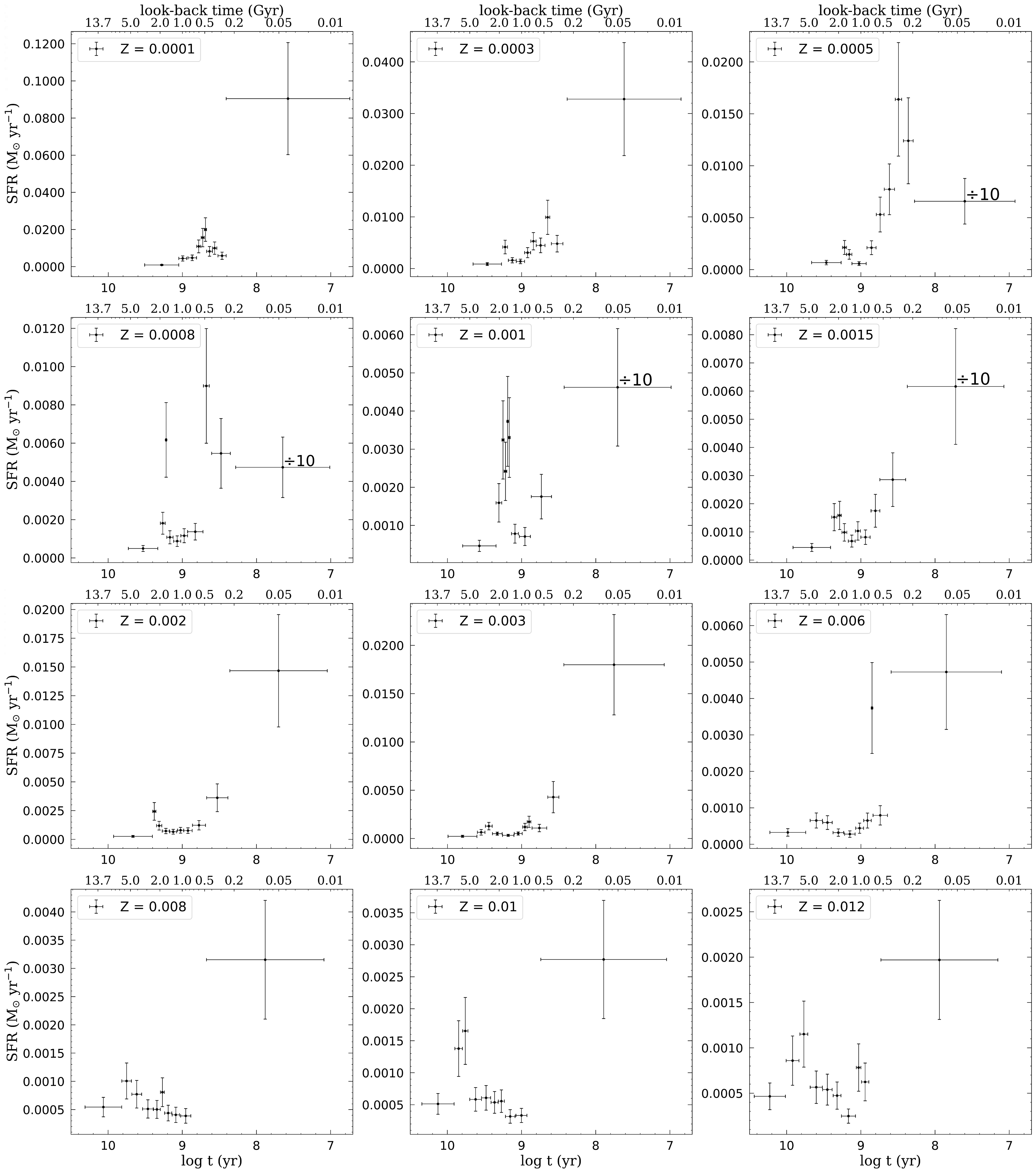,width=180mm,height=200mm}    
   \centering
	}}
	\caption{SFH for 97 LPV stars from \cite{whitelock2013local} located on the bar for 12 metallicity values. The numbers on a bin indicate that the actual SFR value has been divided by that number.}
    \label{fig:Apendix_Fig_1}
\end{figure*}

\begin{figure*}[h!]
\centering
	{\hbox
    { \epsfig{figure=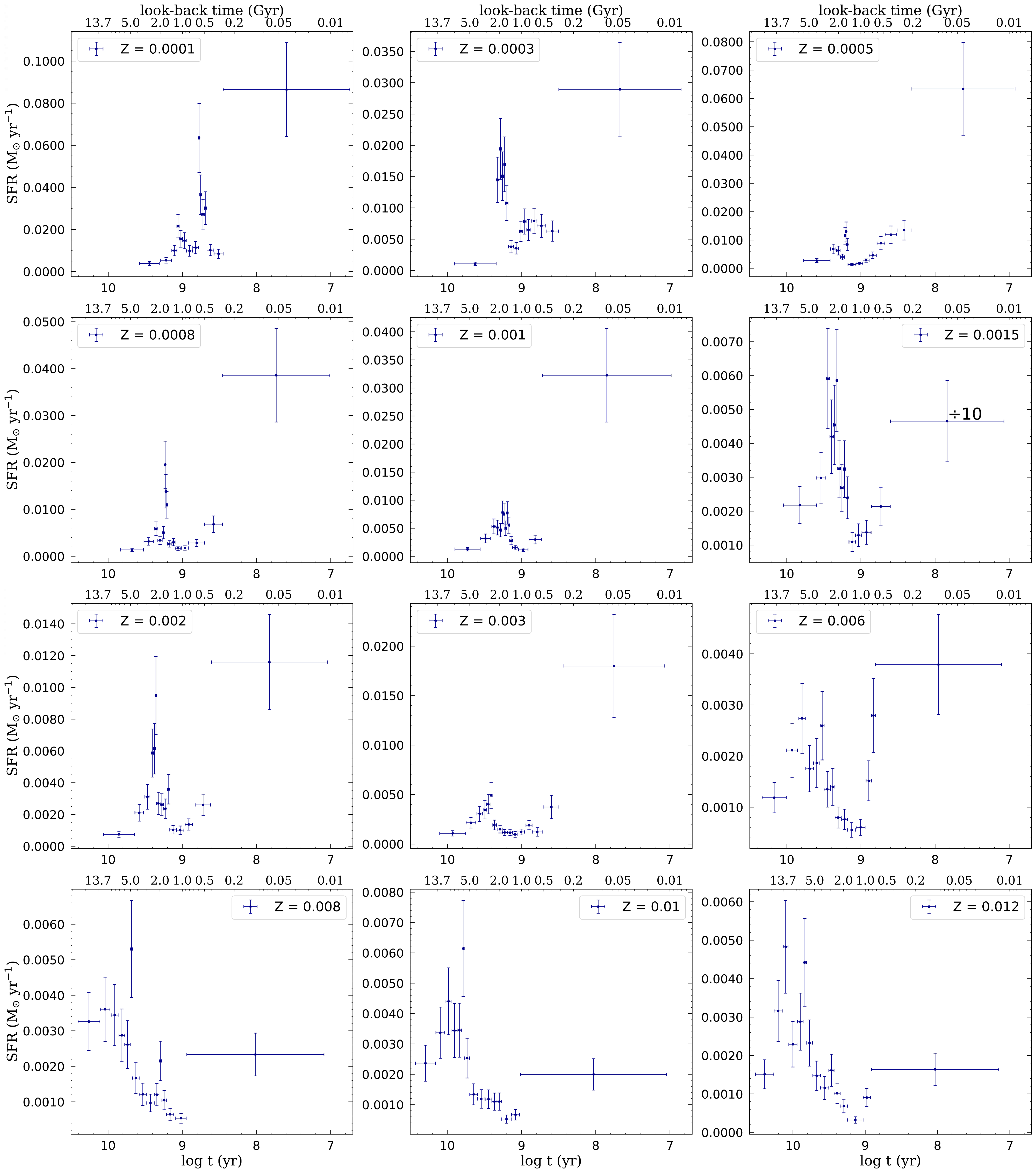,width=180mm,height=200mm}
   \centering
	}}
	\caption{The same as Fig.\ref{fig:Apendix_Fig_1} for the bar region catalog.}
    \label{fig:Apendix_Fig_2}
\end{figure*}

\begin{figure*}[h!]
\centering
	{\hbox
    { \epsfig{figure=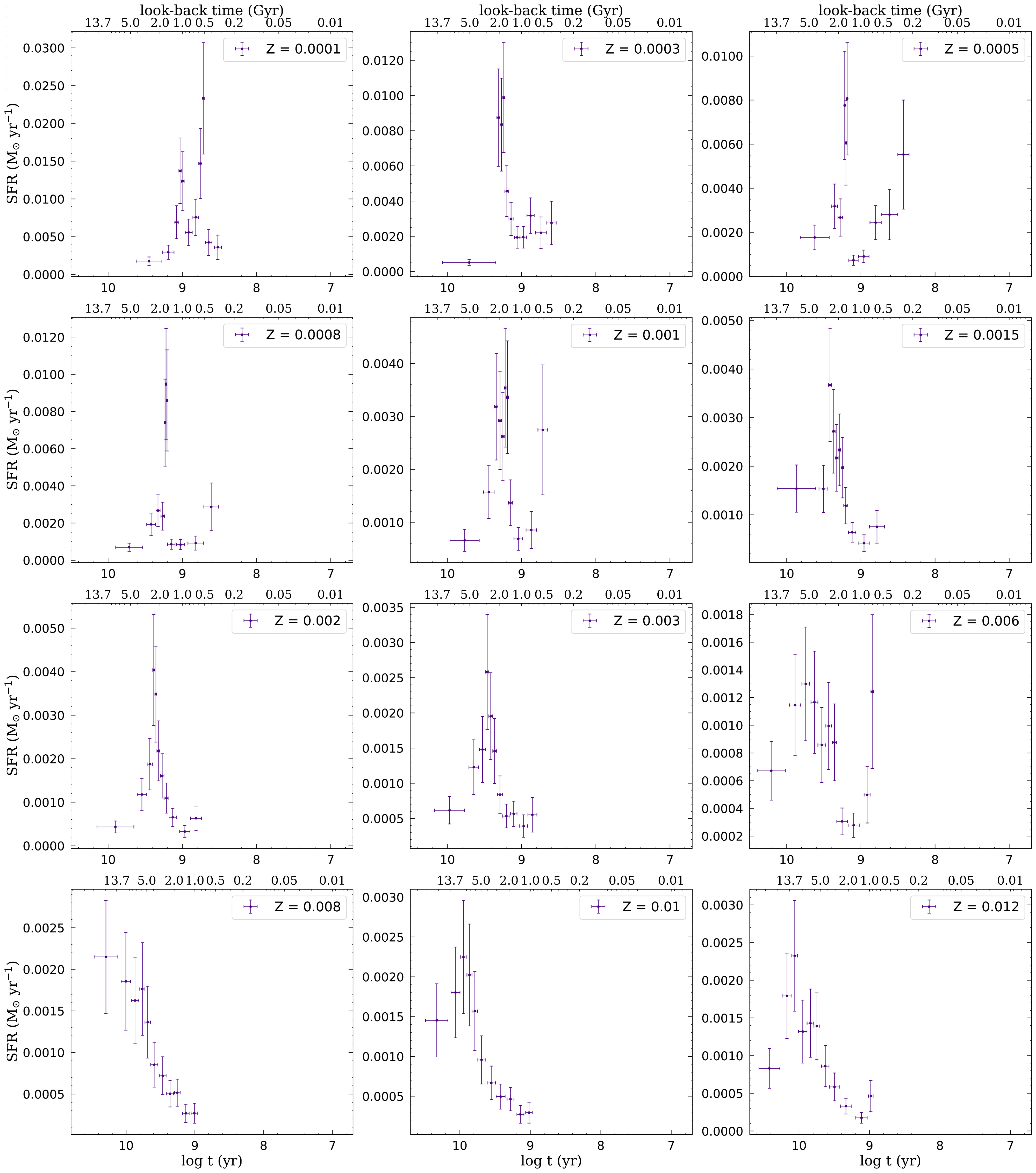,width=180mm,height=200mm}
   \centering
	}}
	\caption{The same as Fig.\ref{fig:Apendix_Fig_1} for the outer region catalog.}
    \label{fig:Apendix_Fig_3}
\end{figure*}

The SFHs derived adopting 12 constant Metallicity values falling within the range of  $-$2.53 $<$ [Fe/H] $<$ $-$0.49 dex for the 97-LPVs, the bar region, and the outer region catalogs (Fig.~\ref{fig:Apendix_Fig_1}, Fig.~\ref{fig:Apendix_Fig_2} and Fig.~\ref{fig:Apendix_Fig_3}) along with the fitting equations of relations to derive birth mass, age, and pulsation durations of the LPVs are presented below (Tabs.~\ref{tab:tab9}--\ref{tab:tab18}). 

\newpage

\newpage

\begin{figure*}[h!]
\centering
	{\hbox
    { \epsfig{figure=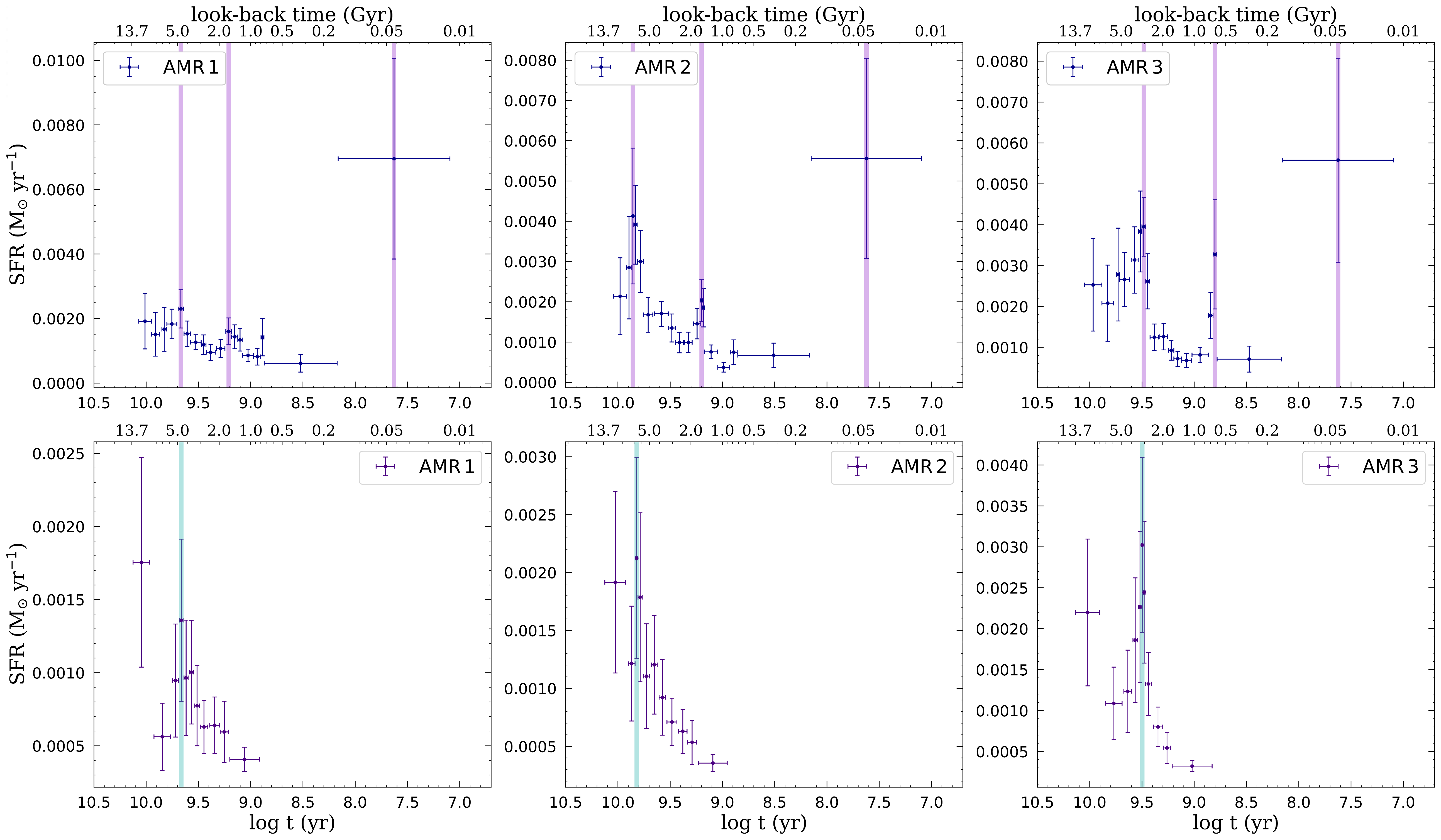,width=180mm,height=120mm}
   \centering
	}}
	\caption{The SFH for the bar region (upper panels) and the outer region (lower panels), assuming variable metallicity by time based
    on the three AMRs from Fig.\,\ref{fig: Fig. 4}. The vertical lines signify the peaks of the distinct star formation epochs.}
    \label{fig:Apendix_Fig_4}
\end{figure*}

\begin{table}[h!]
\addtolength{\tabcolsep}{+7pt}
\renewcommand{\arraystretch}{0.96}
\centering
\caption{Fitting parameters of the relation between birth mass and K$_s$ band magnitude, $\log{\mathrm{(M/M_\sun)}} = a\,\mathrm{K_s} + b$.}
\label{tab:tab9}
\begin{tabular}{@{}ccc@{}}
\toprule
$a$                 & $b$                  & $Validity$ $range$                        \\ \midrule
                  & Z $=$ 0.0001       &                                       \\
\midrule
$-$-0.255 $\pm$ 0.194 & $-$1.293 $\pm$ 2.224 & K$_s$ $\leq$ $-$11.714                      \\
$-$0.106 $\pm$ 0.197     & 0.454 $\pm$ 2.173      & $-$11.714 $<$ K$_s$ $\leq$ $-$11.247 \\
$-$0.223 $\pm$ 0.165     & $-$0.866 $\pm$ 1.745     & $-$11.247 $<$ K$_s$ $\leq$ $-$10.779 \\
$-$0.653 $\pm$ 0.142     & $-$5.492 $\pm$ 1.435     & $-$10.779 $<$ K$_s$ $\leq$ $-$10.312 \\
$-$0.676 $\pm$ 0.136     & $-$5.735 $\pm$ 1.309     & $-$10.312 $<$ K$_s$ $\leq$ $-$9.845  \\
$-$0.705 $\pm$ 0.124     & $-$6.018 $\pm$ 1.139      & $-$9.845 $<$ K$_s$ $\leq$ $-$9.378   \\
$-$0.204 $\pm$ 0.098     & $-$1.317 $\pm$ 0.853     & $-$9.378 $<$ K$_s$ $\leq$ $-$8.910   \\
$-$0.224 $\pm$ 0.107     & $-$1.498 $\pm$ 0.875     & $-$8.910 $<$ K$_s$ $\leq$ $-$8.443   \\
$-$0.106 $\pm$ 0.113     & $-$0.506 $\pm$ 0.878     & $-$8.443 $<$ K$_s$ $\leq$ $-$7.976   \\
$-$0.181 $\pm$ 0.098     & $-$1.098 $\pm$ 0.715     & $-$7.976 $<$ K$_s$ $\leq$ $-$7.509   \\
$-$0.099 $\pm$ 0.106     & $-$0.481 $\pm$ 0.714     & $-$7.509 $<$ K$_s$ $\leq$ $-$7.042   \\
$-$0.233 $\pm$ 0.107     & $-$1.429 $\pm$ 0.685     & $-$7.042 $<$ K$_s$ $\leq$ $-$6.574   \\
$-$0.135 $\pm$ 0.087     & $-$0.785 $\pm$ 0.513     & $-$6.574 $<$ K$_s$ $\leq$ $-$6.107   \\
$-$0.343 $\pm$ 0.089     & $-$2.053 $\pm$ 0.481     & K$_s$ $>$ $-$6.107   \\ 
\hline

                  & Z $=$ 0.0003       &                                       \\
\midrule
$-$0.107 $\pm$ 0.181 & 0.497 $\pm$ 2.124 & K$_s$ $\leq$ $-$11.984                      \\
$-$0.312 $\pm$ 0.182     & $-$1.965 $\pm$ 2.062      & $-$11.984 $<$ K$_s$ $\leq$ $-$11.534 \\
$-$0.249 $\pm$ 0.128     & $-$2.238 $\pm$ 1.390     & $-$11.534 $<$ K$_s$ $\leq$ $-$11.083 \\
$-$0.520 $\pm$ 0.098     & $-$4.239 $\pm$ 1.028     & $-$11.083 $<$ K$_s$ $\leq$ $-$10.633 \\
$-$0.278 $\pm$ 0.087     & $-$1.669 $\pm$ 0.863     & $-$10.633 $<$ K$_s$ $\leq$ $-$10.182  \\
$-$0.660 $\pm$ 0.118     & $-$5.564 $\pm$ 1.112      & $-$10.182 $<$ K$_s$ $\leq$ $-$9.732   \\
$-$0.602 $\pm$ 0.113     & $-$4.993 $\pm$ 1.031     & $-$9.732 $<$ K$_s$ $\leq$ $-$9.281   \\
$-$0.181 $\pm$ 0.083     & $-$1.088 $\pm$ 0.717     & $-$9.281 $<$ K$_s$ $\leq$ $-$8.830   \\
$-$0.310 $\pm$ 0.084     & $-$2.226 $\pm$ 0.687     & $-$8.830 $<$ K$_s$ $\leq$ $-$8.380   \\
$-$0.225 $\pm$ 0.083     & $-$1.514 $\pm$ 0.640     & $-$8.380 $<$ K$_s$ $\leq$ $-$7.929   \\
$-$0.189 $\pm$ 0.129     & $-$1.227 $\pm$ 0.922     & $-$7.929 $<$ K$_s$ $\leq$ $-$7.479   \\
$-$0.074 $\pm$ 0.137     & $-$0.367 $\pm$ 0.943     & $-$7.479 $<$ K$_s$ $\leq$ $-$7.028   \\
$-$0.066 $\pm$ 0.107     & $-$0.310 $\pm$ 0.676     & $-$7.028 $<$ K$_s$ $\leq$ $-$6.578   \\
$-$0.491 $\pm$ 0.091     & $-$3.106 $\pm$ 0.544     & K$_s$ $>$ -6.578   \\
\hline
                  & Z $=$ 0.0005       &                                       \\
\midrule
$-$0.121 $\pm$ 0.227 & 0.309 $\pm$ 2.663 & K$_s$ $\leq$ $-$12.011                      \\
$-$0.281 $\pm$ 0.226     & $-$1.614 $\pm$ 2.558      & $-$12.011 $<$ K$_s$ $\leq$ $-$11.535 \\
$-$0.443 $\pm$ 0.143     & $-$3.475 $\pm$ 1.554     & $-$11.535 $<$ K$_s$ $\leq$ $-$11.060 \\
$-$0.414 $\pm$ 0.123     & $-$3.160 $\pm$ 1.265     & $-$11.060 $<$ K$_s$ $\leq$ $-$10.584 \\
$-$0.231 $\pm$ 0.126     & $-$1.217 $\pm$ 1.247     & $-$10.584 $<$ K$_s$ $\leq$ $-$10.108  \\
$-$0.729 $\pm$ 0.118     & $-$6.259 $\pm$ 1.103      & $-$10.108 $<$ K$_s$ $\leq$ $-$9.632   \\
$-$0.344 $\pm$ 0.126     & $-$2.547 $\pm$ 1.123     & $-$9.632 $<$ K$_s$ $\leq$ $-$9.156   \\
$-$0.151 $\pm$ 0.111     & $-$0.774 $\pm$ 0.941     & $-$9.156 $<$ K$_s$ $\leq$ $-$8.681   \\
$-$0.354 $\pm$ 0.093     & $-$2.538 $\pm$ 0.740     & $-$8.681 $<$ K$_s$ $\leq$ $-$8.205   \\
$-$0.345 $\pm$ 0.152     & $-$2.463 $\pm$ 1.123     & $-$8.205 $<$ K$_s$ $\leq$ $-$7.729   \\
$-$0.045 $\pm$ 0.196     & $-$0.144 $\pm$ 1.383     & $-$7.729 $<$ K$_s$ $\leq$ $-$7.253   \\
$-$0.122 $\pm$ 0.144     & $-$0.710 $\pm$ 0.947     & $-$7.253 $<$ K$_s$ $\leq$ $-$6.778   \\
$-$0.231 $\pm$ 0.140     & $-$1.443 $\pm$ 0.840     & $-$6.778 $<$ K$_s$ $\leq$ $-$6.302   \\
$-$0.139 $\pm$ 0.125     & $-$0.866 $\pm$ 0.706     & K$_s$ $>$ $-$6.302   \\
\hline
\end{tabular}
\end{table}

\begin{table}
\addtolength{\tabcolsep}{+7pt}
\centering
\begin{tabular}{@{}ccc@{}}
\hline
$a$                 & $b$                  & $Validity$ $range$                        \\ \midrule
                  & Z $=$ 0.0008       &                                       \\
\midrule
$-$0.132 $\pm$ 0.203 & 0.184 $\pm$ 2.238 & K$_s$ $\leq$ $-$12.015                      \\
$-$0.237 $\pm$ 0.197     & $-$1.086 $\pm$ 2.208      & $-$12.015 $<$ K$_s$ $\leq$ $-$11.552 \\
$-$0.709 $\pm$ 0.132     & $-$6.537 $\pm$ 1.440     & $-$11.552 $<$ K$_s$ $\leq$ $-$11.089 \\
$-$0.214 $\pm$ 0.122     & $-$1.047 $\pm$ 1.264     & $-$11.089 $<$ K$_s$ $\leq$ $-$10.627 \\
$-$0.278 $\pm$ 0.110     & $-$1.730 $\pm$ 1.093     & $-$10.627 $<$ K$_s$ $\leq$ $-$10.164  \\
$-$0.668 $\pm$ 0.115     & $-$5.686 $\pm$ 1.087      & $-$10.164 $<$ K$_s$ $\leq$ $-$9.701   \\
$-$0.402 $\pm$ 0.124     & $-$3.107 $\pm$ 1.116     & $-$9.701 $<$ K$_s$ $\leq$ $-$9.238   \\
$-$0.114 $\pm$ 0.101     & $-$0.445 $\pm$ 0.864     & $-$9.238 $<$ K$_s$ $\leq$ $-$8.775   \\
$-$0.470 $\pm$ 0.077     & $-$3.570 $\pm$ 0.628     & $-$8.775 $<$ K$_s$ $\leq$ $-$8.312   \\
$-$0.287 $\pm$ 0.110     & $-$2.053 $\pm$ 0.825     & $-$8.312 $<$ K$_s$ $\leq$ $-$7.849   \\
$-$0.035 $\pm$ 0.158     & $-$0.076 $\pm$ 1.128     & $-$7.849 $<$ K$_s$ $\leq$ $-$7.987   \\
$-$0.134 $\pm$ 0.141     & $-$0.808 $\pm$ 0.947     & $-$7.387 $<$ K$_s$ $\leq$ $-$6.924   \\
$-$0.199 $\pm$ 0.119     & $-$1.254 $\pm$ 0.740     & $-$6.924 $<$ K$_s$ $\leq$ $-$6.461   \\
$-$0.205 $\pm$ 0.093     & $-$1.292 $\pm$ 0.544     & K$_s$ $>$ -6.461   \\
\hline
                  & Z $=$ 0.001       &                                       \\
\hline
$-$0.011 $\pm$ 0.303 & $-$1.684 $\pm$ 3.566 & K$_s$ $\leq$ $-$12.033                      \\
$-$0.590 $\pm$ 0.262     & $-$5.274 $\pm$ 2.996      & $-$12.033 $<$ K$_s$ $\leq$ $-$11.573 \\
$-$0.375 $\pm$ 0.176     & $-$2.795 $\pm$ 1.915     & $-$11.573 $<$ K$_s$ $\leq$ $-$11.114 \\
$-$0.347 $\pm$ 0.187     & $-$2.484 $\pm$ 1.948     & $-$11.114 $<$ K$_s$ $\leq$ $-$10.654 \\
$-$0.300 $\pm$ 0.165     & $-$1.979 $\pm$ 1.646     & $-$10.654 $<$ K$_s$ $\leq$ $-$10.195  \\
$-$0.376 $\pm$ 0.150     & $-$2.752 $\pm$ 1.428      & $-$10.195 $<$ K$_s$ $\leq$ $-$9.735   \\
$-$0.561 $\pm$ 0.155     & $-$4.552 $\pm$ 1.399     & $-$9.735 $<$ K$_s$ $\leq$ $-$9.276   \\
$-$0.255 $\pm$ 0.140     & $-$1.714 $\pm$ 1.204     & $-$9.276 $<$ K$_s$ $\leq$ $-$8.816   \\
$-$0.414 $\pm$ 0.115     & $-$3.117 $\pm$ 0.941     & $-$8.816 $<$ K$_s$ $\leq$ $-$8.357   \\
$-$0.280 $\pm$ 0.139     & $-$1.996 $\pm$ 1.056     & $-$8.357 $<$ K$_s$ $\leq$ $-$7.897   \\
$-$0.082 $\pm$ 0.203     & $-$0.432 $\pm$ 1.435     & $-$7.897 $<$ K$_s$ $\leq$ $-$7.438   \\
$-$0.108 $\pm$ 0.208     & $-$0.626 $\pm$ 1.409     & $-$7.438 $<$ K$_s$ $\leq$ $-$6.978   \\
$-$0.216 $\pm$ 0.170     & $-$1.381 $\pm$ 1.070     & $-$6.978 $<$ K$_s$ $\leq$ $-$6.518   \\
$-$0.204 $\pm$ 0.130     & $-$1.305 $\pm$ 0.764     & K$_s$ $>$ $-$6.518   \\
\hline
                  & Z $=$ 0.0015       &                                       \\
\hline
$-$0.225 $\pm$ 0.122 & $-$0.980 $\pm$ 1.430 & K$_s$ $\leq$ $-$12.009                      \\
$-$0.408 $\pm$ 0.106     & $-$3.188 $\pm$ 1.210      & $-$12.009 $<$ K$_s$ $\leq$ $-$11.561 \\
$-$0.533 $\pm$ 0.077     & $-$4.635 $\pm$ 0.841     & $-$11.561 $<$ K$_s$ $\leq$ $-$11.113 \\
$-$0.189 $\pm$ 0.083     & $-$0.812 $\pm$ 0.863     & $-$11.113 $<$ K$_s$ $\leq$ $-$10.665 \\
$-$0.266 $\pm$ 0.074     & $-$1.631 $\pm$ 0.742     & $-$10.665 $<$ K$_s$ $\leq$ $-$10.217  \\
$-$0.613 $\pm$ 0.094     & $-$5.172 $\pm$ 0.890      & $-$10.217 $<$ K$_s$ $\leq$ $-$9.769   \\
$-$0.425 $\pm$ 0.097     & $-$3.334 $\pm$ 0.889     & $-$9.769 $<$ K$_s$ $\leq$ $-$9.322   \\
$-$0.159 $\pm$ 0.066     & $-$0.858 $\pm$ 0.571     & $-$9.322 $<$ K$_s$ $\leq$ $-$8.874   \\
$-$0.497 $\pm$ 0.052     & $-$3.861 $\pm$ 0.424     & $-$8.874 $<$ K$_s$ $\leq$ $-$8.428   \\
$-$0.259 $\pm$ 0.054     & $-$1.850 $\pm$ 0.419     & $-$8.426 $<$ K$_s$ $\leq$ $-$7.978   \\
$-$0.108 $\pm$ 0.080     & $-$0.650 $\pm$ 0.580     & $-$7.978 $<$ K$_s$ $\leq$ $-$7.530   \\
$-$0.099 $\pm$ 0.088     & $-$0.582 $\pm$ 0.607     & $-$7.530 $<$ K$_s$ $\leq$ $-$7.082   \\
$-$0.163 $\pm$ 0.071     & $-$1.030 $\pm$ 0.454     & $-$7.082 $<$ K$_s$ $\leq$ $-$6.634   \\
$-$0.258 $\pm$ 0.054     & $-$1.682 $\pm$ 0.329     & K$_s$ $>$ $-$6.634   \\
\bottomrule
\end{tabular}
\end{table}

\begin{table}
\addtolength{\tabcolsep}{+7pt}
\centering
\begin{tabular}{@{}ccc@{}}
\hline
$a$                 & $b$                  & $Validity$ $range$                        \\ \midrule
                  & Z $=$ 0.002       &                                       \\
\midrule
$-$0.381 $\pm$ 0.151 & $-$2.920 $\pm$ 1.775 & K$_s$ $\leq$ $-$11.965                      \\
$-$0.523 $\pm$ 0.139     & $-$4.618 $\pm$ 1.577      & $-$11.965 $<$ K$_s$ $\leq$ $-$11.527 \\
$-$0.259 $\pm$ 0.132     & $-$1.575 $\pm$ 1.432     & $-$11.527 $<$ K$_s$ $\leq$ $-$11.088 \\
$-$0.234 $\pm$ 0.136     & $-$1.297 $\pm$ 1.417     & $-$11.088 $<$ K$_s$ $\leq$ $-$10.649 \\
$-$0.324 $\pm$ 0.124     & $-$2.257 $\pm$ 1.240     & $-$10.649 $<$ K$_s$ $\leq$ $-$10.211  \\
$-$0.306 $\pm$ 0.116     & $-$2.076 $\pm$ 1.111      & $-$10.211 $<$ K$_s$ $\leq$ $-$9.772   \\
$-$0.623 $\pm$ 0.113     & $-$5.173 $\pm$ 1.031     & $-$9.772 $<$ K$_s$ $\leq$ $-$9.334   \\
$-$0.276 $\pm$ 0.098     & $-$1.932 $\pm$ 0.853     & $-$9.334 $<$ K$_s$ $\leq$ $-$8.895   \\
$-$0.405 $\pm$ 0.087     & $-$7.084 $\pm$ 0.716     & $-$8.895 $<$ K$_s$ $\leq$ $-$8.456   \\
$-$0.269 $\pm$ 0.089     & $-$1.930 $\pm$ 0.697     & $-$8.456 $<$ K$_s$ $\leq$ $-$8.018   \\
$-$0.155 $\pm$ 0.160     & $-$1.017 $\pm$ 1.162     & $-$8.018 $<$ K$_s$ $\leq$ $-$7.579   \\
$-$0.066 $\pm$ 0.185     & $-$0.342 $\pm$ 1.291     & $-$7.579 $<$ K$_s$ $\leq$ $-$7.140   \\
$-$0.168 $\pm$ 0.122     & $-$1.073 $\pm$ 0.793     & $-$7.140 $<$ K$_s$ $\leq$ $-$6.702   \\
$-$0.271 $\pm$ 0.089     & $-$1.764 $\pm$ 0.539     & K$_s$ $>$ $-$6.702   \\
\hline
                  & Z $=$ 0.003       &                                       \\
\hline
$-$0.538 $\pm$ 0.085 & $-$4.810 $\pm$ 0.983 & K$_s$ $\leq$ $-$11.827                      \\
$-$0.449 $\pm$ 0.086     & $-$3.763 $\pm$ 0.967      & $-$11.827 $<$ K$_s$ $\leq$ $-$11.408 \\
$-$0.219 $\pm$ 0.101     & $-$1.134 $\pm$ 1.090     & $-$11.408 $<$ K$_s$ $\leq$ $-$10.990 \\
$-$0.311 $\pm$ 0.101     & $-$2.151 $\pm$ 1.047     & $-$10.990 $<$ K$_s$ $\leq$ $-$10.571 \\
$-$0.211 $\pm$ 0.093     & $-$1.093 $\pm$ 0.926     & $-$10.571 $<$ K$_s$ $\leq$ $-$10.152  \\
$-$0.676 $\pm$ 0.115     & $-$5.818 $\pm$ 1.088      & $-$10.152 $<$ K$_s$ $\leq$ $-$9.733   \\
$-$0.287 $\pm$ 0.110     & $-$2.030 $\pm$ 1.006     & $-$9.733 $<$ K$_s$ $\leq$ $-$9.315   \\
$-$0.311 $\pm$ 0.073     & $-$2.251 $\pm$ 0.639     & $-$9.315 $<$ K$_s$ $\leq$ $-$8.896   \\
$-$0.401 $\pm$ 0.067     & $-$3.049 $\pm$ 0.552     & $-$8.896 $<$ K$_s$ $\leq$ $-$8.477   \\
$-$0.216 $\pm$ 0.067     & $-$1.481 $\pm$ 0.525     & $-$8.477 $<$ K$_s$ $\leq$ $-$8.058   \\
$-$0.235 $\pm$ 0.090     & $-$1.636 $\pm$ 0.66     & $-$8.058 $<$ K$_s$ $\leq$ $-$7.639   \\
$-$0.102 $\pm$ 0.102     & $-$0.623 $\pm$ 0.718     & $-$7.639 $<$ K$_s$ $\leq$ $-$7.221   \\
$-$0.173 $\pm$ 0.085     & $-$1.130 $\pm$ 0.560     & $-$7.221 $<$ K$_s$ $\leq$ $-$6.802   \\
$-$0.218 $\pm$ 0.069     & $-$1.436 $\pm$ 0.431     & K$_s$ $>$ $-$6.802   \\
\hline
                  & Z $=$ 0.006       &                                       \\
\hline
$-$0.459 $\pm$ 0.136 & $-$3.888 $\pm$ 1.602 & K$_s$ $\leq$ $-$11.902                      \\
$-$0.623 $\pm$ 0.127     & $-$5.844 $\pm$ 1.426      & $-$11.902 $<$ K$_s$ $\leq$ $-$11.498 \\
$-$0.159 $\pm$ 0.156     & $-$0.598 $\pm$ 1.698     & $-$11.498 $<$ K$_s$ $\leq$ $-$11.094 \\
$-$0.280 $\pm$ 0.142     & $-$1.581 $\pm$ 1.493     & $-$11.094 $<$ K$_s$ $\leq$ $-$10.689 \\
$-$0.298 $\pm$ 0.146     & $-$2.036 $\pm$ 1.474     & $-$10.689 $<$ K$_s$ $\leq$ $-$10.285  \\
$-$0.355 $\pm$ 0.244     & $-$52.623 $\pm$ 2.344      & $-$10.258 $<$ K$_s$ $\leq$ $-$9.881   \\
$-$0.529 $\pm$ 0.230     & $-$4.347 $\pm$ 2.148     & $-$9.881 $<$ K$_s$ $\leq$ $-$9.477   \\
$-$0.240 $\pm$ 0.112     & $-$1.605 $\pm$ 0.999     & $-$9.477 $<$ K$_s$ $\leq$ $-$9.073   \\
$-$0.397 $\pm$ 0.102     & $-$3.031 $\pm$ 0.862     & $-$9.073 $<$ K$_s$ $\leq$ $-$8.669   \\
$-$0.111 $\pm$ 0.113     & $-$0.555 $\pm$ 0.915     & $-$8.669 $<$ K$_s$ $\leq$ $-$8.265   \\
$-$0.405 $\pm$ 0.110     & $-$2.981 $\pm$ 0.841     & $-$8.265 $<$ K$_s$ $\leq$ $-$7.860   \\
$-$0.198 $\pm$ 0.107     & $-$1.355 $\pm$ 0.785     & $-$7.860 $<$ K$_s$ $\leq$ $-$7.456   \\
$-$0.200 $\pm$ 0.107     & $-$1.371 $\pm$ 0.734     & $-$7.456 $<$ K$_s$ $\leq$ $-$7.052   \\
$-$0.200 $\pm$ 0.095     & $-$1.372 $\pm$ 0.614     & K$_s$ $>$ $-$7.052   \\
\bottomrule
\end{tabular}
\end{table}

\begin{table}
\addtolength{\tabcolsep}{+7pt}
\centering
\begin{tabular}{@{}ccc@{}}
\hline
$a$                 & $b$                  & $Validity$ $range$                        \\ \midrule
                  & Z $=$ 0.008       &                                       \\
\midrule
$-$0.767 $\pm$ 0.131 & $-$7.575 $\pm$ 1.522 & K$_s$ $\leq$ $-$11.767                      \\
$-$0.412 $\pm$ 0.140     & $-$3.402 $\pm$ 1.563      & $-$11.767 $<$ K$_s$ $\leq$ $-$11.381 \\
$-$0.139 $\pm$ 0.165     & $-$0.298 $\pm$ 1.781     & $-$11.381 $<$ K$_s$ $\leq$ $-$10.994 \\
$-$0.346 $\pm$ 0.139     & $-$2.565 $\pm$ 1.449     & $-$10.994 $<$ K$_s$ $\leq$ $-$10.608 \\
$-$0.195 $\pm$ 0.126     & $-$0.972 $\pm$ 1.265     & $-$10.608 $<$ K$_s$ $\leq$ $-$10.221  \\
$-$0.671 $\pm$ 0.173     & $-$5.838 $\pm$ 1.661      & $-$10.221 $<$ K$_s$ $\leq$ $-$9.835   \\
$-$0.200 $\pm$ 0.167     & $-$5.198 $\pm$ 1.550     & $-$9.835 $<$ K$_s$ $\leq$ $-$9.448   \\
$-$0.390 $\pm$ 0.101     & $-$2.999 $\pm$ 0.896     & $-$9.448 $<$ K$_s$ $\leq$ $-$9.061   \\
$-$0.368 $\pm$ 0.089     & $-$2.796 $\pm$ 0.756     & $-$9.061 $<$ K$_s$ $\leq$ $-$8.675   \\
$-$0.307 $\pm$ 0.095     & $-$2.269 $\pm$ 0.770     & $-$8.675 $<$ K$_s$ $\leq$ $-$8.288   \\
$-$0.210 $\pm$ 0.102     & $-$1.461 $\pm$ 0.780     & $-$8.288 $<$ K$_s$ $\leq$ $-$7.902   \\
$-$0.271 $\pm$ 0.110     & $-$1.948 $\pm$ 0.807     & $-$7.902 $<$ K$_s$ $\leq$ $-$7.515   \\
$-$0.161 $\pm$ 0.107     & $-$1.117 $\pm$ 0.740     & $-$7.515 $<$ K$_s$ $\leq$ $-$7.129   \\
$-$0.180 $\pm$ 0.108     & $-$1.258 $\pm$ 0.705     & K$_s$ $>$ $-$7.129   \\
\hline
                  & Z $=$ 0.01       &                                       \\
\hline
$-$0.750 $\pm$ 0.139 & $-$7.390 $\pm$ 1.617 & K$_s$ $\leq$ $-$11.777                      \\
$-$0.402 $\pm$ 0.143     & $-$3.296 $\pm$ 1.599      & $-$11.777 $<$ K$_s$ $\leq$ $-$11.396 \\
$-$0.173 $\pm$ 0.165     & $-$0.677 $\pm$ 1.788     & $-$11.396 $<$ K$_s$ $\leq$ $-$11.016 \\
$-$0.310 $\pm$ 0.141     & $-$2.189 $\pm$ 1.479     & $-$11.016 $<$ K$_s$ $\leq$ $-$10.635 \\
$-$0.217 $\pm$ 0.137     & $-$1.196 $\pm$ 1.382     & $-$10.635 $<$ K$_s$ $\leq$ $-$10.254  \\
$-$0.583 $\pm$ 0.162     & $-$4.950 $\pm$ 1.562      & $-$10.254 $<$ K$_s$ $\leq$ $-$9.873   \\
$-$0.305 $\pm$ 0.150     & $-$2.210 $\pm$ 1.399     & $-$9.873 $<$ K$_s$ $\leq$ $-$9.492   \\
$-$0.309 $\pm$ 0.104     & $-$2.245 $\pm$ 0.929     & $-$9.492 $<$ K$_s$ $\leq$ $-$9.112   \\
$-$0.410 $\pm$ 0.093     & $-$3.168 $\pm$ 0.790     & $-$9.112 $<$ K$_s$ $\leq$ $-$8.731   \\
$-$0.292 $\pm$ 0.108     & $-$2.137 $\pm$ 0.883     & $-$8.731 $<$ K$_s$ $\leq$ $-$8.350   \\
$-$0.249 $\pm$ 0.105     & $-$1.776 $\pm$ 0.816     & $-$8.350 $<$ K$_s$ $\leq$ $-$7.969   \\
$-$0.324 $\pm$ 0.099     & $-$2.376 $\pm$ 0.735     & $-$7.969 $<$ K$_s$ $\leq$ $-$7.589   \\
$-$0.156 $\pm$ 0.109     & $-$1.102 $\pm$ 0.765     & $-$7.589 $<$ K$_s$ $\leq$ $-$7.208   \\
$-$0.166 $\pm$ 0.117     & $-$1.170 $\pm$ 0.776     & K$_s$ $>$ $-$7.208   \\
\hline
                  & Z $=$ 0.012       &                                       \\
\hline
$-$0.771 $\pm$ 0.063 & $-$7.674 $\pm$ 0.731 & K$_s$ $\leq$ $-$11.715                      \\
$-$0.174 $\pm$ 0.067     & $-$0.684 $\pm$ 0.751      & $-$11.715 $<$ K$_s$ $\leq$ $-$11.344 \\
$-$0.243 $\pm$ 0.070     & $-$1.464 $\pm$ 0.757     & $-$11.344 $<$ K$_s$ $\leq$ $-$10.974 \\
$-$0.248 $\pm$ 0.064     & $-$1.526 $\pm$ 0.665     & $-$10.974 $<$ K$_s$ $\leq$ $-$10.603 \\
$-$0.253 $\pm$ 0.069     & $-$1.577 $\pm$ 0.696     & $-$10.603 $<$ K$_s$ $\leq$ $-$10.232  \\
$-$0.660 $\pm$ 0.075     & $-$5.741 $\pm$ 0.723      & $-$10.232 $<$ K$_s$ $\leq$ $-$9.862   \\
$-$0.191 $\pm$ 0.063     & $-$1.119 $\pm$ 0.587     & $-$9.862 $<$ K$_s$ $\leq$ $-$9.492   \\
$-$0.374 $\pm$ 0.043     & $-$2.850 $\pm$ 0.390     & $-$9.492 $<$ K$_s$ $\leq$ $-$9.121   \\
$-$0.388 $\pm$ 0.036     & $-$2.978 $\pm$ 0.309     & $-$9.121 $<$ K$_s$ $\leq$ $-$8.750   \\
$-$0.204 $\pm$ 0.044     & $-$1.370 $\pm$ 0.359     & $-$8.750 $<$ K$_s$ $\leq$ $-$8.380   \\
$-$0.388 $\pm$ 0.045     & $-$2.910 $\pm$ 0.531     & $-$8.380 $<$ K$_s$ $\leq$ $-$8.009   \\
$-$0.294 $\pm$ 0.042     & $-$2.159 $\pm$ 0.312     & $-$8.009 $<$ K$_s$ $\leq$ $-$7.639   \\
$-$0.187 $\pm$ 0.046     & $-$1.342 $\pm$ 0.329     & $-$7.639 $<$ K$_s$ $\leq$ $-$7.268   \\
$-$0.155 $\pm$ 0.058     & $-$1.120 $\pm$ 0.389     & K$_s$ $>$ $-$7.268   \\
\bottomrule
\end{tabular}
\end{table}

\begin{table}
\addtolength{\tabcolsep}{+18pt}
\centering
\begin{tabular}{@{}ccc@{}}
\hline
$a$                 & $b$                  & $Validity$ $range$                        \\ \midrule
                  & AMR\,1       &                                       \\
\midrule
$-$0.766     & $-$7.556   & $-$12.950 $<$ K$_s$ $\leq$ $-$11.750 \\
$-$0.287     & $-$1.936   & $-$11.750 $<$ K$_s$ $\leq$ $-$11.050 \\
$-$0.274     & $-$1.788   & $-$10.050 $<$ K$_s$ $\leq$ $-$10.350 \\
$-$0.520     & $-$4.331   & $-$10.650 $<$ K$_s$ $\leq$ $-$9.950  \\
$-$0.348     & $-$2.620   & $-$9.950  $<$ K$_s$ $\leq$ $-$8.650   \\
$-$0.185     & $-$1.212   & $-$8.650  $<$ K$_s$ $\leq$ $-$6.750   \\
$-$0.218     & $-$1.436   & $-$6.750 $<$ K$_s$ $\leq$ $-$6.550   \\
$-$0.175     & $-$1.156   & $-$6.550 $<$ K$_s$ $\leq$ $-$6.350   \\
$-$0.258     & $-$1.683   & $-$6.350 $<$ K$_s$ $\leq$ $-$4.140   \\

\hline
                  & AMR\,2       &                                       \\
\hline
$-$0.766     & $-$7.556   & $-$12.950 $<$ K$_s$ $\leq$ $-$11.750 \\
$-$0.398     & $-$3.242   & $-$11.750 $<$ K$_s$ $\leq$ $-$11.350 \\
$-$0.140     & $-$0.308   & $-$11.350 $<$ K$_s$ $\leq$ $-$10.950 \\
$-$0.281     & $-$1.857   & $-$10.950 $<$ K$_s$ $\leq$ $-$10.250  \\
$-$0.639     & $-$5.518   & $-$10.250  $<$ K$_s$ $\leq$ $-$9.850   \\
$-$0.198     & $-$1.183   & $-$9.850  $<$ K$_s$ $\leq$ $-$9.450   \\
$-$0.361     & $-$2.723   & $-$9.450 $<$ K$_s$ $\leq$ $-$8.350   \\
$-$0.139     & $-$0.868   & $-$8.350 $<$ K$_s$ $\leq$ $-$7.950   \\
$-$0.223     & $-$1.684   & $-$7.950 $<$ K$_s$ $\leq$ $-$7.050   \\
$-$0.109     & $-$0.730   & $-$7.050 $<$ K$_s$ $\leq$ $-$6.380   \\
$-$0.258     & $-$1.684   & $-$6.380 $<$ K$_s$ $\leq$ $-$4.140   \\

\hline
                  & AMR\,3       &                                       \\
\hline
$-$0.766     & $-$7.556   & $-$12.950 $<$ K$_s$ $\leq$ $-$11.750 \\
$-$0.398     & $-$3.242   & $-$11.750 $<$ K$_s$ $\leq$ $-$11.350 \\
$-$0.140     & $-$0.308   & $-$11.350 $<$ K$_s$ $\leq$ $-$10.950 \\
$-$0.346     & $-$2.566   & $-$10.950 $<$ K$_s$ $\leq$ $-$10.550  \\
$-$0.195     & $-$0.972   & $-$10.550  $<$ K$_s$ $\leq$ $-$10.250   \\
$-$0.639     & $-$5.518   & $-$10.250  $<$ K$_s$ $\leq$ $-$9.850   \\
$-$0.198     & $-$1.183   & $-$9.850 $<$ K$_s$ $\leq$ $-$9.450   \\
$-$0.389     & $-$2.982   & $-$9.450 $<$ K$_s$ $\leq$ $-$8.950   \\
$-$0.306     & $-$2.242   & $-$8.680 $<$ K$_s$ $\leq$ $-$8.280   \\
$-$0.384     & $-$2.817   & $-$8.280 $<$ K$_s$ $\leq$ $-$7.580 \\
$-$0.090     & $-$0.540   & $-$7.580 $<$ K$_s$ $\leq$ $-$7.150 \\
$-$0.173     & $-$1.130   & $-$7.150 $<$ K$_s$ $\leq$ $-$6.850 \\
$-$0.122     & $-$0.779   & $-$6.850 $<$ K$_s$ $\leq$ $-$6.600  \\
$-$0.260     & $-$1.688   & $-$6.600  $<$ K$_s$ $\leq$ $-$4.140   \\

\bottomrule
\end{tabular}
\end{table}

\begin{table}
\addtolength{\tabcolsep}{+18pt}
\centering
\begin{tabular}{@{}ccc@{}}
\hline
$a$                 & $b$                  & $Validity$ $range$                        \\ \midrule
                  & BaSTI - Z = 0.003       &                                       \\
\midrule
$-$0.538     & $-$4.810   & K$_s$ $\leq$ $-$11.827                      \\
$-$0.449    & $-$3.763   & $-$11.827 $<$ K$_s$ $\leq$ $-$11.408 \\
$-$0.219     & $-$1.134   & $-$11.408 $<$ K$_s$ $\leq$ $-$10.990 \\
$-$0.311     & $-$2.151   & $-$10.990 $<$ K$_s$ $\leq$ $-$10.571 \\
$-$0.211     & $-$1.093   & $-$10.571 $<$ K$_s$ $\leq$ $-$10.152  \\
$-$0.676     & $-$5.818   & $-$10.152  $<$ K$_s$ $\leq$ $-$10.081   \\
$-$0.694     & $-$6.053   & $-$10.081  $<$ K$_s$ $\leq$ $-$9.915   \\
$-$0.927     & $-$8.368   & $-$9.915 $<$ K$_s$ $\leq$ $-$9.767   \\
$-$0.458     & $-$3.784   & $-$9.767 $<$ K$_s$ $\leq$ $-$9.492   \\
$-$0.363     & $-$2.884   & $-$9.492 $<$ K$_s$ $\leq$ $-$9.296   \\
$-$0.238     & $-$1.720   & $-$9.296 $<$ K$_s$ $\leq$ $-$9.014 \\
$-$0.225     & $-$1.609   & $-$9.014 $<$ K$_s$ $\leq$ $-$8.454 \\
$-$0.437     & $-$3.395   & $-$8.454 $<$ K$_s$ $\leq$ $-$8.347 \\
$-$2.082     & $-$17.130   & $-$8.347 $<$ K$_s$ $\leq$ $-$8.329  \\
$-$0.322     & $-$2.470   & $-$8.329  $<$ K$_s$ $\leq$ $-$7.976   \\
$-$0.272     & $-$2.070   & $-$7.976 $<$ K$_s$ $\leq$ $-$7.823 \\
$-$0.262     & $-$1.996   & $-$7.823 $<$ K$_s$ $\leq$ $-$7.455 \\
$-$0.102     & $-$0.623   & $-$7.455 $<$ K$_s$ $\leq$ $-$7.221  \\
$-$0.173     & $-$1.130   & $-$7.221  $<$ K$_s$ $\leq$ $-$6.802   \\
$-$0.218     & $-$1.436    & K$_s$ $>$ $-$ 6.802   \\
\bottomrule
\end{tabular}
\end{table}

\begin{table}[h!]
\caption{Fitting parametersof the relation between age and birth mass, $\log t = a \log \mathrm{(M/M_{\sun})} + b$.}
\label{tab:tab12}
\addtolength{\tabcolsep}{+9pt}
\centering
\begin{tabular}{@{}ccc@{}}
\toprule
$a$                 & $b$                  & $Validity$ $range$                        \\ \midrule
                  & Z $=$ 0.0001       &                                       \\
\midrule
$-$3.202 $\pm$ 0.019 & 9.762 $\pm$ 0.004 &  $\log \mathrm{M} \leq$ 0.124                      \\
$-$2.698 $\pm$ 0.017     & 9.700 $\pm$ 0.008      & 0.124 $< \log \mathrm{M} \leq$ 0.366 \\
$-$2.359 $\pm$ 0.018     & 9.575 $\pm$ 0.013     & 0.366 $< \log \mathrm{M} \leq$ 0.608 \\
$-$1.986 $\pm$ 0.020     & 9.349 $\pm$ 0.019     & 0.608 $< \log \mathrm{M} \leq$ 0.850 \\
$-$1.676 $\pm$ 0.022     & 9.085 $\pm$ 0.027     & 0.850 $< \log \mathrm{M} \leq$ 1.091  \\
$-$1.250 $\pm$ 0.025     & 8.621 $\pm$ 0.037      & 1.091 $< \log \mathrm{M} \leq$ 1.333   \\
$-$0.873 $\pm$ 0.029     & 8.118 $\pm$ 0.050     & 1.333 $< \log \mathrm{M} \leq$ 1.575   \\
$-$0.603 $\pm$ 0.035     & 7.693 $\pm$ 0.069     & $\log \mathrm{M} >$ 1.575   \\
\hline
                  & Z $=$ 0.0003       &                                       \\
\midrule
$-$3.202 $\pm$ 0.021 & 9.772 $\pm$ 0.005 & $ \log \mathrm{M} \leq$ 0.128         \\
$-$2.629 $\pm$ 0.019     & 9.699 $\pm$ 0.009      & 0.128 $< \log \mathrm{M} \leq$ 0.370 \\
$-$2.394 $\pm$ 0.020     & 9.612 $\pm$ 0.015     & 0.370 $< \log \mathrm{M} \leq$ 0.612 \\
$-$2.007 $\pm$ 0.022     & 9.375 $\pm$ 0.021     & 0.612 $< \log \mathrm{M} \leq$ 0.854 \\
$-$1.681 $\pm$ 0.024     & 9.096 $\pm$ 0.030     & 0.854 $< \log \mathrm{M} \leq$ 1.096  \\
$-$1.249 $\pm$ 0.028     & 8.623 $\pm$ 0.041      & 1.096 $< \log \mathrm{M} \leq$ 1.338   \\
$-$0.869 $\pm$ 0.033     & 8.115 $\pm$ 0.056     & 1.338 $< \log \mathrm{M} \leq$ 1.579   \\
$-$0.598 $\pm$ 0.040     & 7.688 $\pm$ 0.077     & $\log \mathrm{M} >$ 1.579   \\
\bottomrule
\end{tabular}
\end{table}

\begin{table}[h!]
\addtolength{\tabcolsep}{+9pt}
\centering
\begin{tabular}{@{}ccc@{}}
\toprule
$a$                 & $b$                  & $Validity$ $range$                        \\ \midrule
                  & Z $=$ 0.0005       &                                       \\
\midrule
$-$3.178 $\pm$ 0.023 & 9.777 $\pm$ 0.006 & $\log \mathrm{M} \leq$ 0.130                      \\
$-$2.608 $\pm$ 0.021     & 9.703 $\pm$ 0.010      & 0.130 $< \log \mathrm{M} \leq$ 0.372 \\
$-$2.415 $\pm$ 0.022     & 9.631 $\pm$ 0.016     & 0.372 $< \log \mathrm{M} \leq$ 0.614 \\
$-$2.020 $\pm$ 0.024     & 9.389 $\pm$ 0.024     & 0.614 $< \log \mathrm{M} \leq$ 0.856 \\
$-$1.683 $\pm$ 0.027     & 9.101 $\pm$ 0.033     & 0.856 $< \log \mathrm{M} \leq$ 1.340  \\
$-$1.247 $\pm$ 0.031     & 8.622 $\pm$ 0.045      & 1.098 $< \log \mathrm{M} \leq$ 1.340   \\
$-$0.866 $\pm$ 0.036     & 8.111 $\pm$ 0.062     & 1.340 $< \log \mathrm{M} \leq$ 1.582   \\
$-$0.596 $\pm$ 0.044     & 7.685 $\pm$ 0.086     & $\log \mathrm{M} >$ 1.582   \\
\hline
                  & Z $=$ 0.0008       &                                       \\
\midrule
$-$3.176 $\pm$ 0.025 & 9.791 $\pm$ 0.006 & $ \log \mathrm{M} \leq$ 0.134                      \\
$-$2.587 $\pm$ 0.023     & 9.712 $\pm$ 0.011      & 0.134 $< \log \mathrm{M} \leq$ 0.375 \\
$-$2.452 $\pm$ 0.024     & 9.661 $\pm$ 0.018     & 0.375 $< \log \mathrm{M} \leq$ 0.617 \\
$-$2.005 $\pm$ 0.026     & 9.385 $\pm$ 0.026     & 0.617 $< \log \mathrm{M} \leq$ 0.859 \\
$-$1.684 $\pm$ 0.029     & 9.110 $\pm$ 0.036     & 0.859 $< \log \mathrm{M} \leq$ 1.101  \\
$-$1.250 $\pm$ 0.033     & 8.632 $\pm$ 0.049      & 1.101 $< \log \mathrm{M} \leq$ 1.343   \\
$-$0.864 $\pm$ 0.039     & 8.114 $\pm$ 0.067     & 1.343 $< \log \mathrm{M} \leq$ 1.585   \\
$-$0.605 $\pm$ 0.036     & 7.703 $\pm$ 0.092     & $\log \mathrm{M} >$ 1.585   \\
\hline
                  & Z $=$ 0.001       &                                       \\
\midrule
$-$3.174 $\pm$ 0.026 & 9.797 $\pm$ 0.006 & $ \log \mathrm{M} \leq$ 0.135                      \\
$-$2.579 $\pm$ 0.024     & 9.716 $\pm$ 0.012      & 0.135 $< \log \mathrm{M} \leq$ 0.377 \\
$-$2.466 $\pm$ 0.025     & 9.674 $\pm$ 0.019     & 0.377 $< \log \mathrm{M} \leq$ 0.618 \\
$-$2.020 $\pm$ 0.027     & 9.398 $\pm$ 0.027     & 0.618 $< \log \mathrm{M} \leq$ 0.860 \\
$-$1.692 $\pm$ 0.030     & 9.116$\pm$ 0.037     & 0.860 $< \log \mathrm{M} \leq$ 1.101  \\
$-$1.249 $\pm$ 0.035     & 8.628 $\pm$ 0.051      & 1.101 $< \log \mathrm{M} \leq$ 1.342   \\
$-$0.868 $\pm$ 0.041     & 8.117 $\pm$ 0.070     & 1.342 $< \log \mathrm{M} \leq$ 1.584   \\
$-$0.599 $\pm$ 0.050     & 7.691 $\pm$ 0.097     & $ \log \mathrm{M} >$ 1.584   \\
\hline
                  & Z $=$ 0.0015       &                                       \\
\midrule
$-$3.169 $\pm$ 0.029 & 9.817 $\pm$ 0.007 & $\log \mathrm{M} \leq$ 0.141                      \\
$-$2.563 $\pm$ 0.027     & 9.731 $\pm$ 0.013      & 0.141 $< \log \mathrm{M} \leq$ 0.382 \\
$-$2.515 $\pm$ 0.028     & 9.713 $\pm$ 0.021     & 0.382 $< \log \mathrm{M} \leq$ 0.624 \\
$-$2.012 $\pm$ 0.031     & 9.399 $\pm$ 0.030     & 0.624 $< \log \mathrm{M} \leq$ 0.856 \\
$-$1.658 $\pm$ 0.034     & 9.094$\pm$ 0.042     & 0.865 $< \log \mathrm{M} \leq$ 1.106  \\
$-$1.244 $\pm$ 0.039     & 8.636 $\pm$ 0.058      & 1.106 $< \log \mathrm{M} \leq$ 1.347   \\
$-$0.873 $\pm$ 0.046     & 8.136 $\pm$ 0.078     & 1.347 $< \log \mathrm{M} \leq$ 1.588   \\
$-$0.629 $\pm$ 0.055     & 7.749 $\pm$ 0.108     & $\log \mathrm{M} >$ 1.588   \\
\hline
                  & Z $=$ 0.002       &                                       \\
\midrule
$-$3.177 $\pm$ 0.032 & 9.832 $\pm$ 0.008 & $\log \mathrm{M} \leq$ 0.143                      \\
$-$2.551 $\pm$ 0.029     & 9.742 $\pm$ 0.015      & 0.143 $< \log \mathrm{M} \leq$ 0.382 \\
$-$2.550 $\pm$ 0.030     & 9.742 $\pm$ 0.023     & 0.382 $< \log \mathrm{M} \leq$ 0.620 \\
$-$2.068 $\pm$ 0.033     & 9.443 $\pm$ 0.032     & 0.620 $< \log \mathrm{M} \leq$ 0.859 \\
$-$1.720 $\pm$ 0.037     & 9.144$\pm$ 0.045     & 0.859 $< \log \mathrm{M} \leq$ 1.097  \\
$-$1.260 $\pm$ 0.043     & 8.639 $\pm$ 0.062      & 1.097 $< \log \mathrm{M} \leq$ 1.336   \\
$-$0.865 $\pm$ 0.050     & 8.112 $\pm$ 0.085     & 1.336 $< \log \mathrm{M} \leq$ 1.574   \\
$-$0.627 $\pm$ 0.060     & 7.737 $\pm$ 0.116     & $ \log \mathrm{M} >$ 1.574 \\
\bottomrule
\end{tabular}
\end{table}

\begin{table}[h!]
\addtolength{\tabcolsep}{+9pt}
\centering
\begin{tabular}{@{}ccc@{}}
\toprule
$a$                 & $b$                  & $Validity$ $range$                        \\ \midrule
                  & Z $=$ 0.003       &                                       \\
\midrule
$-$3.209 $\pm$ 0.032 & 9.862 $\pm$ 0.008 & $\log \mathrm{M} \leq$ 0.150                      \\
$-$2.520 $\pm$ 0.029     & 9.759 $\pm$ 0.015      & 0.150 $< \log \mathrm{M} \leq$ 0.385 \\
$-$2.620 $\pm$ 0.030     & 9.797 $\pm$ 0.022     & 0.385 $< \log \mathrm{M} \leq$ 0.620 \\
$-$2.091 $\pm$ 0.033     & 9.469 $\pm$ 0.032     & 0.620 $< \log \mathrm{M} \leq$ 0.855 \\
$-$1.685 $\pm$ 0.037     & 9.122 $\pm$ 0.045     & 0.855 $< \log \mathrm{M} \leq$ 1.090  \\
$-$1.269 $\pm$ 0.042     & 8.669 $\pm$ 0.061      & 1.090 $< \log \mathrm{M} \leq$ 1.325   \\
$-$0.937 $\pm$ 0.049     & 8.228 $\pm$ 0.083     & 1.325 $< \log \mathrm{M} \leq$ 1.560   \\
$-$0.698 $\pm$ 0.057     & 7.856 $\pm$ 0.110     & $\log \mathrm{M} >$ 1.560\\
\hline
                  & Z $=$ 0.006       &                                       \\
\midrule
$-$3.252 $\pm$ 0.033 & 9.929 $\pm$ 0.009 & $\log \mathrm{M} \leq$ 0.164                      \\
$-$2.479 $\pm$ 0.030     & 9.802 $\pm$ 0.015      & 0.164 $< \log \mathrm{M} \leq$ 0.392 \\
$-$2.753 $\pm$ 0.031     & 9.910 $\pm$ 0.023     & 0.392 $< \log \mathrm{M} \leq$ 0.621 \\
$-$2.194 $\pm$ 0.033     & 9.562 $\pm$ 0.032     & 0.621 $< \log \mathrm{M} \leq$ 0.849 \\
$-$1.800 $\pm$ 0.037     & 9.228 $\pm$ 0.044     & 0.849 $< \log \mathrm{M} \leq$ 1.077  \\
$-$1.338 $\pm$ 0.043     & 8.730 $\pm$ 0.061      & 1.077 $< \log \mathrm{M} \leq$ 1.305   \\
$-$0.898 $\pm$ 0.051     & 8.156 $\pm$ 0.085     & 1.305 $< \log \mathrm{M} \leq$ 1.534   \\
$-$0.756 $\pm$ 0.058     & 7.938 $\pm$ 0.110     & $\log \mathrm{M} >$ 1.534   \\
\hline
                  & Z $=$ 0.008       &                                       \\
\midrule
$-$3.263 $\pm$ 0.038 & 9.695 $\pm$ 0.010 & $\log \mathrm{M} \leq$ 0.172                      \\
$-$2.485 $\pm$ 0.034     & 9.830 $\pm$ 0.017      & 0.172 $< \log \mathrm{M} \leq$ 0.398 \\
$-$2.817 $\pm$ 0.034     & 9.692 $\pm$ 0.025     & 0.398 $< \log \mathrm{M} \leq$ 0.623 \\
$-$2.226 $\pm$ 0.037     & 9.594 $\pm$ 0.036     & 0.623 $< \log \mathrm{M} \leq$ 0.849 \\
$-$1.831 $\pm$ 0.041     & 9.258$\pm$ 0.049     & 0.849 $< \log \mathrm{M} \leq$ 1.074  \\
$-$1.354 $\pm$ 0.048     & 8.746 $\pm$ 0.068      & 1.074 $< \log \mathrm{M} \leq$ 1.300   \\
$-$0.908 $\pm$ 0.057     & 8.166 $\pm$ 0.095     & 1.300525 $< \log \mathrm{M} \leq$ 1.584   \\
$-$0.801 $\pm$ 0.067     & 8.004 $\pm$ 0.125     & $\log \mathrm{M} >$ 1.525   \\
\hline
                  & Z $=$ 0.01       &                                       \\
\midrule
$-$3.256 $\pm$ 0.039 & 9.993 $\pm$ 0.011 & $\log \mathrm{M} \leq$ 0.179                      \\
$-$2.482 $\pm$ 0.035     & 9.852 $\pm$ 0.018      & 0.179 $< \log \mathrm{M} \leq$ 0.403 \\
$-$2.865 $\pm$ 0.035     & 10.007 $\pm$ 0.026     & 0.403 $< \log \mathrm{M} \leq$ 0.626 \\
$-$2.272 $\pm$ 0.038     & 9.635 $\pm$ 0.037     & 0.626 $< \log \mathrm{M} \leq$ 0.850 \\
$-$1.852 $\pm$ 0.042     & 9.278$\pm$ 0.051     & 0.850 $< \log \mathrm{M} \leq$ 1.073  \\
$-$1.351 $\pm$ 0.049     & 8.740 $\pm$ 0.070      & 1.073 $< \log \mathrm{M} \leq$ 1.297   \\
$-$0.955 $\pm$ 0.057     & 8.227 $\pm$ 0.094     & 1.297 $< \log \mathrm{M} \leq$ 1.520   \\
$-$0.819 $\pm$ 0.069     & 8.020 $\pm$ 0.129     & $\log \mathrm{M} >$ 1.520   \\
\hline
                  & Z $=$ 0.012       &                                       \\
\midrule
$-$3.266 $\pm$ 0.039 & 10.019 $\pm$ 0.011 & $\log \mathrm{M} \leq$ 0.186                      \\
$-$2.499 $\pm$ 0.035     & 9.876 $\pm$ 0.018      & 0.186 $< \log \mathrm{M} \leq$ 0.408 \\
$-$2.898 $\pm$ 0.035     & 10.039 $\pm$ 0.026     & 0.408 $< \log \mathrm{M} \leq$ 0.630 \\
$-$2.280 $\pm$ 0.038     & 9.650 $\pm$ 0.037     & 0.630 $< \log \mathrm{M} \leq$ 0.852 \\
$-$1.771 $\pm$ 0.043     & 9.216$\pm$ 0.052     & 0.852 $< \log \mathrm{M} \leq$ 1.074  \\
$-$1.387 $\pm$ 0.049     & 8.803 $\pm$ 0.069      & 1.074 $< \log \mathrm{M} \leq$ 1.296   \\
$-$1.080 $\pm$ 0.054     & 8.405 $\pm$ 0.088     & 1.296 $< \log \mathrm{M} \leq$ 1.519   \\
$-$0.779 $\pm$ 0.067     & 7.949 $\pm$ 0.125     & $\log \mathrm{M} >$ 1.519 \\
\bottomrule
\end{tabular}
\end{table}

\begin{table}[h!]
\caption{Fitting parameters of the relation between magnitude and age, $\log t = a\, \mathrm{K_s} + b$.}
\label{tab:tab15}
\addtolength{\tabcolsep}{18pt}
\centering
\begin{tabular}{@{}ccc@{}}
\toprule
$a$                 & $b$                  & $Validity$ $range$                        \\ \midrule
                  & AMR\,1       &                                       \\
\midrule
0.620     & 14.143   & $-$12.950 $<$ K$_s$ $\leq$ $-$11.750 \\
0.385     & 11.382   & $-$11.750 $<$ K$_s$ $\leq$ $-$11.350 \\
0.189     & 9.163    & $-$11.350 $<$ K$_s$ $\leq$ $-$10.950 \\
0.469     & 12.221   & $-$10.950 $<$ K$_s$ $\leq$ $-$10.550 \\
0.338     & 10.844   & $-$10.550 $<$ K$_s$ $\leq$ $-$10.250 \\
1.247     & 20.164   & $-$10.250 $<$ K$_s$ $\leq$ $-$9.850  \\
0.442     & 12.227   & $-$9.850  $<$ K$_s$ $\leq$ $-$9.450 \\
0.991     & 17.421   & $-$9.450  $<$ K$_s$ $\leq$ $-$8.550 \\
0.199     & 10.650   & $-$8.550  $<$ K$_s$ $\leq$ $-$8.130 \\
0.786     & 15.421   & $-$8.130  $<$ K$_s$ $\leq$ $-$7.630 \\
0.168     & 10.708   & $-$7.630  $<$ K$_s$ $\leq$ $-$7.260 \\
0.426     & 12.577   & $-$7.260  $<$ K$_s$ $\leq$ $-$7.050 \\
2.564     & 13.550   & $-$7.050  $<$ K$_s$ $\leq$ $-$6.350 \\
0.694     & 14.376   & $-$6.350  $<$ K$_s$ $\leq$ $-$6.240 \\
0.818     & 14.376   & $-$6.240  $<$ K$_s$ $\leq$ $-$4.140 \\
\hline
                  & AMR\,2       &                                       \\
\midrule
0.620     & 14.143  & $-$12.950 $<$ K$_s$ $\leq$ $-$11.750 \\
0.385     & 11.382  & $-$11.750 $<$ K$_s$ $\leq$ $-$11.350 \\
0.189     & 9.163   & $-$11.350 $<$ K$_s$ $\leq$ $-$10.950 \\
0.469     & 12.221  & $-$10.950 $<$ K$_s$ $\leq$ $-$10.550 \\
0.338     & 10.844  & $-$10.550 $<$ K$_s$ $\leq$ $-$10.250 \\
1.247     & 20.164  & $-$10.250 $<$ K$_s$ $\leq$ $-$9.850  \\
0.442     & 12.227  & $-$9.850  $<$ K$_s$ $\leq$ $-$9.450  \\
0.991     & 17.421   & $-$9.450 $<$ K$_s$ $\leq$ $-$8.550   \\
0.699     & 14.920   & $-$8.550 $<$ K$_s$ $\leq$ $-$8.250   \\
0.183     & 10.663   & $-$8.250 $<$ K$_s$ $\leq$ $-$7.950 \\
0.978     & 16.663   & $-$7.950 $<$ K$_s$ $\leq$ $-$7.850 \\
0.622     & 14.190   & $-$7.850 $<$ K$_s$ $\leq$ $-$7.050 \\
0.143     & 10.813   & $-$7.050 $<$ K$_s$ $\leq$ $-$6.570  \\
0.365     & 12.269   & $-$6.570 $<$ K$_s$ $\leq$ $-$6.380 \\
0.706     & 14.446   & $-$6.380 $<$ K$_s$ $\leq$ $-$6.270 \\
0.818     & 15.147   & $-$6.270 $<$ K$_s$ $\leq$ $-$4.140 \\
\bottomrule
\end{tabular}
\end{table}

\begin{table}[h!]
\addtolength{\tabcolsep}{+18pt}
\centering
\begin{tabular}{@{}ccc@{}}
\toprule
$a$                 & $b$                  & $Validity$ $range$                        \\ \midrule
                  & AMR\,3       &                                       \\
\midrule
0.620    & 14.143   & $-$12.950 $<$ K$_s$ $\leq$ $-$11.750 \\
0.385     & 11.382  & $-$11.750 $<$ K$_s$ $\leq$ $-$11.350 \\
0.189     & 9.163   & $-$11.350 $<$ K$_s$ $\leq$ $-$10.950 \\
0.469     & 12.221  & $-$10.950 $<$ K$_s$ $\leq$ $-$10.550  \\
0.338     & 10.844  & $-$10.550 $<$ K$_s$ $\leq$ $-$10.250  \\
1.247     & 20.164  & $-$10.250 $<$ K$_s$ $\leq$ $-$9.850   \\
0.442     & 12.227  & $-$9.850 $<$ K$_s$ $\leq$ $-$9.450   \\
1.017     & 17.666  & $-$9.450 $<$ K$_s$ $\leq$ $-$8.850   \\
0.563     & 13.649  & $-$8.850 $<$ K$_s$ $\leq$ $-$8.580   \\
0.278     & 11.202  & $-$8.580 $<$ K$_s$ $\leq$ $-$8.280 \\
0.935     & 16.643  & $-$8.280 $<$ K$_s$ $\leq$ $-$7.980 \\
0.687     & 14.660  & $-$7.980 $<$ K$_s$ $\leq$ $-$7.580 \\
0.149     & 10.584  & $-$7.580 $<$ K$_s$ $\leq$ $-$7.150  \\
0.555     & 13.488  & $-$7.150  $<$ K$_s$ $\leq$ $-$6.850 \\
0.274     & 11.562  & $-$6.850 $<$ K$_s$ $\leq$ $-$6.600 \\
0.816     & 15.143  & $-$6.600 $<$ K$_s$ $\leq$ $-$4.140 \\
\bottomrule
\end{tabular}
\end{table}

\begin{table}[h!]
\caption{Fitting parameters of the relation between age and birth mass, $\log t = a \log \mathrm{(M/M_{\sun})} + b$, based on BaSTI model.}
\label{tab:tab17}
\addtolength{\tabcolsep}{+9pt}
\centering
\begin{tabular}{@{}ccc@{}}
\toprule
$a$                 & $b$                  & $Validity$ $range$                        \\ \midrule
                  & Z $=$ 0.003       &                                       \\
\midrule
$-$3.602   & 9.859   & $\log \mathrm{M} \leq$ 0.057                      \\
$-$3.273   & 9.840    & 0.057 $< \log \mathrm{M} \leq$ 0.149 \\
$-$2.779   & 9.767    & 0.149 $< \log \mathrm{M} \leq$ 0.212 \\
$-$2.090   & 9.620    & 0.212 $< \log \mathrm{M} \leq$ 0.296 \\
$-$2.677   & 9.794    & 0.296 $< \log \mathrm{M} \leq$ 0.521  \\
$-$2.406   & 9.653    & 0.521 $< \log \mathrm{M} \leq$ 0.687   \\
$-$2.184   & 9.500    & 0.687 $< \log \mathrm{M} \leq$ 0.824   \\
$-$1.926   & 9.288    & 0.824 $< \log \mathrm{M} \leq$ 0.940 \\
$-$1.685   & 9.122    & 0.640 $< \log \mathrm{M} \leq$ 1.090 \\
$-$1.269   & 8.669    & 1.090 $< \log \mathrm{M} \leq$ 1.325  \\
$-$0.937   & 8.288    & 1.325 $< \log \mathrm{M} \leq$ 1.560   \\
$-$0.698   & 7.856    & $\log \mathrm{M} >$ 1.560   \\
\bottomrule
\end{tabular}
\end{table}

\begin{table}[h!]
\addtolength{\tabcolsep}{+9pt}
\caption{Fitting parameters of the relation between pulsation duration and birth mass, log ($\delta t/t$) $=$ $D$ $+$ $\Sigma_{i=1}^{4} a_i$exp[$-$(log M[M$_\sun$] $-$ $b_i$)$^2$/$c_i ^2$].}
\centering
\label{tab:tab18}
\begin{tabular}{@{}ccccc@{}}
\toprule
$D$      & $i$ & $a$             & $b$     & $c$     \\ \midrule
       &   & Z $=$ 0.0001 &       &       \\ \midrule
$-$3.727 & 1 & 2.895         & 1.787 & 0.087 \\
       & 2 & 0.773         & 0.569 & 0.225 \\
       & 3 & $-$0.626        & 1.226 & 0.193 \\
       & 4 & $-$0.738        & 0.080 & 0.103 \\
\hline
       &   & Z $=$ 0.0003 &       &       \\ \midrule
$-$6.926 & 1 & 3.848         & 0.620 & 0.876 \\
       & 2 & 0.513         & 0.248 & 0.070 \\
       & 3 & 4.833        & 1.821 & 0.432 \\
       & 4 & 342.189        & $-$1.779 & 0.009 \\
\hline
       &   & Z $=$ 0.0005 &       &       \\ \midrule
$-$5.054 & 1 & 1.810         & 0.586 & 0.211 \\
       & 2 & 1.882         & 0.248 & 0.159 \\
       & 3 & 3.526        & 1.666 & 0.271 \\
       & 4 & 1.496        & 1.169 & 0.197 \\
\hline
       &   & Z $=$ 0.0008 &       &       \\ \midrule
$-$4.489 & 1 & 1.255         & 0.587 & 0.169 \\
       & 2 & 0.636         & 1.138 & 0.088 \\
       & 3 & 3.410        & 1.714 & 0.389 \\
       & 4 & 1.483        & 0.266 & 0.155 \\
\hline
       &   & Z $=$ 0.001 &       &       \\ \midrule
$-$4.570 & 1 & 1.336         & 0.589 & 0.166 \\
       & 2 & 0.0949         & 1.148 & 0.099 \\
       & 3 & 3.481        & 1.513 & 0.252 \\
       & 4 & 1.610        & 0.265 & 0.160 \\
\hline
       &   & Z $=$ 0.0015 &       &       \\ \midrule
$-$6.101 & 1 & 3.100         & 1.596 & 0.474 \\
       & 2 & 3.436         & 0.383 & 0.440 \\
       & 3 & 3.283        & 1.150 & 0.277 \\
       & 4 & 0.759        & 0.697 & 0.092 \\
\hline
       &   & Z $=$ 0.002 &       &       \\ \midrule
$-$6.154 & 1 & 16.347         & 2.288 & 0.344 \\
       & 2 & 7.870         & 0.872 & 0.717 \\
       & 3 & $-$5.799        & 0.864 & 0.382 \\
       & 4 & $-$0.986        & 0.465 & 0.117 \\
\hline
       &   & Z $=$ 0.003 &       &       \\ \midrule
$-$5.323 & 1 & 1.531         & 1.826 & 0.151 \\
       & 2 & 3.421         & 1.337 & 0.405 \\
       & 3 & 2.566        & 0.306 & 0.290 \\
       & 4 & 1.288        & 0.647 & 0.127 \\
\bottomrule
\end{tabular}
\end{table}

\begin{table}[h!]
\addtolength{\tabcolsep}{+9pt}
\centering
\begin{tabular}{@{}ccccc@{}}
\toprule
$D$      & $i$ & $a$             & $b$     & $c$     \\ \midrule
       &   & Z $=$ 0.006 &       &       \\ \midrule
$-$5.809 & 1 & 3.149         & 0.336 & 0.366 \\
       & 2 & 4.238         & 1.290 & 0.444 \\
       & 3 & 0.676        & 0.662 & 0.095\\
       & 4 & 2.149        & 1.575 & 0.116 \\
\hline
       &   & Z $=$ 0.008 &       &       \\ \midrule
$-$10.000 & 1 & 7.386         & 0.375 & 0.676 \\
       & 2 & 0.637         & 0.695 & 0.088 \\
       & 3 & 7.380        & 1.538 & 0.289 \\
       & 4 & 5.073        & 1.133 & 0.261 \\
\hline
       &   & Z $=$ 0.01 &       &       \\ \midrule
$-$8.225 & 1 & 3.056         & 1.062 & 0.209 \\
       & 2 & 0.697         & 0.711 & 0.095 \\
       & 3 & 6.275        & 1.444 & 0.333 \\
       & 4 & 5.593        & 0.385 & 0.598 \\
\hline
       &   & Z $=$ 0.012 &       &       \\ \midrule
$-$6.060 & 1 & 4.495         & 1.346 & 0.436 \\
       & 2 & 3.393         & 0.372 & 0.420 \\
       & 3 & 1.027        & 1.077 & 0.122 \\
       & 4 & 0.655        & 0.708 & 0.106 \\
\bottomrule
\end{tabular}
\end{table}

\end{document}